\documentclass[12pt, a4paper]{book}		
\usepackage[utf8x]{inputenc}

\usepackage[T1]{fontenc}

\usepackage[english]{babel}
\usepackage{amsfonts}
\usepackage{graphicx}				
\usepackage{epsf}
\usepackage{verbatim}
\usepackage{amsfonts}
\usepackage{nicefrac}
\usepackage{slashed}
\usepackage{amssymb}
\usepackage{amsmath}
\usepackage{mathtools}
\usepackage{cancel}
\usepackage{bbold}
\usepackage{amssymb}
\usepackage{titlesec}
\usepackage{titletoc}
\usepackage[titletoc]{appendix}
 \usepackage{pdfpages}
 \usepackage{hyperref}		
 \usepackage{mciteplus}
 \usepackage{tocloft}
 \usepackage{microtype}
 \usepackage{scrextend}%

\usepackage[labelsep=period,justification=RaggedRight, tableposition=top,font=small,labelfont=bf]{caption}
\usepackage[caption=false]{subfig}
\usepackage{hyperref}

\usepackage{mciteplus}
\usepackage[comma,square,numbers,compress]{natbib}

\usepackage{chngcntr}
\hypersetup{%
  pdfauthor={Jarkko Peuron},%
  pdftitle={Quasiparticle properties of nonequilibrium gluon plasma}
  pdfsubject={PhD thesis}
}

\newcommand{\dtmax}{\Delta t_{\text{max}}}
\newcommand{\wplas}{\omega_{\mrm{pl}}}
\newcommand{\HTL}{\mrm{HTL}}
\newcommand{\ud}{\mathrm{d}}
\newcommand{\nr}[1]{(\ref{#1})} 
\newcommand{\der}{\mathrm{d}}  % Derivaatta-d

\newcommand{\pInit}{p_0}
\newcommand{\nc}{N_\mathrm{c}}

\newcommand{\ah}{\mathrm{ah}}
\newcommand{\cl}{\mrm{cl}}
\newcommand{\mrm}{\mathrm}

\renewcommand\Re{\operatorname{Re}}

\newcommand\tr{\operatorname{Tr}}

\newcommand\PO{\operatorname{P}}  % Path ordering

\newcommand{\I}{\hat{\imath}}
\newcommand{\J}{\hat{\jmath}}
\newcommand{\oik}{\right)}
\newcommand{\vas}{\left(}
\newcommand{\Tr}{\mathrm{Tr}}
\newcommand{\nicehalf}{\nicefrac{1}{2}}
\newcommand{\half}{\frac{1}{2}}
\newcommand{\equref}[1]{Eq.~(\ref{#1})}

\newcommand{\tcent}{\bar{t}}
\newcommand{\fig}{Fig.~}

\newcommand{\eq}{Eq.~}
\newcommand{\se}{Sec.~}

\newcommand{\re}{Ref.~}
\newcommand{\res}{Refs.~}
\newcommand{\eqs}{Eqs.~}
\newcommand{\ch}{Chapter~}

\newcommand{\ktt}{k_\perp}
\newcommand{\Q}{Q}
\newcommand{\Qeff}{Q_{\mathrm{eff}}}
\newcommand{\neff}{n_0^{\mathrm{eff}}}

\newcommand{\xx}{{\mathbf{x}}}

\newcommand{\commutator}[2]{\left[#1,#2\right]}

\newcommand{\mbf}{\mathbf}
\newcommand{\tpert}{t_{\text{pert}}}
\newcommand{\uj}{\mrm{j}}
\newcommand{\rel}{\mrm{rel}}
\newcommand{\fit}{\mrm{fit}}

\newcommand{\ylatila}{\mbox{}\\ \mbox{}\\\mbox{}\\ \mbox{}\\ \mbox{}\\ \mbox{}\\ \mbox{}\\ \mbox{}\\ }

\begin{document}
\pagestyle{empty}

%\vspace*{-40mm}

\centerline{DEPARTMENT OF PHYSICS}
\centerline{UNIVERSITY OF JYV\"ASKYL\"A}
\centerline{RESEARCH REPORT No. 7/2018 }

\vspace{25mm}

\centerline{\bf QUASIPARTICLE PROPERTIES OF NONEQUILIBRIUM GLUON PLASMA  }
\vspace{13mm}

\centerline{\bf BY}
\centerline{\bf JARKKO PEURON}

\vspace{13mm}

\centerline{Academic Dissertation}
\centerline{for the Degree of}
\centerline{Doctor of Philosophy}

\vspace{13mm}

\centerline{To be presented, by permission of the}
\centerline{Faculty of Mathematics and Natural Sciences}
\centerline{of the University of Jyv\"askyl\"a,}
\centerline{for public examination in Auditorium YAA303 of the}
\centerline{University of Jyv\"askyl\"a on August 7th, 2018}
\centerline{at 12 o'clock noon}

\vspace{20mm}

\begin{figure}[!h]
\center
\includegraphics[height=27mm]{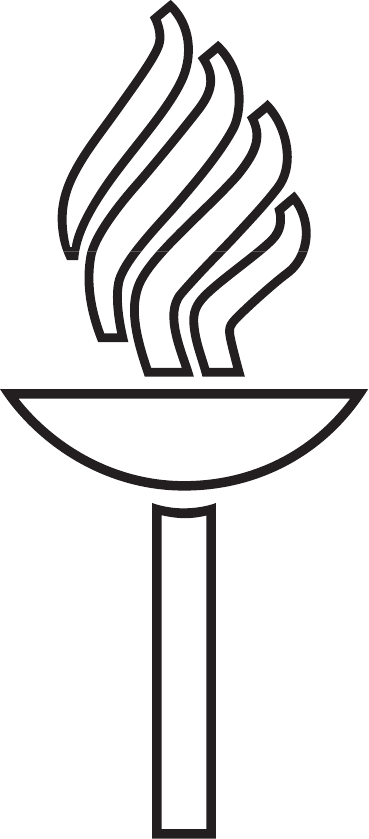}
\end{figure}
%\centerline{\picture{19mm}{50mm}{bwteksti.EPS}}
\centerline{Jyväskylä, Finland}
\centerline{August 2018}
\pagebreak

%DRAFT \today

\newpage
\null
\vfill
\noindent
ISBN: 978-951-39-7498-5 (Printed version) \\
ISBN: 978-951-39-7499-2 (Electronic version)\\
ISSN: 0075-465X

\frontmatter
% näytä sivunumerointi myös tyhjillä sivuilla
\KOMAoptions{cleardoublepage=plain}  
\chapter*{Preface}
The work presented in this thesis has been carried out at University of Jyväskylä and partially at CERN theory division between 2014 and 2018. I would like to thank CERN-TH for hospitality and providing a scientifically stimulating atmosphere during my various visits over the years. 

First and foremost I would like to express my gratitude to Prof. Tuomas Lappi for his friendly and competent instruction and advice over the years. Prof. Aleksi Kurkela also deserves special thanks for his guidance. I would also like to thank Prof. Kari J. Eskola for running the research group flawlessly together with Tuomas. I  also acknowledge Dr. Kirill Boguslavski for smooth and efficient collaboration. I am grateful to Mr. David Müller for various valuable discussions concerning the discretization of the fluctuations of the classical fields. 

My fellow current and former PhD students deserve a special thanks, in particular Dr. Heikki Mäntysaari and Ms. Andrecia Ramnath. I would  like to thank all my friends and colleagues (current and past) at the Department of Physics, with whom I have had countless, sometimes even physics related, discussions.

I would like to thank Prof. Jürgen Berges and Dr. Björn Schenke for reviewing the manuscript of this thesis and providing useful comments and Prof. Anders Tranberg for promising to be my opponent.

I gratefully acknowledge the financial support from Jenny and Antti Wihuri Foundation and support for travel from Magnus Ehrnrooth foundation. I also wish to acknowledge  CSC – IT Center for Science, Finland, for providing computational resources to carry out this work. 

Last but not least I would like to thank my family and Johanna for unconditional love and support. 

\vfill
\noindent Jyväskylä, July 2018 \\
Jarkko Peuron

%\newpage

%\begin{titlepage}
%\cleardoublepage

\chapter*{Abstract}

We apply classical gluodynamics to early stages of ultrarelativistic heavy-ion collisions. We start by giving a brief overview of QCD. Then we proceed to the space-time evolution of ultrarelativistic heavy-ion collisions in the color glass condensate framework and go through the basics of real-time gluodynamics on the lattice in the temporal gauge. 

We study the plasmon mass scale in three- and two-dimensional systems by comparing three different methods to measure the mass scale. The methods are a formula which can be derived from Hard Thermal Loop effective theory at leading order (HTL), the effective dispersion relation (DR) and measurement of the plasma oscillation frequency triggered by the introduction of a uniform electric field (UE) into the system. We observe that in both systems the plasmon mass scale decreases like a power law after an occupation number dependent initial transient time. In 3 dimensions we observe the power law to be $\omega_{pl}^2 \sim t^{\nicefrac{-2}{7}},$ which is predicted by the literature. In 2 dimensions the observed power law is $\omega_{pl}^2 \sim t^{\nicefrac{-1}{3}}.$ In both cases the UE and HTL methods are in rough agreement, and in the three-dimensional case the two agree in the continuum limit. 

As a second way to study the quasiparticle properties, we derive, implement and test an algorithm which can be used to simulate linearized fluctuations on top of the classical background. The algorithm is derived by requiring conservation of Gauss' law and gauge invariance. We then apply the algorithm to spectral properties of overoccupied gluodynamics using linear response theory. We establish the existence of transverse and longitudinal quasiparticles by extracting their spectral functions. We also extract the dispersion relation, effective mass, plasmon mass and damping rate of the quasiparticles. Our results are consistent with the HTL effective theory, but we also observe effects beyond leading order HTL.

\ylatila

\begin{tabular}{ll}

{\bf Author} & Jarkko Peuron \\ 

 & Department of Physics \\ 

 & University of Jyväskylä \\ 

 & Finland \\
 
 & \\ 

{\bf Supervisor} & Prof. Tuomas Lappi \\ 

 & Department of Physics \\ 
 
 & University of Jyväskylä \\ 
 
 & Finland \\ 
 & \\
 
{\bf Reviewers}  & Prof. Jürgen Berges \\ 
 
 & Institut für Theoretische Physik \\ 
 
 & University of Heidelberg\\ 
 
 & Germany \\ 
 
 &  \\ 

& Dr. Björn Schenke \\ 
 
 & Nuclear Theory Group  \\ 
 
 & Brookhaven National Laboratory\\ 
 
 & Upton, New York \\ 
 
 & USA \\ 
 
\\
 
{\bf Opponent} & Prof. Anders Tranberg\\ 
 
 & Department of Mathematics and Physics \\ 
 
 & University of Stavanger \\ 
 
 & Norway \\ 
 
\end{tabular} 

%\end{titlepage}

%\setcounter{page}{1}
%\pagenumbering{roman}

\chapter*{List of Publications}
\vspace*{-0.15cm}
This PhD thesis consists of an introduction and the following publications:
\begin{enumerate} 
\item[{\bf \cite{Kurkela:2016mhu}}]
A.~Kurkela, T.~Lappi, J.~Peuron, {\it {Time evolution of linearized gauge
  field fluctuations on a real-time lattice}},
  \href{http://dx.doi.org/10.1140/epjc/s10052-016-4523-9}{{\em Eur. Phys. J.}
  {\bf C76} (2016)~no.~12 688} [\href{http://arXiv.org/abs/1610.01355}{{\tt
  arXiv:1610.01355 [hep-lat]}}].
 \item[\bf \cite{Lappi:2016ato}]
T.~Lappi, J.~Peuron, {\it {Plasmon mass scale in classical nonequilibrium
  gauge theory}},  \href{http://dx.doi.org/10.1103/PhysRevD.95.014025}{{\em
  Phys. Rev. D} {\bf 95} (2017) 014025}
  [\href{http://arXiv.org/abs/1610.03711}{{\tt arXiv:1610.03711 [hep-ph]}}].
\item[{\bf \cite{Lappi:2017ckt}}]
T.~Lappi, J.~Peuron, {\it Plasmon mass scale in two dimensional classical nonequilibrium gauge theory},
\href{https://doi.org/10.1103/PhysRevD.97.034017}{{\em Phys. Rev.}
  {\bf D97} (2018)~no.3,~034017}  [\href{https://arxiv.org/abs/1712.02194}{{\tt
  arXiv:1712.02194 [hep-lat]}}] 
\item[{\bf \cite{Boguslavski:2018beu}}]
K. Boguslavski, A. Kurkela, T.~Lappi, J.~Peuron, {\it Spectral function for overoccupied gluodynamics from real-time lattice simulations}, 
\href{https://doi.org/10.1103/PhysRevD.98.014006}{{\em Phys. Rev.}
  {\bf D98} (2018) 014006} 
[\href{https://arxiv.org/abs/1804.01966}{{\tt
  arXiv:1804.01966 [hep-ph]}}] 
\end{enumerate}

The author did all the numerical computations in publications \cite{Kurkela:2016mhu,Lappi:2016ato,Lappi:2017ckt}, and wrote the original draft of papers \cite{Lappi:2016ato,Lappi:2017ckt}. The author participated in development of methods and writing of the publication \cite{Boguslavski:2018beu}.
\pagebreak

%\newpage
%\phantom{Tyhjaa}
%\newpage

%\pagenumbering{arabic}

\cleardoublepage

% Ei sivunumeroa viimeiselle tyhjälle sivulle sisällysluettelon jälkeen
\KOMAoptions{cleardoublepage=empty}
\setcounter{tocdepth}{1}
\tableofcontents
\pagestyle{plain}

\mainmatter

\pagenumbering{arabic}
\chapter{Introduction}
\label{intro}
According to our current understanding, all natural phenomena are explained by four fundamental interactions: the gravitational interaction, the weak interaction, the electromagnetic interaction and the strong interaction. The standard model of particle physics describes three of these fundamental interactions (the electromagnetic, weak  and strong interactions). Two of these, the electromagnetic and weak interactions are unified into electroweak interaction  \cite{Glashow:1961tr,Weinberg:1967tq,Salam:1968rm}. Unlike electromagnetism, which has only one charge, the strong interaction has three distinct charges called colors. The only elementary particles charged under color are spin $\nicefrac{1}{2}$ quarks, and the force mediators, gluons, with spin 1. The strong interaction is the dominant interaction at the length scales of the atomic nucleus. It binds quarks into hadrons and the residual strong interaction between color neutral nucleons prevents the atomic nucleus from dissociating in spite of the Coulomb repulsion between the nucleons. Unlike other interactions, the force exerted on a static quark-antiquark pair by the strong interaction does not vanish when the separation between the two quarks increases. To a good approximation, over long distances the force binding the two quarks together is constant. Due to this, it is energetically more favourable to create a new pair of particles when separating two strongly interacting particles. This provides an explanation for the observation that we do not see any free strongly interacting particles at everyday energy scales. Instead, they are always bound to composite particles forming a color neutral state. This phenomenon is called confinement. Mathematically the theory of strong interaction is Quantum ChromoDynamics (QCD) \cite{Fritzsch:1973pi,Greenberg:1964pe,Han:1965pf,Yang:1954ek}. QCD is a nonabelian gauge theory, which is symmetric under local $SU(3)$ gauge transformations. The nonabelian nature of strong interactions manifests itself as gluonic self-interactions, which makes understanding QCD a tremendous task. 

%One of the most striking features of QCD is the running of the QCD coupling \textbf{Viitteitä}, leading into asymptotic freedom \cite{Gross:1973id,Politzer:1973fx}. At higher energies the QCD coupling becomes weaker, and in this regime perturbation theory can be applied to understand QCD, leading into perturbative QCD (pQCD). Based on asymptotic freedom, one might expect that at high energies the atomic nucleus actually does dissociate when the coupling becomes weak enough. At this point a phase transition happens, and instead of observing hadrons, one observes that quarks and gluons become the relevant degrees of freedom. This has also been experimentally confirmed at RHIC and LHC \cite{Arsene:2004fa,Adams:2005dq,Adcox:2004mh,Back:2004je}.

One of the striking predictions of QCD is that at high energies we expect to observe deconfined state of matter \cite{Hagedorn:1965st,Collins:1974ky,Cabibbo:1975ig} consisting of free quarks and gluons. This matter is also known as Quark-Gluon Plasma (QGP).  It has been shown numerically that there indeed is a phase transition from the confined to the deconfined phase \cite{Creutz:1980zw,Creutz:1980wj,Susskind:1979up,Kuti:1980gh,McLerran:1981pb,Engels:1980ty}. The existence of QGP has also been verified experimentally \cite{Arsene:2004fa,Adams:2005dq,Adcox:2004mh,Back:2004je}.  Thus at high energies the atomic nucleus ``melts'', and instead of hadrons, the appropriate degrees of freedom are the individual partons, quarks and gluons. The deconfinement of matter is a consequence of asymptotic freedom  \cite{Gross:1973id,Politzer:1973fx}, which  predicts that the strong interaction grows weaker at higher energy scales. In the high energy (weak coupling) limit  perturbation theory becomes applicable, leading to perturbative QCD (pQCD) \cite{Brock:1993sz}. However, to fully understand QCD one also needs tools which work in the nonperturbative regime. Currently the most established nonperturbative method is the lattice formulation of QCD \cite{Wilson:1974sk}. Lattice discretization renders the QCD path integral calculable, allowing one to study thermodynamical properties of QCD nonperturbatively from first principles. 
%
%However, lattice QCD fails at \textbf{At what regime, finite chemical potential?}. The most recently developed tool to study QCD in the nonperturbative regime is the Ads/CFT correspondence. \textbf{Mieti mitä sanot ADSCFT hommista, vai sanotko mitään?}

%One of the predictions of QCD is that at high energies we should observe deconfined state of matter \cite{Hagedorn:1965st,Collins:1974ky,Cabibbo:1975ig} also known as Quark-Gluon Plasma (QGP). Numerically using lattice QCD \cite{Wilson:1974sk} it has been shown that there indeed is a phase transition from confined to deconfined phase \cite{Creutz:1980zw,Creutz:1980wj,Susskind:1979up,Kuti:1980gh,McLerran:1981pb,Engels:1980ty}. Existence of QGP has also been verified experimentally \cite{Arsene:2004fa,Adams:2005dq,Adcox:2004mh,Back:2004je}. % and it seems that its source is  Color Glass Condensate (CGC)\cite{Gyulassy:2004zy}.

%QCD is a gauge field theory which describes one of the four fundamental interactions interaction between quarks and gluons. QCD shares certain similarities with QED, which describes electromagnetism. However, one of the most striking differences between the two theories is, that in QCD the gluons, the mediators of the strong interaction, interact with each other. 
\section{Quantum chromodynamics}
In this section we give a brief introduction to QCD. For a more complete introduction to QCD we refer the reader to the following books: \cite{Muta:1987mz,Field:1989uq,Ellis:1991qj,Smilga:2001ck} and articles and lecture notes \cite{Brock:1993sz,Skands:2012ts}. 

QCD is a $SU(N)$ (with $N=3$) gauge theory, which is invariant under local gauge transformations. The conserved quantity corresponding to this gauge symmetry is the color charge. The theory is defined by the Lagrangian
\begin{equation}
\label{eq:QCDaction}
\mathcal{L}_{QCD} = \sum_q \left(i \bar{\psi}_q \gamma^\mu D_\mu \psi_q - m_q \bar{\psi}_q \psi \right) - \dfrac{1}{2} \mathrm{Tr} \vas F_{\mu \nu} F^{\mu \nu} \oik,
\end{equation} 
where the sum over $q$ runs over 6 quark flavors.  The mass of the quark of flavor $q$ is given by $m_q$ and $\gamma^\mu$ is the Dirac gamma matrix.
The field strength tensor is defined as 
\begin{equation}
F_{\mu \nu} = \partial_\mu A_\nu - \partial_\nu A_\mu + i g \left[A_\mu , A_\nu \right],
\end{equation} 
with the gluon field being denoted by  $A_\mu$. Thus the field strength tensor contains gluonic self interactions. 
In the QCD Lagrangian (\ref{eq:QCDaction}) the fermion field $\psi_q$ is a four component spinor describing quarks of flavor q.  The covariant derivative is defined as 
\begin{equation}
D_\mu = \partial_\mu + i g A_\mu.
\end{equation}
Due to the fact that quarks and gluons are also charged under color, the fields in \eq (\ref{eq:QCDaction}) are not only Lorentz 4-vectors (or tensors) but they are also vectors in color space.  The color structure of the gluon field (and that of the quark fields and the field strength tensor) is the following
\begin{equation}
\label{eq:gluoncolor}
A_\mu = A_\mu^a t^a.
\end{equation} 
The trace appearing in \eq (\ref{eq:QCDaction}) is taken over the color structure of the field strength tensor. 

The matrices $t^a$ appearing in \equref{eq:gluoncolor} are generators of $SU(N)$ in the fundamental representation and they are elements of $\mathfrak{su(n)}$ (the Lie algebra of $SU(N)$). For $N=3$ the $t$ matrices are the Gell-Mann matrices divided by two, and for $N=2$ they are the Pauli matrices divided by two. 
The generator matrices obey the following trace relations
\begin{align}
\label{eq:2trace}
\mathrm{Tr}\left(t^a t^b \right) &= \dfrac{1}{2}\delta^{ab} \\
\label{eq:3trace}
\mathrm{Tr}\left(t^a t^b t^c\right) & = \dfrac{1}{4}\left( d^{abc} + i f^{abc}\right),
\end{align}
The structure constants of the group are denoted by $f^{abc}$, and they are totally antisymmetric under exchange of indices. 
The generator matrices also obey the commutation relation
\begin{equation}
\left[t^a, t^b\right] = i f^{abc} t^c.
\end{equation}
The $d^{abc}$ are called the symmetric structure constants and they are defined by the relation
\begin{equation}
\{ t^a , t^b \} = \dfrac{1}{N}\delta^{ab} + d^{abc} t^c.
\end{equation}
In particular for $N=2, f^{abc} = \varepsilon^{abc}$, with  $\varepsilon^{abc}$ being the Levi-Civita symbol and the symmetric structure constants vanish, i.e. $d^{abc} = 0.$ 
The generator matrices also obey a Fierz identity
\begin{equation}
\label{eq:Fierz}
\left(t^a\right)_{ij}\left(t^a\right)_{kl} = \dfrac{1}{2} \left(\delta_{il} \delta_{jk} - \dfrac{1}{N}\delta_{ij}\delta_{kl}\right).
\end{equation}

%The color structure of the covariant derivative is 
%\begin{equation}
%D_\mu^{ij} = \partial_\mu \mathbb{1}^{ij}_{N \times N} + ig A_\mu^{ij},
%\end{equation}
%where the unit matrix and gauge field are matrices in color space.
The covariant derivative and the field strength tensor are related by the identity
\begin{equation}
F_{\mu \nu} = \dfrac{-i}{g} \left[D_\mu , D_\nu \right],
\end{equation}
which can be verified by a straightforward computation. 

The local gauge symmetry of the theory manifests itself as an invariance under local color rotations. The local gauge transformations are of the form
\begin{equation}
V\left(x\right) = e^{i \alpha^a\left(x\right) t^a},
\end{equation}
where $\alpha^a\left(x\right)$ are  arbitrary real functions, and $V\left(x\right)$ is an $SU(N)$ matrix. The quark field transforms as 
\begin{equation}
\psi^\prime\left(x\right) = V\left(x\right) \psi\left(x\right),
\end{equation}
where $\psi^\prime$ is the gauge transformed quark spinor. The gauge transformation of the gluon field is given by 
\begin{equation}
A_\mu^\prime\left(x\right) = V\left(x\right) A_\mu\left(x\right) V^\dagger\left(x\right) + \dfrac{i}{g} \left(\partial_\mu V\left(x\right)\right) V^{-1}\left(x\right).
\end{equation}
Consequently the transformation law for the covariant derivative is 
\begin{equation}
D_\mu^\prime\left(x\right) = V\left(x\right) D_\mu\left(x\right) V^{-1}\left(x\right).
\end{equation}
And similarly for the field strength tensor 
\begin{equation}
F_{\mu \nu}^\prime\left(x\right) = V\left(x\right) F_{\mu \nu }\left(x\right) V^{-1}\left(x\right).
\end{equation}

Armed with these definitions it is straightforward to verify that the Lagrangian (\ref{eq:QCDaction}) is invariant under local gauge transformations. Actually the gauge and quark terms are also gauge invariant individually. This means that the pure Yang-Mills Lagrangian 
\begin{equation}
\label{eq:YMLagrange}
\mathcal{L}_{YM} = - \dfrac{1}{2} \mathrm{Tr} \left(F_{\mu \nu} F^{\mu \nu} \right),
\end{equation}
is gauge invariant due to the cyclicity of the trace and we can also consider a theory without quarks. From now on we will exclusively focus on gluonic interactions. 

Similarly as in classical electrodynamics, the (color) electric and magnetic fields are defined as 
\begin{align}
\label{eq:electricfielddef}
E^i = F^{i0} \\
B^i = \varepsilon^{ijk} \dfrac{F_{jk}}{2}.
\end{align}
This allows us to write the Yang-Mills Lagrangian in a more simple form 
\begin{equation}
\mathcal{L}_{YM} = -\dfrac{1}{2}\left(E^2-B^2\right),
\end{equation}
where we have dropped out the color indices for brevity. This is somewhat analogous to classical mechanics: The role of the kinetic energy in this system is played by the electric fields, and the potential energy is represented by the magnetic fields. 

The classical Euler-Lagrange equations of motion for pure glue QCD without sources read
\begin{equation}
\label{eq:YMEOM}
\left[ D^\mu , F_{\mu \nu} \right] = 0.
\end{equation}
This equation encodes information of two familiar equations from classical electrodynamics. By choosing $\nu = 0$ one obtains the nonabelian Gauss's law
\begin{equation}
\partial_i E_i + i g \left[A_i , E_i \right] = 0,
\end{equation}
which takes care of the color charge conservation. The spatial components ($\nu = 1,2,3$) give the nonabelian counterpart of Ampere's law. 

\subsection{Hamiltonian formalism}
So far we have derived the equation of motion in the Lagrangian formalism. However, one can also do Hamiltonian field theory. In the following we show how classical Yang-Mills equations of motion are derived in the Hamiltonian formalism. 
In order to obtain the Yang-Mills Hamiltonian, we have to find the canonical conjugate momentum to the gluon field. It can be obtained as 
\begin{equation}
\pi^{\mu b} = \dfrac{\partial \mathcal{L}}{\partial \dot{A}_\mu^b} = F^{\mu 0 b}.
\end{equation}

Thus it turns out that the canonical conjugate momentum for the spatial gauge field is the electric field. However, for the temporal component the conjugate momentum does not exist, and therefore Hamiltonian field theory is applicable only in the temporal gauge. 
The Hamiltonian is obtained by a Legendre transformation, which reads 
\begin{equation}
\mathcal{H} = -\mathcal{L} + \dot{A}_i \dfrac{\partial \mathcal{L}}{\partial \dot{A}_i}.
\end{equation}
The resulting Hamiltonian is
\begin{equation}
\label{eq:conthamilton}
 H =\int \ud^3\xx \left[ \tr E^i E^i + \half \tr F_{ij} F_{ij}\right] = \half \int \ud^3\xx \left(E_i^2+ B_i^2 \right),
\end{equation}
where on the right hand side we have suppressed the color indices for brevity. Thus the Yang-Mills Hamiltonian allows for a simple interpretation in terms of electric and magnetic energies.

The corresponding equations of motion are given by the Hamilton's equations
\begin{align}
\dot{A}^{i a} &= \dfrac{\delta \mathcal{H}}{\delta \pi_i^a} \\
\dot{E}^{i a} &= - \dfrac{\delta  \mathcal{H}}{\delta A_i^a}.
\end{align} 
Here the functional derivatives with respect to field $\eta$ are evaluated as
\begin{equation}
\dfrac{\delta}{\delta \eta_i^a} = \dfrac{\partial}{\partial \eta_i^a} - \partial_\mu \dfrac{\partial}{\partial \left( \partial_\mu \eta_i^a \right)}.
\end{equation}
Using this and Hamilton's equations we find
\begin{align}
\label{eq:hamaupdate}
\dot{A}_i  &= E^i
\\
\label{eq:hameupdate}
\dot{E}^i  &=  \left[ D_j,F_{ji} \right].
\end{align}

As one would expect, in the Hamiltonian formalism we obtain coupled first order differential equations, whereas the Lagrangian approach gave us only one second order differential equation \equref{eq:YMEOM}. The main difference between the Hamiltonian and Lagrangian formalism is the nature of the electric field \equref{eq:electricfielddef}. In Lagrangian formalism it is a quantity which we define analogously to the classical electrodynamics. In the Hamiltonian formalism the electric field naturally arises as the canonical conjugate momentum to the gluon field. The Hamiltonian formalism also poses a problem in terms of Gauss' law. In the Lagrangian formalism Gauss' law arose when we considered a variation of the action with respect to temporal links, and thus it was one of the equations of motion. In the Hamiltonian formalism we were forced to adopt temporal gauge in order to derive the equations of motion, and because of this Gauss' law can not be derived in this scheme. Fortunately the Hamiltonian equations of motion \equref{eq:hamaupdate} and \equref{eq:hameupdate} also do satisfy Gauss' law, even though this is not a priori as clear as in the Lagrangian case.

\section{Thermalization in the weak coupling framework}
\subsection{Path to equilibrium in the weak coupling framework}
\begin{figure}
\centerline{\includegraphics[scale=0.75]{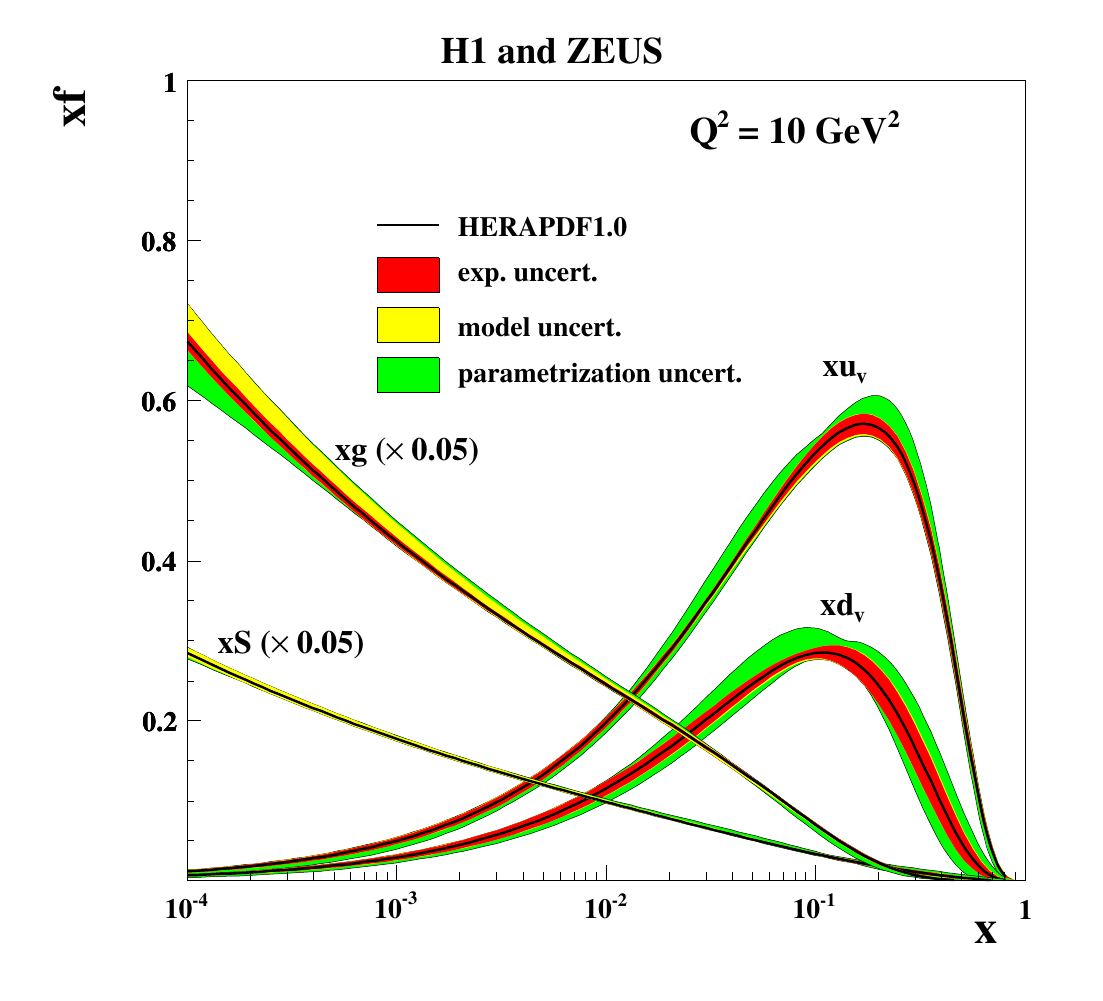}}
\caption{Parton distribution functions extracted from HERA data. Here $x$ denotes the momentum fraction carried by individual partons and $f$ is the density of that parton species. The symbols $u$ and $v$ refer to the u and d quark densities and S refers to the strange quark density. The gluon density is denoted by g.  The figure is taken from \cite{Aaron:2009aa}}
\label{fig:PDF}
\end{figure}
In order to understand how quark-gluon plasma is created in an ultrarelativistic heavy-ion collision, it is beneficial to go through the initial stages of the space-time evolution of an ultrarelativistic heavy-ion collision in the weak coupling framework. For a more thorough picture of thermalization see e.g. the recent reviews \cite{Fukushima:2016xgg,Gelis:2016rnt}. 

The Color Glass Condensate (CGC)\cite{Iancu:2003xm,Gelis:2010nm} is an effective theory of QCD at high energy. The development of CGC was inspired by Deep Inelastic Scattering (DIS) measurents made by HERA (Hadron Electron Ring Accelerator). At HERA the structure of the proton was probed with electrons and positrons in electron-proton and positron-proton collisions. The collision is mediated by a virtual photon. When we go to higher energy, also the short lived excitations within the proton become visible to the photon, and provided that the energy is high enough the lifetime of these excitations is longer than the duration of the collision due to time dilation. This means that the photon indeed also sees the exchanged gluons and the sea quarks inside the proton as free particles. One would expect the sea quark and gluon contributions to rise at small Bjorken $x$, which corresponds to the momentum fraction of the total momentum carried by an individual parton.  Measurements at HERA \cite{Aaron:2009aa} revealed that at small $x$ most of the observed particles are gluons as can be seen in \fig \ref{fig:PDF}. This is at the heart of the formulation of CGC, where the soft gluons at large densities are described as strong classical color fields and the gluons with large momentum are described as color sources. The observed growth in the gluon density does not continue indefinitely. When the gluon occupation number becomes $\mathcal{O}\vas \nicefrac{1}{\alpha_s} \oik,$ the gluon recombination effects start to curb the growth of the gluon distribution. The momentum scale, at which this happens is known as the saturation scale $Q_s \vas x \oik.$

\begin{figure}
\centerline{\includegraphics[scale=1]{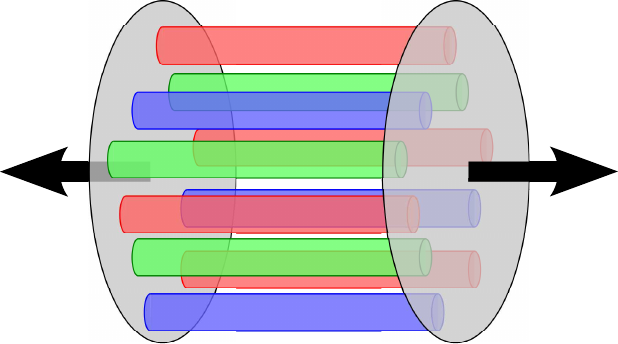}}
\caption{Illustration of the color-flux tubes after the collision. Figure taken from \cite{Fukushima:2011ca}}
\label{fig:tubes}
\end{figure}

In the CGC picture, the two colliding hadrons are described as two thin sheets of CGC. A very short time after the collision the outgoing nuclei are connected by longitudinal boost invariant color flux tubes \cite{Dumitru:2008wn,Lappi:2006fp}, whose transverse size is rougly $\nicefrac{1}{Q_s}$ \cite{Dumitru:2014nka,Dumitru:2013koh}. An illustration of these fluxtubes is shown in \fig \ref{fig:tubes}. Furthermore, the energy density of these color flux tubes is positive, and thus work is done while stretching them. This means that in the initial state the longitudinal pressure of the system is negative.

Ultimately this strongly interacting matter becomes describable by relativistic hydrodynamics \cite{Adams:2005dq,Adcox:2004mh,Kolb:2000fha,Hirano:2005xf,Huovinen:2001cy,Kolb:2003dz,Romatschke:2007mq,Song:2007ux,Dusling:2007gi}. Thus the matter must also thermalize, or be sufficiently close to thermal equilibrium so that the hydrodynamical description works. It seems that a state close to thermal equilibrium is reached in a time $\tau = 0.2-1.5 \mathrm{fm}$ \cite{Heinz:2013th}. The question of how the strongly interacting matter reaches isotropy and thermal equilibrium has been one of the most challenging questions in the field of ultrarelativistic heavy-ion collisions.

%The quantum fluctuations are relevant at the very early stages of the space-time evolution of the ultrarelativistic heavy-ion collision. 
The matter formed at the very early stages of the space-time evolution of the ultrarelativistic heavy-ion collision can be described using classical fields \cite{Lappi:2003bi}, since the occupation numbers of the gluons are nonperturbatively large ($\sim \nicefrac{1}{\alpha_s}$). In general the classical statistical field theory description is valid when the occupation number per mode is large and quantum fluctuations are much smaller than the statistical fluctuations \cite{Berges:2004yj}. During the early stages of the evolution the occupation number of the system begins to fall, and at some point the regime $\nicefrac{1}{g^2} \gg f \gg 1,$ is reached. Within this regime the classical theory and kinetic theory \cite{Arnold:2002zm} are both valid descriptions of the system \cite{Mueller:2002gd,Jeon:2004dh,Berges:2004yj}. Earlier similar studies have also been performed in a context outside of QCD \cite{Skullerud:2003ki,Arrizabalaga:2004iw,Aarts:2001yn,Berges:2008wm,Berges:2013lsa}. For a complete understanding of the thermalization process it is crucial to understand how these two approaches can fit together. In a recent study kinetic theory has been matched to relativistic hydrodynamics and classical Yang-Mills calculations with promising results: hydrodynamization was reached at $\tau < 1 \mathrm{fm}$ \cite{Kurkela:2015qoa}. It has also been shown that kinetic theory and classical theory are equivalent, even quantitatively \cite{York:2014wja} (in the case of  an overoccupied, far from equilibrium and isotropic system).

\begin{figure}
\centerline{\includegraphics[scale=1]{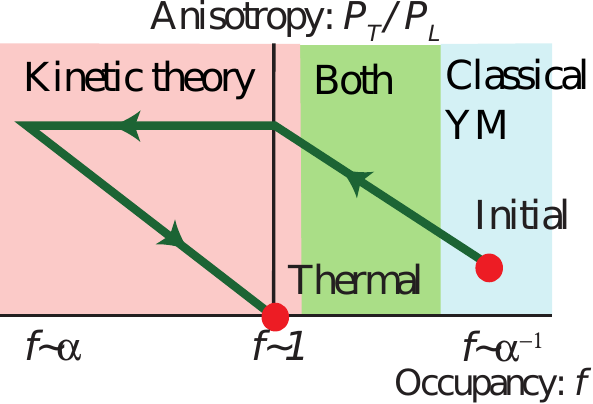}}
\caption{Illustration of the system's path to equilibrium in the occupancy-anisotropy plane. The validity regimes of kinetic theory and classical theory have been marked in the figure. The figure is taken from \cite{Kurkela:2016vts}}
\label{fig:equilibriumpath}
\end{figure}

The path to equilibrium starting from an overoccupied ($f \sim \nicefrac{1}{\alpha} $) initial condition is demonstrated in \fig \ref{fig:equilibriumpath}, and it proceeds as follows (for a more detailed discussion see \cite{Kurkela:2016vts}). In the absense of longitudinal expansion the system goes straight from anisotropy to isotropy \cite{Kurkela:2011ti,Kurkela:2012hp,Schlichting:2012es,Kurkela:2014tea}. However the Yang-Mills evolution starting from an overoccupied initial condition with longitudinal expansion will never reach thermal equilibrium \cite{Berges:2013fga}, instead the system undergoes self-similar evolution \cite{Berges:2013lsa} and the anisotropy of the system will keep on growing \cite{Berges:2013lsa}. When the overlapping range of validity of the classical theory and the kinetic theory ends, the system will follow the kinetic theory evolution which significantly deviates from the classical one. The trajectory of the system goes through an underoccupied region with an approximately constant anisotropy following a bottom-up type thermalization scenario \cite{Baier:2000sb}. The hard gluons first radiate soft gluons, which quickly form a thermal bath. Eventually the occupation number becomes of the order of   $f \sim \mathcal{O}\vas \alpha \oik.$ \cite{Kurkela:2015qoa}. After this the system starts to move towards isotropy and thermal occupation number via radiative breakup, i.e. the hard gluons lose energy to the thermal bath by emitting soft gluons which then thermalize \cite{Baier:2000sb}. Thermal equilibrium is reached when the hard gluons have lost all their energy to the thermal bath.

\subsection{Boost invariance breaking and equilibration}
In the previous section we briefly described how the system reaches equilibrium in the weak coupling framework. However, we did not address the boost invariance breaking, which must take place before thermalization. The initial state is dominated by nonperturbatively strong, leading order boost invariant gluon fields forming the color flux tubes \cite{Krasnitz:1998ns,Kovner:1995ts,Kovner:1995ja,Lappi:2006fp}. At finite $\sqrt{s}$ the boost invariance is broken by the longitudinal structure of the colliding nuclei \cite{Gelfand:2016yho,Schenke:2016ksl,Ipp:2017lho}, and  by rapidity dependent quantum fluctuations \cite{Fukushima:2006ax,Dusling:2010rm,Epelbaum:2011pc,Dusling:2012iga,Epelbaum:2013waa}.

In order to isotropize the system, these fluctuations must grow very rapidly. The QCD plasma exhibits plasma instabilities, such as Weibel instability \cite{Weibel:1959zz} and Nielsen-Olesen instability \cite{Nielsen:1978rm}, which could make the rapid growth possible. Plasma instabilities and their contribution to thermalization of QGP has been subject of intensive research for a long time, see e.g \cite{Mrowczynski:2016etf,Mrowczynski:2005ki,Mrowczynski:1994xv,Mrowczynski:1993qm,Mrowczynski:1988dz,Romatschke:2003ms,Dumitru:2005gp,Mrowczynski:1996vh,Mrowczynski:2004kv,
Kurkela:2011ub,Arnold:2003rq,Romatschke:2004jh,Arnold:2004ti,
Arnold:2004ih,Rebhan:2005re,Rebhan:2004ur,Kurkela:2011ti,Nara:2005fr,
Bodeker:2007fw,Rebhan:2008uj,Attems:2012js,Strickland:2007fm,Arnold:2005qs,Randrup:2003cw}.  Studies on the contribution of the quantum fluctuations to the thermalization process have been carried out in scalar theory \cite{Dusling:2010rm,Epelbaum:2011pc,Dusling:2012ig} and also by classical field simulations \cite{Gelis:2013rba}, where a rapid pressure isotropization was observed. However, the numerical treatment of the fluctuations causes concern here. In \cite{Gelis:2013rba} the quantum fluctuations are included in the equations of motion of the background field. This approach is justified only for modes which become classical due to their growth \cite{Khlebnikov:1996wr,Khlebnikov:1996mc}. For quantum fluctuations with an UV-divergent spectrum this problem becomes especially severe. The UV-divergence will dominate the entire simulation in the continuum limit. This means that a better numerical framework is needed for the treatment of these quantum fluctuations. We will present our numerical framework in \ch \ref{flucts}. By linearizing the classical Yang-Mills equations we can exclude the backreaction from the fluctuations to the classical fields.

\chapter{Classical statistical lattice gauge theory}
\label{latticeintro}
\section{Introduction}
%Next we will go through basics of lattice formulation of QCD. Since we are interested in how the overoccupied strongly interacting system out of equilibrium gluons evolves into a thermal state, we will focus on real time methods. Unfortunately quantized QCD cannot be solved using real time lattice formulation \textbf{Why?}. However, in the  classical approximation \textbf{Tarkemmin, mikä se on tässä?}, the time-evolution becomes straigthforward to compute. Fortunately, the initial stage in ultrarelativistic heavy-ion collisions is characterized by overoccupied gluon states, which do admit to a classical field description. Since the occupation numbers of gluons are nonperturbatively large $\sim \nicefrac{1}{\alpha_s}$ we can neglect the quarks since their occupancies are parametrically smaller due to the Pauli principle.  However, as we saw in the previous chapter, the classical theory cannot follow the expanding system all the way to equilibrium, and at some point one has to switch the degrees of freedom to the ones employed by the effective kinetic theory description.
%A similar real time formalism is also applicable to wide range of problems appearing in the context of ultrarelativistic heavy-ion collisions. The real time formalism with dynamical fermions has been used in studying chiral magnetic effect \cite{Mace:2016shq} \textbf{lisää}, sphaleron transitions both in early universe \cite{Ambjorn:1995xm,Ambjorn:1997jz,DOnofrio:2012phz,Moore:1999fs,Tranberg:2003gi,Arrizabalaga:2004iw} and non-equilibrium glasma \cite{Mace:2016svc} \textbf{Onko muuta?}
In the previous chapter we went through the basics of thermalization in the weak coupling framework. Since we are interested in the dynamical phenomena happening during the very early stages of the time-evolution of an ultrarelativistic heavy-ion collision, where the nonperturbatively large gluon fields dominate, we need a method which is also valid in the regime where the physics is nonperturbative. Lattice regularization of QCD is the most well established nonperturbative method available. It permits first principles computations concerning properties of QCD matter. Among other things lattice QCD can be used to predict hadron spectrum, QCD phase structure, bulk properties of QCD matter (pressure, energy density and entropy density..), QCD phase transition temperature and QCD equation of state (for a more thorough review see e.g. \cite{Ding:2015ona}). These computations involve highly sophisticated Monte Carlo integration methods used to evaluate the QCD path integral. 

In order to get a grasp on the quantities evaluated in the lattice field theory. Consider the QCD action. It is given by 
\begin{equation}
\mathcal{S}_{QCD} = \int \mathrm{d}^4 x \mathcal{L}_{QCD},
\end{equation}
where the QCD Lagrangian is given by \equref{eq:QCDaction}.
The partition function can be computed as an integral
\begin{equation}
\label{eq:QCDpartitionfunction}
\mathcal{Z}_{QCD} = \int \prod_\mu \mathcal{D} A_\mu \prod_{q = u, d, s} \mathcal{D} \psi_q \mathcal{D} \bar{\psi}_q  e^{-i S_{QCD}}.
\end{equation}
And consequently expectation values of physical observables are given by
\begin{equation}
\label{eq:QCDexpectationvalue}
\left< \mathcal{O} \right> =  \dfrac{1}{\mathcal{Z}}  \int \prod_\mu \mathcal{D} A_\mu \prod_{q = u, d, s} \mathcal{D} \psi_q \mathcal{D} \bar{\psi}_q \mathcal{O} e^{-i S_{QCD}}.
\end{equation}
Lattice discretization allows one to compute these integrals, which are intractable in the continuum. However, the integrals appearing in \equref{eq:QCDpartitionfunction} and \equref{eq:QCDexpectationvalue} are highly oscillatory functions (due to the complex weight $e^{-i S_{QCD}}$) of the fields. In practice evaluation of integrals \equref{eq:QCDpartitionfunction} and \equref{eq:QCDexpectationvalue} is impossible. Due to the high dimensionality one has to resort to Monte Carlo techniques. However there is no way to sample these fields. If one wants to make sure that the complex phases are properly taken into account one should perform all the integrals with very high accuracy. This is known as the sign problem, and it prevents us from evaluating real time quantities starting from the first principles. 

This problem, however, concerns only the lattice formulation of QCD in Minkowski space. Rotating to Euclidean time removes the factor of $i$ and removes the sign problem entirely. However, in this case the Euclidean time parameter turns out to play a similar role as inverse temperature in statistical mechanics and we are able to extract only equilibrium properties of QCD in the Euclidean approach. The sign problem emerges also in the Euclidean formulation of lattice QCD when one tries to perform computations with finite chemical potential, since it introduces an imaginary phase to the action. 

Thus evaluating time dependent observables by evaluating the QCD path integral is not an option. However, since the early stages of ultrarelativistic heavy-ion collisions are dominated by classical color fields, classical field theory can be used to evaluate time dependent observables. In this case the path integral is dominated by the classical path, and the evaluation of real-time observables can be achieved by averaging over a set of initial conditions. Formally this can be written as
\begin{equation}
\label{eq:classicalexpectationvalue}
\left< \mathcal{O}(\left( x \right) \right> = \int \mathcal{D} A_0 \mathcal{D} E_0 W_0\left[A_0 , E_0 \right] \mathcal{O}\left[A_0^{cl} , E_0^{cl} , x\right].
\end{equation}
Here the integration over $A_0$ and $E_0$ refers to integration over initial condition with a weight $W$ which is a functional of the initial conditions. The superscripts $cl$ refer to the fact that the observables are evaluated only for the field configurations which are solutions of the classical equations of motion. The connection of \equref{eq:classicalexpectationvalue} to the path integral formulation is that here we integrate only over the solutions of classical equations of motion which dominate the path integral when the system is close to classical limit.

In the following we will first consider the lattice discretization of QCD. Then we will go through how classical Yang-Mills theory is discretized and how its equations of motion are solved on the lattice.

\section{Lattice formulation of QCD}
For a more complete introduction to lattice formulation of QCD we refer the reader to \res \cite{Lepage:1998dt,Gupta:1997nd,Rothe:1992nt,Creutz:1984mg,DeGrand:2006zz}.

The lattice discretization of QCD involves changing the degrees of freedom from Lie algebra valued gluon fields to group valued link matrices in order to preserve  gauge invariance. The link matrices are defined as the discretized version of the path ordered exponential of the line integral of the gauge field
\begin{equation}
\label{eq:linkdef}
U = \PO e^{\int_x^y i g A_\mu(x) \der x_\mu}.
\end{equation}
On the lattice we approximate this path ordered product with link variables
\begin{equation}
\label{eq:linkvariable}
U_\mu\left(x\right) = e^{i a_\mu g A_\mu\left(x+ \nicefrac{\hat{\mu}}{2}\right)},
\end{equation}
where $\hat{\mu}$ stands for a unit vector in the $\mu$ direction.
By expanding the link matrix as a power series in the lattice spacing $a_\mu$, one finds that the leading term in the lattice spacing is captured by the expression
\begin{equation}
A^a_\mu\left(x+ \nicefrac{\hat{\mu}}{2}\right) = \dfrac{2}{a_\mu g}\mathfrak{Im} \mathrm{Tr}\left(t^a U_\mu \left(x\right)\right) + \mathcal{O}(a^4).
\end{equation}

Using the \eq\ref{eq:linkdef} one can show that the gauge transformation of a link matrix is 
\begin{equation}
U_\mu\left(x\right) \rightarrow V\left(x\right) U_\mu\left(x\right) V^\dagger\left(x+\mu \right)
\end{equation}

Knowing the gauge transformation rule of the link matrix, we can guess what is the easiest gauge invariant observable we can construct on the lattice.
The plaquette is defined as 
\begin{equation}
U_{\mu \nu}\left(x\right) = U_\mu\left(x\right) U_\nu\left(x+\hat{\mu}\right)U^\dagger_\mu\left(x + \hat{\nu}\right)U^\dagger_\nu\left(x\right).
\end{equation}
Using the definition one instantly observes that the plaquette transforms locally
\begin{equation}
U_{\mu \nu}\left(x\right) \rightarrow V\left(x\right) U_{\mu \nu}\left(x\right) V^\dagger\left(x\right). 
\end{equation}
Thus the trace of a plaquette is a gauge invariant. 
The plaquette is connected to the field strength tensor via the relation
\begin{equation}
U_{\mu \nu}\left(x\right)  = e^{i g a_\mu a_\nu F_{\mu \nu}\left(x + \nicefrac{\hat{\mu}}{2}+\nicefrac{\hat{\nu}}{2}\right) + \mathcal{O}\left(a^3\right) }.
\label{eq:plaqcomp}
\end{equation}
Similarly as for the gauge fields, the leading term of the power series is captured by 
\begin{equation}
\label{eq:fmunudef}
 F_{\mu \nu}^a(x) = \dfrac{2}{ a_\mu a_\nu g} \mathfrak{Im}\mathrm{Tr} \left( t^a U_{\mu \nu}\left(x\right) \right).
\end{equation}

This enables us to define the counterparts of the chromoelectric and chromomagnetic fields on the lattice:
\begin{equation}
\label{eq:elatdef}
E_i^a(x) = \dfrac{2}{a_s a_t g} \mathfrak{Im}\mathrm{Tr}(t^a U_{i0}\left(x\right))
\end{equation}
\begin{equation}
B_i^a(x) = -\dfrac{ \varepsilon_{ijk}}{a_s^2 g} \mathfrak{Im}\mathrm{Tr} \left( t^a U_{jk}\left(x\right) \right)
\end{equation}

The plaquette can also be taken in the $(\mu, -\nu)$ direction, we denote it by
\begin{align}
W_{\mu \nu }\vas x \oik  & = U_\mu \vas x \oik U^\dagger_\nu \vas x + \hat{\mu} - \hat{\nu} \oik U^\dagger_\mu \vas x - \hat{\nu} \oik U_\nu \vas x - \hat{\nu}  \oik  \\ & = e^{i g a_\mu a_\nu F_{\mu \nu } \vas x+ \nicefrac{\hat{\mu}}{2} - \nicefrac{\hat{\nu}}{2} \oik + \mathcal{O} \vas a^3 \oik} \nonumber ,
\end{align}
which will repeatedly appear in the equations of motion.

\section{Classical action and equations of motion in the temporal gauge}
\label{sec:lagrangianlattice}
Next we will construct the lattice action and derive the equations of motion. We eliminate the gauge freedom by adopting the temporal gauge ($A_0 = 0$ or on the lattice $U_0 = \mathbb{1}$). It turns out that this gauge choice also simplifies the time-evolution. The linear term in $F_{\mu \nu}$ was captured by the imaginary part of the trace in \eq(\ref{eq:fmunudef}). The quadratic term can be obtained by taking the real part of the trace. It turns out that in the continuum limit the Yang-Mills action \eq (\ref{eq:YMLagrange}) is reproduced by the Wilson action 
\begin{equation}
\mathcal{S}  =  -\beta_0 \sum_{x,i} \left( \frac{1}{N} \mathfrak{Re}\mathrm{Tr}\left(U_{0i}\left(x\right)\right) - 	1 \right)   +  \beta_s \sum_{x,i<j} \left( \frac{1}{N} \mathfrak{Re}\mathrm{Tr}\left(U_{ij}\left(x\right)\right) -1 \right) 
\label{eq:wilsonaction}
\end{equation}
 where $N$ is the number of colors $\beta_0 = \frac{2 N \gamma}{g^2},$ $\beta_s = \frac{2 N}{g^2 \gamma},$ and  $\gamma = \frac{a_s}{a_t}.$ \eq \ref{eq:wilsonaction} is the action of classical pure glue QCD on the lattice. In standard lattice QCD the action is rotated to Euclidian space, and then used to evaluate the path integral to compute thermodynamical properties of QCD. 

The classical equations of motion are obtained by varying the action as is usually done in Lagrangian mechanics. In practice the variation is carried out by performing the substitution 
\begin{equation}
U_i\left(x\right) \rightarrow e^{i \delta A_i \left(x\right)}.
\end{equation}
Then we expand to linear order in variation $\delta A_i\left(x\right)$, and obtain the equations of motion by demanding that the coefficient of the variation equals zero. 
 Varying the action with respect to the spatial links gives the equation of motion of the electric field
\begin{equation}
\label{eq:Eevolution}
E_j (t , x)   =  E_j(t-a_t , x) + \dfrac{a_t}{ a_s^3 g} \sum_k \left[  U_{jk}\left(x\right)  + W_{kj}\vas x \oik \right]_\ah. 
\end{equation}
The antihermitian traceless part is denoted by $[]_\ah.$ It is given by 
\begin{equation}
[V]_\ah \equiv \frac{-i}{2} \left[ V - V^\dag - \frac{\mathbb{1}}{N}\tr(V-V^\dag)\right].
\end{equation}
It is beneficial to keep in mind, that the electric fields appearing in \equref{eq:Eevolution}, arise when we vary the temporal plaquettes with respect to the spatial links. It turns out they coincide with the definition \equref{eq:elatdef}. Defining the electric field serves two separate purposes. Firstly, it has a clear physical interpretation in terms of its analogue in classical electrodynamics. The second reason is numerical convenience. While solving second order partial differential equations in time, it is advisable to decompose the equations into two first order differential equations in time. In \se \ref{sec:lathamilton} we will consider the Hamiltonian equations of motion on the lattice. There the electric field arises as the canonical conjugate momentum of the gauge field. 

In order to compute the real time evolution we also need to know how to update the links to the next time step. In temporal gauge the temporal plaquette simplifies to
\begin{equation}
U_{i0}\left(x\right) = U_i\left(x\right) U_i^\dagger\left(x+ \hat{t}\right),
\end{equation}
where the notation $x + \hat{t}$ refers to the link matrix at position $x$ at the next time step.
On the other hand we can use the definition of the electric field to solve for the temporal plaquette. We also make use of a matrix decomposition specific to $\mathrm{SU(2)}$ 
\begin{equation} \label{eq:decomposition}
U_{i0}\left(x\right) = c_0 \mathbb{1} + 2ic^at^a,
\end{equation}
where the parameter $c_0$ is eliminated by the constraint $1 = \sqrt{c_0^2+c_a c_a}.$ Making use of this and the Fierz identity \equref{eq:Fierz} and the definition of the electric field \equref{eq:elatdef} we can solve for the temporal plaquette corresponding to the electric field
\begin{equation}
\label{eq:linkevolution}
U_{i0}\left(x\right)  = \sqrt{1 - \left(\dfrac{a_s a_t g}{2} E_a \right)^2}\mathbb{1} + ia_s a_t g E^a t^a.
\end{equation}
Using this the link at the next time-step is easy to solve
\begin{equation}
\label{eq:linknextstep}
U_i^\dagger\left(x+ \hat{t}\right) = U_{i0}\left(x\right)^\dagger U_i\left(x\right).
\end{equation}

We can also vary the action \equref{eq:wilsonaction} with respect to the temporal links. This gives us a nondynamical constraint which is analogous to Gauss' law in classical electrodynamics
\begin{equation}
\sum_j \left( E_j(x) - U_{x-j, j }^\dagger E_j(x-j) U_{x-j, j } \right) = 0.
\label{eq:gauss}
\end{equation}
The typical way to simulate a classical Yang-Mills system is to evolve the links and the electric fields using \eqs (\ref{eq:Eevolution}) and (\ref{eq:linkevolution}) in a leapfrog scheme, which is a time translationally invariant integration scheme (guaranteeing second order accuracy in $\ud t$), where the electric fields and links live on interleaved timesteps $\nicefrac{\ud t}{2}$ apart from each other. 

\section{Hamiltonian equations of motion on the lattice}
\label{sec:lathamilton}
In this section we will show the corresponding Hamiltonian equations of motion on the lattice. Throughout this section we will also be using a slightly different convention for the lattice fields: here we absorb the factors of $a_s$ and $g$ into the definitions of the  fields. Thus the relationship between the lattice electric field and the continuum electric field is $E^i_\textup{lat} \approx a_s g E^i_\textup{cont}$, and similarly for the gauge field.

When one defines the lattice fields in this way, the gauge field becomes dimensionless, but the electric field has dimensions of $\mathrm{GeV}$. In principle we could make the electric field dimensionless by simply multiplying it by a factor of $a_s.$ However, \equref{eq:plaqcomp} suggests that the natural definition of the lattice electric field would involve a multiplication by a factor of $a_t.$ However, in order to keep the timestep explicitly visible in our equations we will stick to the dimensionful lattice electric field. 

The lattice counterpart of the Hamiltonian (\ref{eq:conthamilton}) is the Kogut-Susskind Hamiltonian~\cite{Kogut:1974ag}
\begin{equation}\label{eq:ksh}
H =\frac{a_s^3}{g^2} \sum_x \Bigg\{ \tr \big[ a_s^{-2} E^i(x) E^i(x) \big]
+ \frac{2}{ a_s^4} \sum_{i<j} \Re\tr\big[ \mathbb{1}  -  U_{ij}(x) \big]
  \Bigg\}.
\end{equation}
Here time is treated as a continuous variable while the space has been discretized as in the Lagrangian case. 
The corresponding equations of motion read
\begin{eqnarray}
\dot{U}_i(x) &=& i E^i(x) U_i(x) \qquad  (\text{no sum over i}) \label{eq:Udot}
\\
a_s^2 \dot{E}^i(x) &=& - \sum_{j\neq i } \left[U_{ij}(x) + W_{ij}(x) \right]_{\ah}. \label{eq:Edot}
\end{eqnarray}

For numerical computations the time has to be discretized. In order to guarantee energy conservation and second order accuracy in $\mathrm{d}t$ we utilize leapfrog discretization 
\begin{align}
 U(t+\ud t) &= e^{i E^i(t+\ud t/2) \ud t} U_i(t)
\label{eq:cllinkstep}
\\
\label{eq:clestep}
a^2 E^i(t+\ud t) &=a^2  E^i(t) -\ud t \sum_{j\neq i }  \left[U_{ij}\left(t+\frac{\ud t}{2}\right) + W_{ij}\left(t+\frac{\ud t}{2} \right) \right]_\ah.
\end{align}
It is a straightforward exercise to check that these equations of motion separately conserve the discretized Gauss' law \eq (\ref{eq:gauss}).

The main difference between the Hamiltonian equations of motion \equref{eq:cllinkstep} and \equref{eq:clestep} and the Lagrangian equations of motion \equref{eq:Eevolution} and \equref{eq:linknextstep} is the timestep for link matrices, in both approaches the equation for the electric fields is the same. Interestingly both time-evolution equations for the link matrices also satisfy Gauss' law exactly (Gauss' law is actually left unchanged by time-evolution equations for $U$ and $E$ separately).

\section{Quasiparticle spectrum}
Next we want to understand to what extent our system can be understood in a quasiparticle picture and how to derive the quasiparticle spectrum. Our system is a plasma, which is a gas of charged particles. In a plasma, interactions have longer range than in neutral gases, since in a neutral gas the interaction between the constituents happens through weak van der Waals forces. In a charged plasma the interactions take place through electromagnetic or strong interaction, and they have a lot greater range. Thus perturbing a single constituent of a plasma will induce a response from multiple particles giving rise to collective behavior \cite{Piel:plasmafysiikka}. If these collective excitations have particle-like properties they are referred to as quasiparticles.

Before deriving an expression for the quasiparticle spectrum we will briefly discuss Debye screening (in plasma physics this phenomenon is also known as Debye shielding). Consider an electromagnetic plasma as an example. Inserting a test charge into the plasma repels charges of the same type, and attracts charges of the oppositely charged particles. From a distance, one can no longer see only the original charge, but one also sees the effect of the surrounding charges screening the original charge. As a result the Coulomb potential 
\begin{equation}
\label{eq:SIcoulomb}
V = \dfrac{1}{4 \pi} \dfrac{Q}{r}
\end{equation}
gets modified due to the screening effects and becomes 
\begin{equation}
\label{eq:debyescreening}
\dfrac{Q}{4 \pi r} e^{-\nicefrac{r}{\lambda_D}}.
\end{equation}
Here $Q$ is the charge of the constituent and $r$ is the distance from the charged particle. The Debye screening length $\lambda_D$ is in general a function of temperature and density of charged particles and gives the characteristic length scale of the plasma. The inverse of the Debye screening length is the Debye mass $m_D = \nicefrac{1}{\lambda_D}$.  As a consequence of the screening effects, the electric potential dies off a lot faster in a plasma than in vacuum. 

This characteristic length scale also has implications for the kinetic theory description of the ultrarelativistic plasma. For modes which have $k < m_D,$ or alternatively $\lambda > \lambda_D,$ the kinetic theory description becomes problematic due to medium effects. If we excite the plasma with a perturbation with a wavelength $\lambda \gg \lambda_D$ it interacts within a region whose size is comparable to its wavelength. However, if we would be able to describe this using the particle description, the particle would interact with other particles which lie beyond $\lambda_D$.

Since our dynamical variables are fields, we need to find a way to map these into particle degrees of freedom. Let us start with the energy density on the lattice
\begin{equation} \label{eq:edensity}
\epsilon =  2\left(\nc^2-1\right)\int \dfrac{\mathrm{d}^3 k}{\left(2 \pi \right)^3} \omega\left(k\right) f\left(k\right),
\end{equation}
where $\omega\left(k\right)$ is the energy of the mode with momentum $k$ and $f\left(k\right)$ is the quasiparticle spectrum per degree of freedom per unit volume.
Here the factor $2\left(\nc^2-1\right)$ is the number of transverse polarization states. Physically gluons have three available polarization states (since we will later find out that they acquire a mass from interactions). However, the longitudinal polarization of the gauge potential is eliminated by our gauge choice. 

The occupation number interpretation of a system of classical gauge fields is most straightforward in Coulomb gauge. The reason is that it takes the number of polarization states properly into account. In other gauges the interpretation of the longitudinal mode for large momenta becomes very difficult.

In practice we impose the Coulomb gauge condition only at readout times. As a consequence the longitudinal mode is present in the electric field. However the longitudinal mode contributes only for the modes close to the Debye scale, and we do not expect it to contribute significantly to the energy density of the system. Most of the energy of the system resides at the hard scale $Q$ instead of the soft mass scale. Thus, introducing the longitudinal polarization in the degrees of freedom in \equref{eq:edensity} would significantly overestimate the total energy density. 

The total energy of the system is given by the Yang-Mills Hamiltonian \equref{eq:conthamilton}.
Next we keep only the terms which are quadratic in the gluon field, and write this integral in momentum space in the Coulomb gauge. We get
\begin{equation}
\label{eq:hamiltonianenergy}
\mathcal{H} = \int \frac{\mathrm{d}^3k}{\left( 2 \pi \right)^3} \left(E^i_C E^i_C + \left|k\right|^2 \left|A_{C,i}\right|^2\right),
\end{equation}
where the subscript $C$ refers to Coulomb gauge.
Now we require that the energy densities of \eqs (\ref{eq:edensity}) and (\ref{eq:hamiltonianenergy}) are the same for each mode of the system. This allows us to solve for the quasiparticle spectrum. The result is
\begin{equation} \label{eq:discF}
f_{A+E}\left(k\right) = \dfrac{1}{4\left(\nc^2-1 \right)} \dfrac{1}{V} \left(\dfrac{\left| E_C\left(k\right)\right|^2}{\omega\left(k\right)} + \dfrac{k^2}{\omega\left(k\right)} \left|A_C\left(k\right)\right|^2\right).
\end{equation}
Unless otherwise stated, we will be using the massless dispersion $\omega = k$ when extracting the quasiparticle spectrum, since it is not obvious whether we should keep the factor of $k^2$ in \equref{eq:discF} or replace it with $k^2+m^2.$ The replacement would correspond to an estimate of the higher order terms in the gauge potential (which we have already neglected) in the energy.

Alternatively the quasiparticle spectrum can be extracted using only gauge fields or electric fields. This method relies on the fact that the classical system obeys equipartition of energy, and thus the energy tends to be equally divided between electric and magnetic modes after an initial transient time. Even though the system is not yet in thermal equlibrium, in practice one observes that the energies of electric and magnetic modes tend to even out very quickly (see e.g. \cite{Kurkela:2012hp} fig. 2).
Thus, above the Debye scale we expect these definitions to be equivalent to \equref{eq:discF}
\begin{equation} \label{eq:discFE}
f_E\left(k\right) =  \dfrac{1}{2\left(\nc^2-1 \right)} \dfrac{1}{V} \left(\dfrac{\left| E_C\left(k\right)\right|^2}{\omega\left(k\right)} \right)
\end{equation}
for the electric estimator and
\begin{equation} \label{eq:discFA}
f_A\left(k\right) =  \dfrac{1}{2\left(\nc^2-1 \right)} \dfrac{1}{V} \left( \dfrac{k^2}{\omega\left(k\right)} \left|A_C\left(k\right)\right|^2\right)
\end{equation}
for the magnetic estimator.
It is  also possible to use a combination of electric and magnetic fields
\begin{equation} \label{eq:discFEA}
f_{EA}\left(k\right) =  \dfrac{1}{2\left(\nc^2-1 \right)} \dfrac{1}{V} \left( \sqrt{\left|A_C\left(k\right)\right|^2\left| E_C\left(k\right)\right|^2}\right).
\end{equation}
This expression can be derived by assuming that $f_A = f_E,$ which should hold above the Debye scale. 

Based on thermal field theory, we are expecting the results for the longitudinal modes to be different from the results for the transverse modes. Thus we wish to be able to separate these two. The projections operators on the lattice are 
\begin{align}
\label{eq:proj1}
P_T^{ij} & = \delta_{ij} - \dfrac{p_{B}^{i*} p_{B}^j}{\left|p\right|^2} \\
P_L^{ij} & = \dfrac{p_{B}^i p_{B}^{j*}}{\left|p\right|^2}.
\label{eq:proj2}
\end{align}
Here the momenta correspond to the complex eigenvalues of the backward difference operator, since we are using backward derivatives in e.g. Coulomb gauge condition and Gauss's law. In principle one is free to choose the discretization of the derivative in whichever way one desires. This choice is done when one writes down the lattice action, and the choice of derivative therein propagates to the equations of motion and Gauss' law. However, the backward derivative is a natural choice for the derivative. Consider the 
(backward) divergence of the gauge fields
\begin{equation}
\label{eq:latdivergence}
\dfrac{1}{a_s}\sum_i \left[ A_i \left(x + \nicefrac{\I}{2} \right) - A_i\left(x - \nicefrac{\I}{2} \right) \right] = 0.
\end{equation}
Now bearing in mind the connection between the link matrix and the gauge field given by \equref{eq:linkvariable}, we can write the lattice version of \equref{eq:latdivergence} as 
\begin{equation}
\sum_i \left( \mathfrak{Im}\mathrm{Tr}\left(U_i\left(x\right) \right) - \mathfrak{Im}\mathrm{Tr}\left(U_i\left(x-\I\right) \right) \right) = 0.
\end{equation}
Thus the backward difference between two links actually evaluates the derivative at the correct position. If one chooses to use forward difference or central difference, one immediately faces a problem concerning the point at which the derivative is defined. This is why we consider the backward difference to be the superior choice in this case.

When the projection operators are defined as in \equref{eq:proj1} and \equref{eq:proj2} we have $P_T P_L = 0.$ The simple derivation of these eigenvalues is the following
\begin{equation}
\partial_{B,i} e^{i k x} = e^{i k x} \dfrac{1}{a_s} \vas 1- e^{i k_i a_s} \oik.
\end{equation}
Comparing this with the continuum result
\begin{equation}
\partial_i e^{i k x} = i k_i e^{i k x},
\end{equation}
allows us to identify the complex momenta as 
\begin{equation}
p_{B,i} = \dfrac{-i}{a_s}\vas 1- e^{i k_i a} \oik.
\end{equation}

%When separating the longitudinal and the transverse modes, one has to also keep in mind that the actual dynamical variables (the electric fields and gluon fields), are not defined on the lattice points, but halfway between the two points. This introduces an additional complex phase, which vanishes when taking the absolute value of the field, but not when defining transverse and longitudinal fields. Let us compute the Fourier transform of the gauge field as an example
%\begin{equation}
%A_i \vas k \oik = \int d^3x \dfrac{e^{- i \boldsymbol{k} \cdot \boldsymbol{x} }}{\vas 2 \pi \oik^{\nicefrac{3}{2}}} A_i \vas x + \dfrac{\I}{2} \oik = e^{i \nicefrac{k_i}{2}} \int dx \dfrac{e^{- i \boldsymbol{k} \cdot \boldsymbol{x} }}{\vas 2 \pi \oik^{\nicefrac{3}{2}}} A_i \vas x\oik.
%\end{equation}
%Thus, when handling Fourier transforms of the fields, it is necessary to also introduce this additional complex phase, unless one is computing absolute value of the field. 

Finally we want to demonstrate that when computing the Fourier transform of the lattice fields one must also introduce a nontrivial phase factor. 
With the lattice discretized gauge fields (defined between the lattice points) the Fourier transform can be written as
\begin{equation}
\label{eq:funnyfourier}
A_i^L\left(k\right) = \sum_x e^{i k \left(x+\nicefrac{\I}{2} \right)} A_i\left(x+\nicefrac{\I}{2}\right).
\end{equation}
We can now replace the gauge field on the right hand side of \equref{eq:funnyfourier} with the gauge field extracted from the link variable. We get
\begin{equation}
\label{eq:FFTlink}
A_i^L\left( k \right) = e^{i \nicefrac{k_i}{2}} \sum_x e^{i k x}  \dfrac{2}{a_s g }\mathfrak{Im}\mathrm{Tr}\left(U_i\left(x\right)\right).
\end{equation}
This means that on the lattice the extracted gauge fields must be multiplied by an extra phase factor of $e^{i \nicefrac{k_i}{2}}$ in momentum space. The right hand side of \equref{eq:FFTlink} has been written in this way on purpose: the typical way to extract the gauge field from the links is using the expression $\mathfrak{Im}\mathrm{Tr}\left(U_i\left(x\right)\right)$ multiplied by constant factors. One has to keep this mismatch in mind when computing the numerical Fourier transform of the fields, since the libraries compute the Fourier transforms by convoluting by $\sum_x e^{i k x}$. %However, in order to account for the mismatch between the points in which the links and gauge fields are defined, we must introduce an additional phase factor of $e^{i \nicefrac{k_i}{2}}$.

\chapter{Basic results of Hard Thermal Loop perturbation theory}
In this section we will briefly introduce Hard Thermal Loop perturbation theory and review some of its predictions which we will need in later chapters.  We will only go through the main aspects of the calculations. For more details we refer the reader to \cite{Blaizot:2001nr}, and for an introduction to perturbative thermal field theory calculations see e.g. \cite{Bellac:2011kqa,Kapusta:2006pm,Laine:2016hma}. 

Hard Thermal Loop perturbation theory is based on a separation of scales. At high temperature the coupling $g = g\left(T\right)$ is small. This creates a separation of scales between the hard scale given by temperature $T$ and the soft scale given by $gT$. Sometimes one also considers the so called ultra soft scale $g^2 T$. The power of the HTL perturbation theory arises from the fact that the system is dominated by hard particles with characteristic momentum scale $T$ and their masses are parametrically smaller, of the order of the soft scale. As a consequence the relevant loop diagrams will be dominated by the hard particles. 

%In our classical nonequilibrium simulations the temperature is not a well defined concept. Also the coupling $g$ can be scaled out of the classical simulations be redefinition of the fields. Thus one can not interpret classical simulations in the standard HTL framework. However, HTL like scale separation given by the dominant hard momentum scale and the soft scale given by the effective mass of the quasiparticles. 

We will start by introducing the retarded propagator in the HTL formalism. Then we will derive the spectral function and the dispersion relation of the collective excitations in the plasma.

\begin{figure}
\includegraphics[scale=1.25]{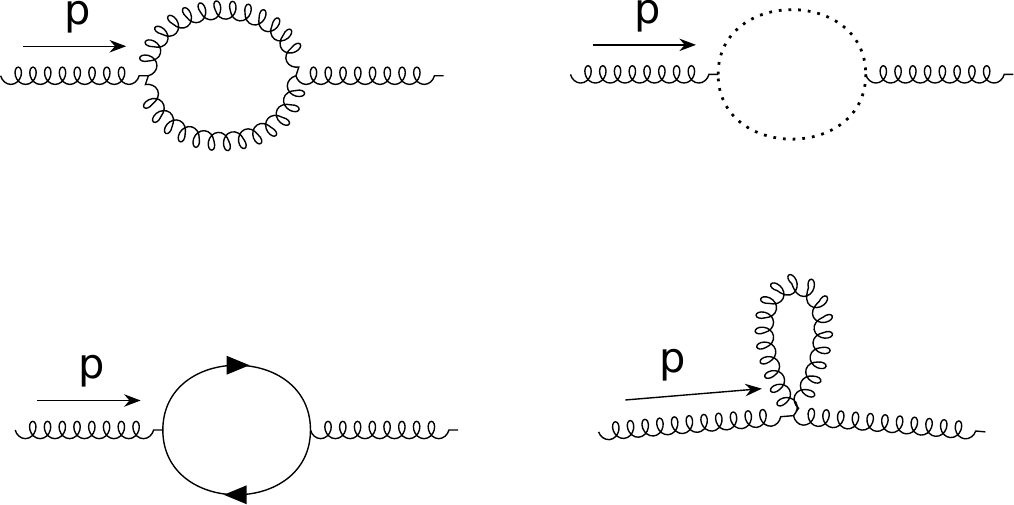}
\caption{The relevant loop diagrams contributing to the gluon self energy. The diagrams are drawn using TikZ-Feynman \cite{Ellis:2016jkw}}
\label{fig:HTLgraphs}
\end{figure}

\label{sec:HTLbasics}
\section{The retarded propagator}
The computation of the retarded propagator in the HTL framework starts by the evaluation of the gluon self-energy. The self-energy is given by the four graphs shown in \fig \ref{fig:HTLgraphs}. We neglect the external momentum $p,$ since we assume that the momentum in the loop is hard. Summing the contributions to the gluon self-energy yields
\begin{equation}
\Pi_{\mu \nu} =  m_D^2 \left( - \delta_\mu^0 \delta_\nu^0 + \omega \int\dfrac{\mathrm{d}\Omega}{4 \pi} \dfrac{v_\mu v_\nu}{\omega - \boldsymbol{v}\cdot \boldsymbol{p} + i \eta }\right),
\end{equation}
where $v$ stands for the four velocity of the particle and $+i \eta$ implements the retarded boundary condition. The self-energy tensor is transverse to the four momentum
\begin{equation}
p_\mu \Pi^{\mu \nu} = 0.
\end{equation}
Because of the transversality condition, the self-energy tensor can be characterized using two scalar functions. It is customary to choose these functions as the longitudinal and transverse (with respect to the three momentum) parts
\begin{align}
\Pi_{00} &= -\Pi_L \\
\Pi_{0i} &= -\dfrac{p_0 p_i }{\left|\boldsymbol{p}\right|^2} \Pi_L \\
\Pi_{ij} &= \left(\delta_{ij} - \dfrac{p_i p_j}{\left|\boldsymbol{p}\right|^2} \right) \Pi_T - \dfrac{p_i p_j}{\left|\boldsymbol{p}\right|^2} \dfrac{\omega^2}{\left|\boldsymbol{p}\right|^2} \Pi_L.
\end{align}
The longitudinal and transverse components are given by
\begin{align}
\Pi_L &= 2 m^2\left(1-Q_0\left(x\right)\right) \\
\Pi_T &=  m^2 \dfrac{\omega^2}{p^2} \left(1 - \dfrac{\omega^2 - p^2}{p^2} Q_0\left(x\right)\right),
\end{align}
where $x = \omega/p.$ The function $Q_0\left(x\right)$ is given by 
\begin{align}
 Q_0(x) = \frac{1}{2}\ln \frac{x+1}{x-1} = \frac{1}{2}\ln \left|\frac{x+1}{x-1}\right| -\frac{i\pi}{2} \theta (1-x^2).
\end{align}
Here $\theta$ refers to the Heaviside step function.
The retarded propagator in the HTL formalism in the temporal gauge is given by
\begin{align}
 \label{eq_GR_HTL}
 G_T^{\HTL}(\omega, p) &= \frac{-1}{\omega^2 - p^2 - \Pi_T} \nonumber \\
 G_L^{\HTL}(\omega, p) &= \frac{-1}{\omega^2 - \Pi_L}.
\end{align}

\begin{comment}
The denominators of the propagators are given by 
\begin{align}
\label{eq_HTL_denom}
 \omega^2 - p^2 - \Pi_T &= p^2 \left( x^2 - 1 - \frac{m^2}{p^2}\,x\left[ x + (1-x^2)Q_0(x) \right] \right) \nonumber \\
 \omega^2  -\Pi_L &= \omega^2 \left( 1 + \frac{2m^2}{p^2}\left[1 - x\,Q_0(x)\right] \right),
\end{align}
where we use the shorthand notation $x = \omega/p$. The function $Q_0\left(x\right)$ is given by 
\begin{align}
 Q_0(x) = \frac{1}{2}\ln \frac{x+1}{x-1} = \frac{1}{2}\ln \left|\frac{x+1}{x-1}\right| -\frac{i\pi}{2} \theta (1-x^2). 
\end{align}
\end{comment}

\section{Spectral function}
The transverse spectral function is obtained as the imaginary part of the retarded propagator
\begin{align}
 \label{eq_rhoHTL_GT_rel}
 \rho_T^{\HTL} (\omega, p) = 2\, \mrm{Im}\,G_T^{\HTL}(\omega, p). 
\end{align}
%For longitudinal modes the expression of the spectral function is modified with an additional prefactor 
%\begin{align}
% \label{eq_rhoHTL_GL_rel}
% \rho_L^\HTL (\tcent,\omega, p) = \frac{\omega^2}{p^2}\, 2\,\mrm{Im}\,G_L^\HTL(\tcent,\omega, p),
%\end{align}
%where the factor cancels the unphysical pole at $\omega=0$ which arises from the choice of temporal gauge.
The physical interpretation of the spectral function (in frequency space) is that it gives the possible excited frequencies $ \omega $ related to momentum $p$. The spectral function can reveal whether the system exhibits quasiparticle excitations: in frequency space plasmon excitations appear as damped oscillations with damping rate $ \gamma $. See also the discussion in \re \cite{Blaizot:2001nr}.

In order to better understand the shape the spectral function takes in momentum space, let us compute the Fourier transform of a damped oscillator of the form 
\begin{equation}
h\left(t\right) = e^{-\gamma_T t } e^{-\omega_T t} \theta\left(t\right),
\end{equation}
where $\theta\left(t\right)$ is the Heaviside step function. Fourier transforming this yields
\begin{equation}
\hat{F}\left(h\left(t\right)\right) = \dfrac{\gamma_T}{\sqrt{2 \pi} \left( \gamma_T^2 + \left(\omega - \omega_T\right)^2\right)} + i \dfrac{\left(\omega - \omega_T\right)}{\sqrt{2 \pi } \left( \gamma_T^2 + \left( \omega - \omega_T\right)^2\right)}.
\end{equation}
The real part of the Fourier transform corresponds to Lorentzian distribution
\begin{align}
 \label{eq_lor_curve}
 g_{\mrm{Lor}}(\omega) = \frac{A}{\pi}\, \frac{\gamma_T}{ \gamma_T^2 + (\omega - \omega_T)^2}.
\end{align}

Instead of exhibiting only a single quasiparticle peak, the spectral function also involves low frequency ($\omega < p$) excitations, which are usually refered to as the Landau cut and denoted by $\beta$ here. These low frequency excitations arise from the imaginary part of the polarization tensor $\Pi.$ The functional form of the Landau cut can be computed analytically using \equref{eq_rhoHTL_GT_rel}. For the transverse modes the spectral function for $p > \omega$ is given by
\begin{align}
\label{eq:transverselandau}
 &\beta_T(\omega, p) = \pi m^2  x (1-x^2)\, \theta (1-x^2) \nonumber \\
 &\times \Bigg[\left( p^2 (x^2 - 1) - m^2 \left( x^2 + \frac{1}{2}\, x (1-x^2) \ln \left|\frac{1+x}{1-x}\right| \right) \right)^2 \nonumber \\
 &\qquad \qquad + \left( \frac{\pi}{2}\, m^2 x (1-x^2) \right)^2 \Bigg]^{-1}.
\end{align}
For the longitudinal excitations the Landau cut turns out to be
\begin{align} 
 \label{eq_Landau_cut_long}
\beta_L(\omega, p)  &= 2\pi m^2  x \, \theta (1-x^2) \nonumber \\
 &\times \Bigg[\left( p^2 + 2m^2 \left( 1 - \frac{x}{2}\, \ln \left|\frac{1+x}{1-x}\right| \right) \right)^2 + \left( \pi\, m^2 x \right)^2 \Bigg]^{-1}.
\end{align}
For frequencies $\omega < p$ we have $\rho = \beta.$ In the leading order the quasiparticle peak is a delta function for $\omega = p$ and has zero width. The analytical  form of the spectral function  is shown together with our data in the frequency domain for transverse excitations in \fig \ref{fig_Landau_cut} and for the longitudinal modes in \fig \ref{fig_rho_dt_w_long}. Especially for the transverse modes one can see that the spectral function for fixed $\omega$ clearly consists of two distinct parts.

\section{Dispersion relation}
The quasiparticle peaks of the spectral function in frequency space are given by the zeroes of the denominators of the propagators \equref{eq_GR_HTL}. The equations can only be solved numerically, but one can analytically find their low and high momentum limits. They are given by 
\begin{align}	
 \label{eq_wHTL_Taylor_trans}
 \omega_{T}^{\HTL} &\overset{p \ll m}{\simeq} \sqrt{(\wplas^\HTL)^2 + \nicefrac{6}{5}\,p^2} \nonumber \\
 \omega_{T}^{\HTL} &\overset{p \gg m}{\simeq} \sqrt{m_\HTL^2 + p^2}, \nonumber \\
 \omega_L^{\HTL}  &\overset{p \gg m}{\simeq} p\left(1 +  \exp{\left(-\dfrac{p^2+m_\HTL^2}{m_\HTL^2}\right)}\right), \nonumber \\
 \omega_{T}^{\HTL} &\overset{p \ll m}{\simeq} \sqrt{\left(\omega_{pl}^\HTL \right)^2 + \nicefrac{3}{5} p^2} . 
\end{align}
Here the sub- and superscripts $\HTL$ refer to the results obtained from the HTL perturbation theory. This notation will become useful later, when we compare these values to our numerical results. The dispersion relation actually features two distinct masses. At low momenta, the mass scale is called the plasmon mass $\omega_{pl}$. However, at higher momenta we observe a mass scale with a different value. We call this mass the asymptotic mass $m,$ which is sometimes also denoted as $m_\infty$ in the literature. In HTL the two are connected by a constant factor
\begin{align}
 \label{eq_wplas_m_rel}
 \wplas^\HTL = \sqrt{\nicefrac{2}{3}} m_\HTL
\end{align}
At low momenta the transverse and longitudinal modes become inseparable, and they will both converge to the same value $\omega_{pl}.$

\begin{figure}[t]
\centerline{\includegraphics[width=0.7\textwidth]{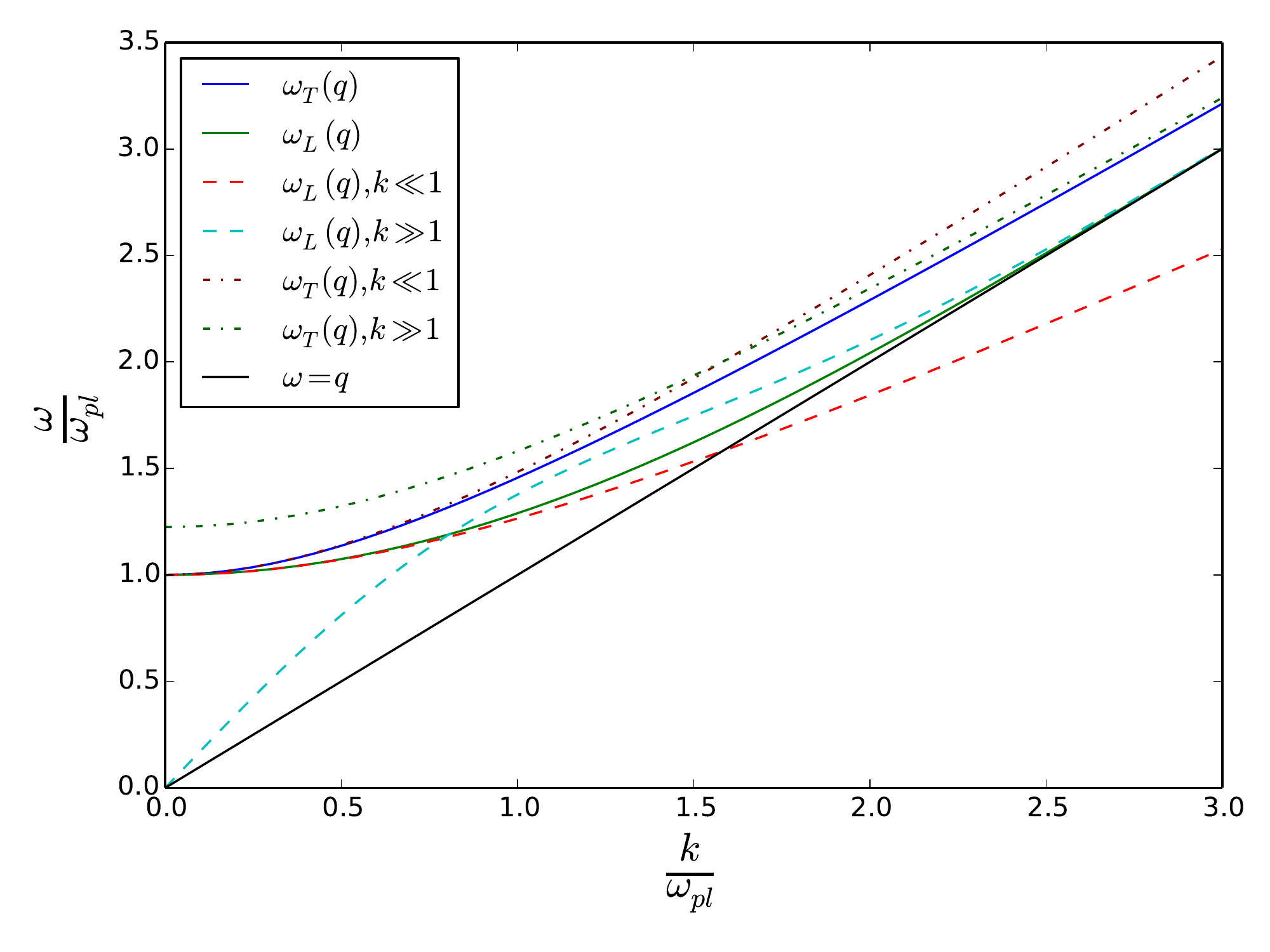}}
 \caption{Numerical solutions of the longitudinal and transverse dispersion relations and their low and high momentum approximations given by \equref{eq_wHTL_Taylor_trans}. The numerical solutions are shown with solid lines. The straight solid line corresponds to ultrarelativistic dispersion relation. The low and high momentum approximations of the longitudinal dispersion relation are shown with dashed lines and the corresponding approximation of the transverse dispersion relation are shown with dot-dashed lines. }
\label{fig:HTLDRsol}
\end{figure}

\subsection{Summary of basic concepts}
We will briefly summarize the basic concepts which we will be using repeatedly throughout the rest of this PhD thesis.
\begin{itemize}
\item The dispersion relation $\omega\left(k\right)$ tells us the relationship between energy and momentum of an excitation with momentum $k$. It is obtained by solving for the poles of the propagator. 
\item The plasmon mass $\omega_{pl}$ corresponds to the dispersion relation at $k=0,$ thus $\omega_{pl} = \omega\left( 0 \right).$ If the system exhibits quasiparticle excitations this will also be the mass of the quasiparticles. 
\item The asymptotic mass $m$ is the mass scale that is observed at large momenta in the HTL dispersion relation. There the dispersion relation is approximately relativistic $\omega = \sqrt{k^2 + m^2}.$
\item The Debye mass $m_D$ is conceptually different from the two mass scales mentioned above. The Debye mass is connected to screening of charges in the plasma. On the perturbative level HTL predicts $m_D^2 = 3 \omega_{pl}^2.$ We will be measuring $m$ and $\omega_{pl}$ in this thesis instead of focusing on screening effects in the plasma.
\item The spectral function $\rho$ tells us the possible spectrum of frequencies $\omega$ which correspond to an excitation with momentum $k$. If the spectral function is sufficiently peaked in frequency space (i.e. it exhibits the Lorentzian form shown in \equref{eq_lor_curve}) the system exhibits quasiparticle excitations. In the HTL framework the spectral function also features a continuum of low frequency (compared to $k$ of the mode) excitations which are known as the Landau cut.
\end{itemize}
%\pagenumbering{arabic}
\chapter{Plasmon mass scale in classical Yang-Mills theory}
\label{plasmon}
In a classical field simulation the degrees of freedom are fields, and not particles. When these fields oscillate with a well defined dispersion relation, such that for a specific momentum $p$ we observe an excitations at frequency $\omega \left( p \right)$, we call these field modes plasmons. Thus plasmons are   quasiparticles describing plasma oscillations. These quasiparticles correspond to the particle degrees of freedom in the kinetic theory description of the strongly interacting matter. Thus we would expect the quasiparticle picture to be valid in the regime where the occupation number of the system is no longer nonperturbatively large (when both the classical field and the kinetic theory description of the system are expected to be valid). Here we want to study the quasiparticle picture in more detail in order to understand its limits in the description of strongly interacting gauge fields.

In this chapter we will mostly focus on extracting the quasiparticle mass, which corresponds to the oscillation frequency of the zero momentum modes in the fields. The spectal properties of the classical theory are studied in more detail in  \ch \ref{response}. We will start by introducing different methods used to estimate the plasmon mass scale. Our aim  is to compare different methods used to estimate this quantity. The results presented in this chapter were originally published in papers \cite{Lappi:2016ato} and \cite{Lappi:2017ckt}.

%Physically the plasmon mass scale determines the plasma instability growth rate \cite{Romatschke:2005pm, Romatschke:2006nk}, which may contribute to the isotropization process of the glasma. We also want to study to what extent the strongly interacting overoccupied system of gauge fields can be understood in a quasiparticle picture. However, we can address this question only for the three dim,ensional system, since to our knowledge it is nontrivial whether a kinetic theory description exists in the two-dimensional case.

\section{Methods}
We start by introducing the three different methods we use to estimate the plasmon mass scale. 

\subsection{Uniform electric field (UE)}
This method was first described in \re \cite{Kurkela:2012hp}. In this method we introduce a spatially homogeneous electric field into the system corresponding to a perturbation with zero momentum. This triggers an oscillation of energy between electric and magnetic energy, and the frequency of this oscillation corresponds to the plasmon mass. Intuitively this can be understood as follows: introducing the uniform electric field corresponds to introducing a large amount of coherent quasiparticles (plasma oscillations of the zero mode). The energy of these quasiparticles oscillates between the electric and the magnetic sectors. When the amount of these quasiparticles we introduced is so large that the oscillations between the electric and magnetic sectors overwhelm the noise in electric and magnetic energies we can measure their frequency from the oscillation frequency of the electric and magnetic energies.

We choose the magnitude of the introduced field in such a way that the amount of energy we inject into the system is roughly 10 \% of the total energy of the system. This amount has been chosen to satisfy two mutually exclusive requirements. Firstly, one wants to keep the amount of injected energy as small as possible in order not to perturb the system too much with the introduced field. However when one increases the amount of energy introduced in the system, the oscillations between electric and magnetic energies become clearer, increasing the signal to noise ratio. However the price to pay for the increased resolution is that the perturbation grows stronger. In practice, in the three-dimensional case we have observed that when the amount of energy introduced lies between 5 and 30 \% the change in measured plasmon mass is about 5 \%. 

The extraction of the plasmon mass is done by doing a damped oscillation fit to the data. In principle this also allows for the extraction of the damping rate. In the two-dimensional case we measure the plasmon mass also by computing the autocorrelation of the energy, and measure the frequency. The correlation function is defined as 
\begin{equation}
c_{av}[k] = \sum_n a[n+k] v[n]^*,
\end{equation}
where $a$ and $v$ are sequences, padded with zeros wherever necessary to keep the sum well defined. In practice $a$ and $v$ will be electric or magnetic energies measured on each time step.

The main drawbacks of this method are the computational cost, the fact that it breaks Gauss' law and its destructivity to the system. The high computational cost arises from the fact that when one introduces the uniform electric field, one also drastically perturbs the system. After the uniform electric field is introduced we can no longer perform other measurements since we have greatly altered the system.  In order to get an estimate for the plasmon mass, one must follow the full time-evolution of the system over a period proportional to a few inverse plasmon mass scales. Introducing a spatially uniform electric field would not break Gauss' law in classical electrodynamics. However, in a nonabelian theory Gauss' law involves parallel transport. This means that in the presence of a nonzero gluon field, a uniform electric field will break Gauss' law, leading to creation of unphysical charges. This charge can be smeared  in a diffusion like process to restore Gauss' law as illustrated in \re \cite{Moore:1996qs}. 

\begin{figure}
\centerline{\includegraphics[width=0.65\textwidth]{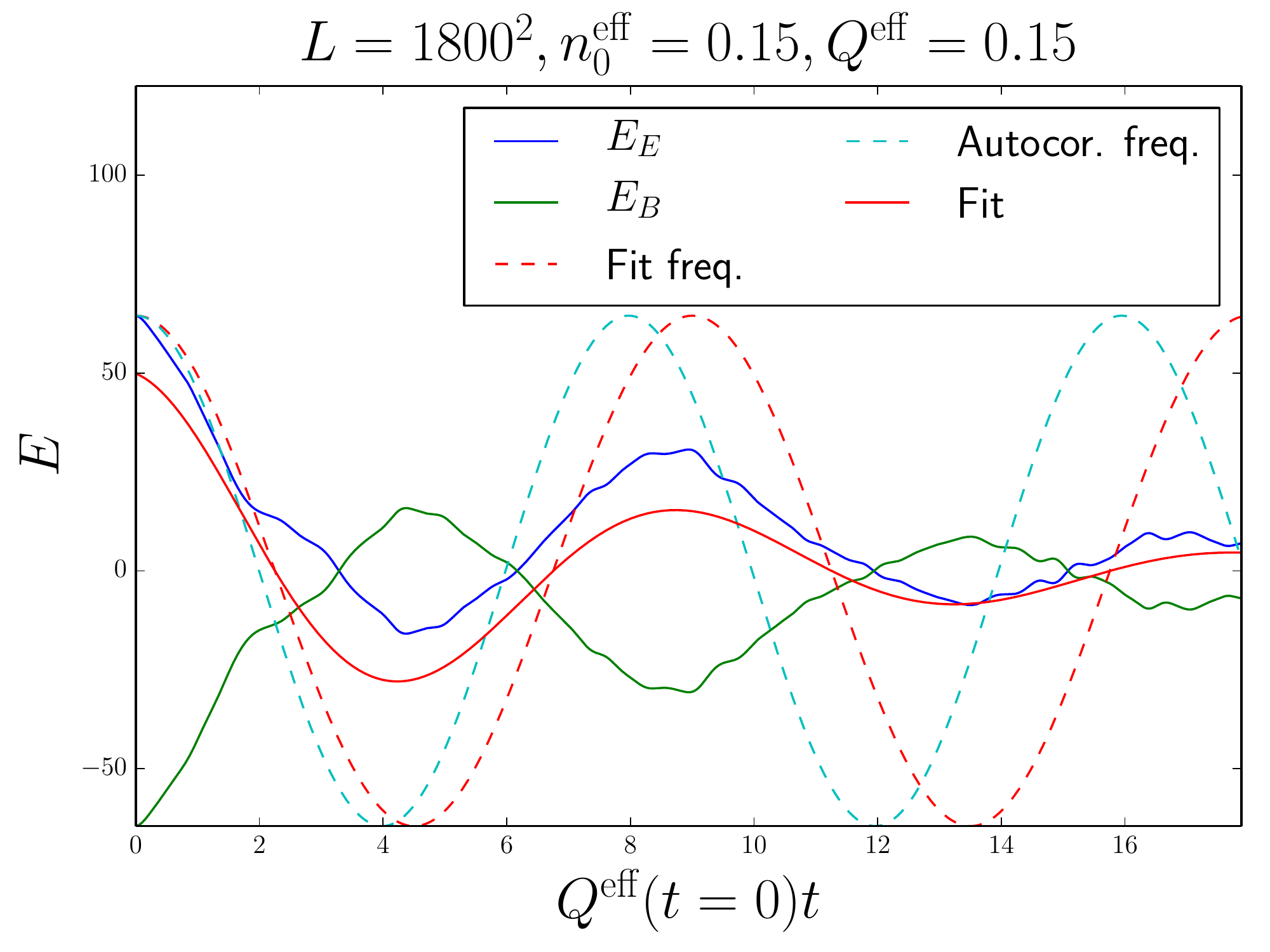}}
 \caption{The measured electric and magnetic energies after the addition of the uniform chromoelectric field. The signal has been moved to oscillate around zero by subtracting the time average in order to make the frequency extraction easier. The curve labeled as ``fit'' shows the damped oscillator fit. The frequency of the fit (the same curve without damping) is shown by the dashed red curve, and the frequency obtained from the autocorrelation measurement is shown by the dashed light blue curve.  The uniform electric field was introduced at $Q^{\mathrm{eff}}t = 160.$
}
 \label{fig:ueosc}
\end{figure}

\begin{figure}
\centerline{\includegraphics[width=0.7\textwidth]{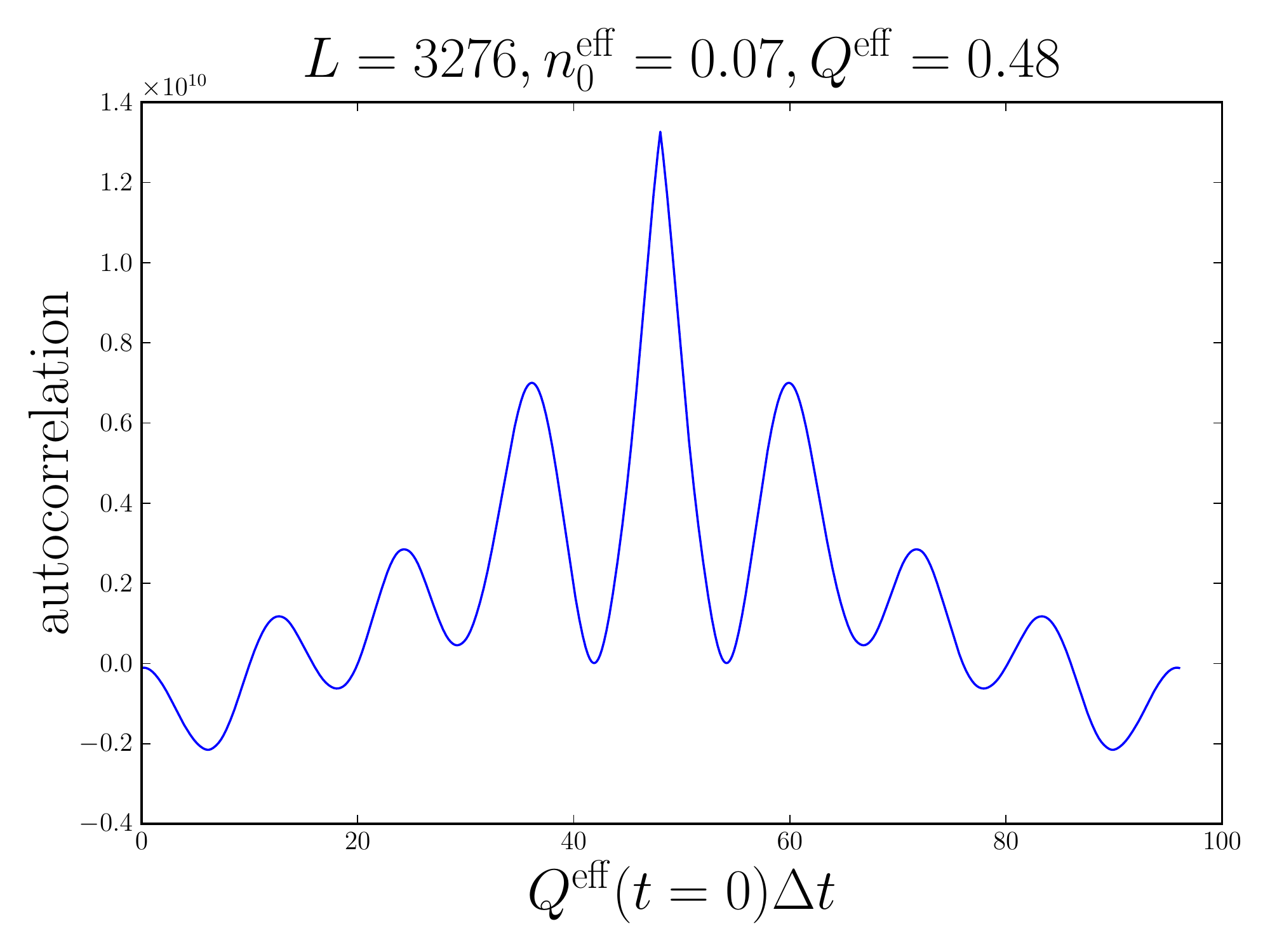}}
 \caption{Autocorrelation function of the electric energy after the introduction of the uniform electric field is shown. The distance between the peak and the first maxima gives the period of the oscillation divided by 2. 
}
 \label{fig:autocor}
\end{figure}

Typical oscillations between the electric and magnetic energies are shown in \fig \ref{fig:ueosc} for a two-dimensional simulation. We also show a fitted damped oscillator function and the extracted frequencies for the fit and the autocorrelation function. A typical autocorrelation function of the electric energy is shown in \fig \ref{fig:autocor}. The figure also demonstrates the ability of the autocorrelation function to eliminate noise from the signal. 

\subsection{Effective dispersion relation (DR)}
\label{subsec:effdr}
As we already explained, we expect the quasiparticle excitations to appear as damped oscillating field modes of the gauge field. Thus in the case of  quasiparticle excitations we expect the gauge field to take the functional form 
\begin{equation}
A_i\left(x \right) \sim \exp{\left( i \boldsymbol{k} \cdot \boldsymbol{x} -  i \omega t\right)} \exp{\left( -\gamma t \right)},
\end{equation}
where $\omega$ is the frequency of the wave, $\boldsymbol{k}$ is the wavevector and $\gamma$ is the damping rate. Remembering that the electric field is the time derivative of the gauge field, we expect that the squared sum of the damping rate (which we assume to be negligible compared to the frequency) and the frequency is captured by the expression  
\begin{equation}
\omega^2\left(k\right) = \dfrac{\left<\left|E_{i}^a\left(k\right)\right|^2 \right>}{\left< \left|A_{i}^a\left(k\right)\right|^2 \right>},
\label{eq:huonodispersiorelaatio}
\end{equation}
where the subscript C stands for Coulomb gauge. We will shortly address a way to measure the damping rate and the frequency separately, shedding more light on the concerns whether the damping rate is small compared to the frequency. The expression \equref{eq:huonodispersiorelaatio} has been used to measure the dispersion relation in (2+1) dimensional gauge theory \cite{Krasnitz:2000gz}, and more recently in e.g. \cite{Berges:2012ev}. It turns out that this estimate undershoots the plasmon mass (this problem is especially severe in three dimensions, in two dimensions the problem is not as bad as in three dimensions). In order to alleviate this problem we consider another similar estimate
\begin{equation}
\omega^2_{T,L}\left(k\right) = \dfrac{\left<\left|\dot{E}_{i,T,L}^a\left(k\right)\right|^2 \right>}{\left< \left|E_{i,T,L}^a\left(k\right)\right|^2 \right>},
\label{eq:dispersiorelaatio}
\end{equation}
where the dot refers to the time derivative, and the subscripts T and L refer to the longitudinal and transverse components. Thus we can study the longitudinal and transverse components separately, which is not possible using \eq (\ref{eq:huonodispersiorelaatio}), since the Coulomb gauge gluon field is always transverse. The plasmon mass is then extracted by performing a fit of the form $\omega^2 = \omega_0^2 + a k^2$, with $\omega_0$ and $a$ as free parameters, to the data obtained using \eqs (\ref{eq:dispersiorelaatio}) and (\ref{eq:huonodispersiorelaatio}). The choice of the fit cutoff is a delicate matter, since with too large cutoff the fit is dominated by high momentum modes, which easily overwhelm the data in the infrared region. A too small cutoff leads to a fit which may not have a physical long wavelength behaviour (i.e. the parameter $a$ considerably deviates from unity). 

If we assume that the electric fields are damped oscillators of the form $e^{i\omega t - \gamma t},$ the DR given by equation (\ref{eq:dispersiorelaatio}) actually gives us  $\omega^2 + \gamma^2.$ 
With this approach one can solve for the damping rate and dispersion relation. The dispersion relation is given by
\begin{equation}
\omega^2\left(p\right) = \dfrac{\left<\left(\mathfrak{Re}\dot{E}\right)^2\right>}{\left<\left(\mathfrak{Re}E\right)^2\right>} - \left|\dfrac{\left<\mathfrak{Re}E \mathfrak{Re}\dot{E}\right>}{\left<\left(\mathfrak{Re}E\right)^2\right>}\right|^2.
\label{eq:omega2b}
\end{equation}
The damping rate can be obtained using
\begin{equation}
\gamma^2\left(p\right) =  \left|\dfrac{\left<\mathfrak{Re}E \mathfrak{Re}\dot{E}\right>}{\left<\left(\mathfrak{Re}E\right)^2\right>}\right|^2.
\label{eq:gamma2b}
\end{equation}
In three dimensional simulations we have observed that \eq (\ref{eq:omega2b}) gives roughly similar values as equation (\ref{eq:dispersiorelaatio}). We have also observed that the damping rate given by \eq (\ref{eq:gamma2b}) is negligible compared to the plasmon mass. Thus we can use equation (\ref{eq:dispersiorelaatio}) to determine the plasmon mass scale.

\begin{figure}[t]
\centerline{\includegraphics[width=0.65\textwidth]{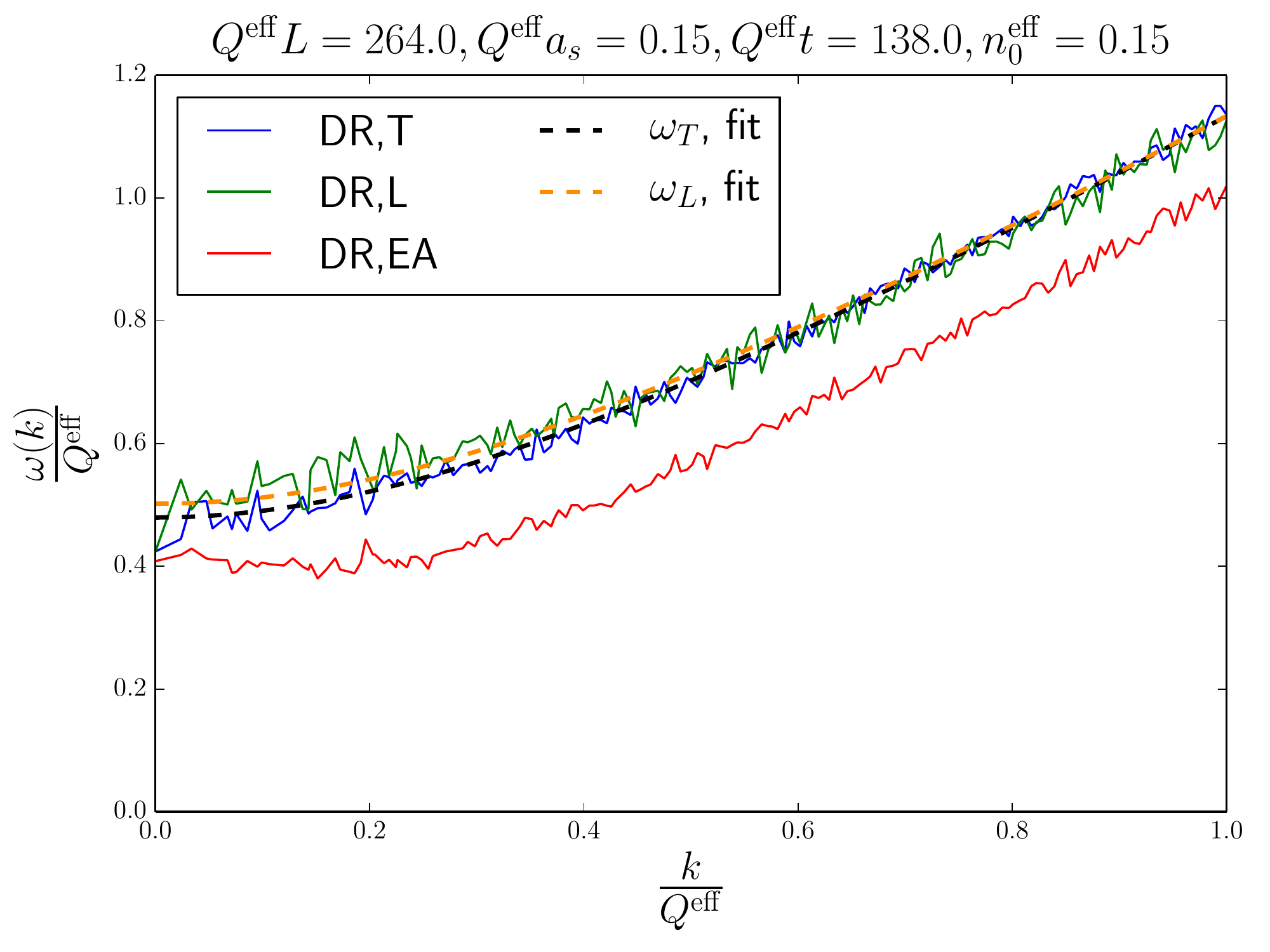}}
 \caption{Effective dispersion relations extracted from a two dimensional simulation. The cutoff used in the fit is $\nicefrac{k}{Q^{\mathrm{eff}}} < 0.25$ here. Averaged over 20 configurations.}
 \label{fig:DR}
\end{figure}

Figure \ref{fig:DR} features a typical dispersion relation extracted from a two-dimensional simulation using different estimates for the transverse and longitudinal dispersion relations. We also show fits to the transverse dispersion relations. 

\subsection{HTL resummed approximation (HTL)}
The third method is to use a formula, which one can derive from thermal field theory in the hard thermal loop approximation. Provided that the HTL type separation of scales between the soft scale $\omega_{pl}$ and the dominant hard  momentum scale $Q$ is valid in our case, the plasmon mass is given by 
\begin{equation} \label{eq:htlintegral}
\omega_{pl}^2 = \dfrac{4}{3} g^2 N_c \int \dfrac{\mathrm{d}^3k}{\left(2 \pi \right)^3} \dfrac{f\left(k \right)}{k}.
\end{equation}
On the lattice we discretize the integral in the usual way
\begin{equation}
\int\dfrac{ \mathrm{d}^3k}{\left(2\pi \right)^3} \rightarrow \sum_k\dfrac{1}{V}.
\end{equation}

This is probably the most widely used measurement in the literature, e.g., \res \cite{Epelbaum:2011pc,Berges:2013fga,Mace:2016svc}. Because the occupation number distribution is used as an input in \eq (\ref{eq:htlintegral}), the result we get for the plasmon mass naturally depends on the definition of the occupation number distribution. In three dimensions this turns out to be not so important, after a few $\nicefrac{1}{Q}$  the different definitions become equivalent above the Debye scale. The contribution of the modes below the Debye scale is not very important in three dimensions. In two dimensions, because of the different phase space, one observes large variations between the different methods even at late times.

The fact that \eq (\ref{eq:htlintegral}) uses the quasiparticle spectrum as an input, and that the occupation number (\ref{eq:discF}) also needs the dispersion relation as an input raises the question how can they both be extracted self-consistently. However, it is not obvious how the plasmon mass should be introduced into \eq (\ref{eq:discF}). And this mass correction would be of the same order as the higher order terms in gluon field, which are neglected anyway in \eq (\ref{eq:discF}). Thus we will be using a massless dispersion relation here.

The HTL formalism is typically applied for systems in thermal equilibrium. However, we expect it to be valid also out of equilibrium as long as the fundamental building block of scale separation is valid. In HTL perturbation theory the scales are set by the hard scale, temperature $T$, and the soft scale is given by the screening scale $gT$. In classical theory the coupling can be completely scaled out of the computation by redefinition of the fields, and thus one can not expand in its powers. In classical simulations the hard scale is given by $Q$  and the soft scale is given by the mass $m$. For non-Abelian plasma close to the self-similar scaling solution we expect to have a separation between these scales. These scales also evolve in time, the hard scale increases as $t^{\nicefrac{1}{7}}$ and the soft scale decreases as $t^{\nicefrac{-1}{7}}$ \cite{Kurkela:2012hp}. This means that the scale separation increases in time, improving the validity of the HTL approach.

\section{Plasmon mass scale in three-dimensional isotropic system}
We start by studying the three-dimensional isotropic case, where the HTL expectation is better understood than in the two-dimensional system. However, the two-dimensional system is physically more relevant, since it mimics the boost invariant 2+1 dimensional system occurring in the initial stages of ultrarelativistic heavy-ion collisions.% For the three-dimensional isotropic system the comparison to the thermal system is the most straightforward. 

\subsection{Initial conditions}
The initial conditions are sampled from a distribution which corresponds to the following quasiparticle spectrum
\begin{equation}
g^2 f\left(k, t=0 \right) = n_0 \dfrac{k}{\Delta} \exp{\left( \dfrac{-k^2}{2 \Delta^2}\right)}.
\label{eq:initdist3d}
\end{equation}
Here the parameter $n_0$ determines the typical occupation number of a mode at the momentum scale $\Delta$. For the classical approximation to be valid we should have $\nicefrac{n_0}{g^2} \gg 1$. Originally this kind of initial condition was first used in \res \cite{Berges:2012ev,Berges:2007re}. At $t=0$ we initialize the system using only gauge fields in order to satisfy Gauss' law. 

\begin{figure}
\centerline{\includegraphics[width=0.7\textwidth]{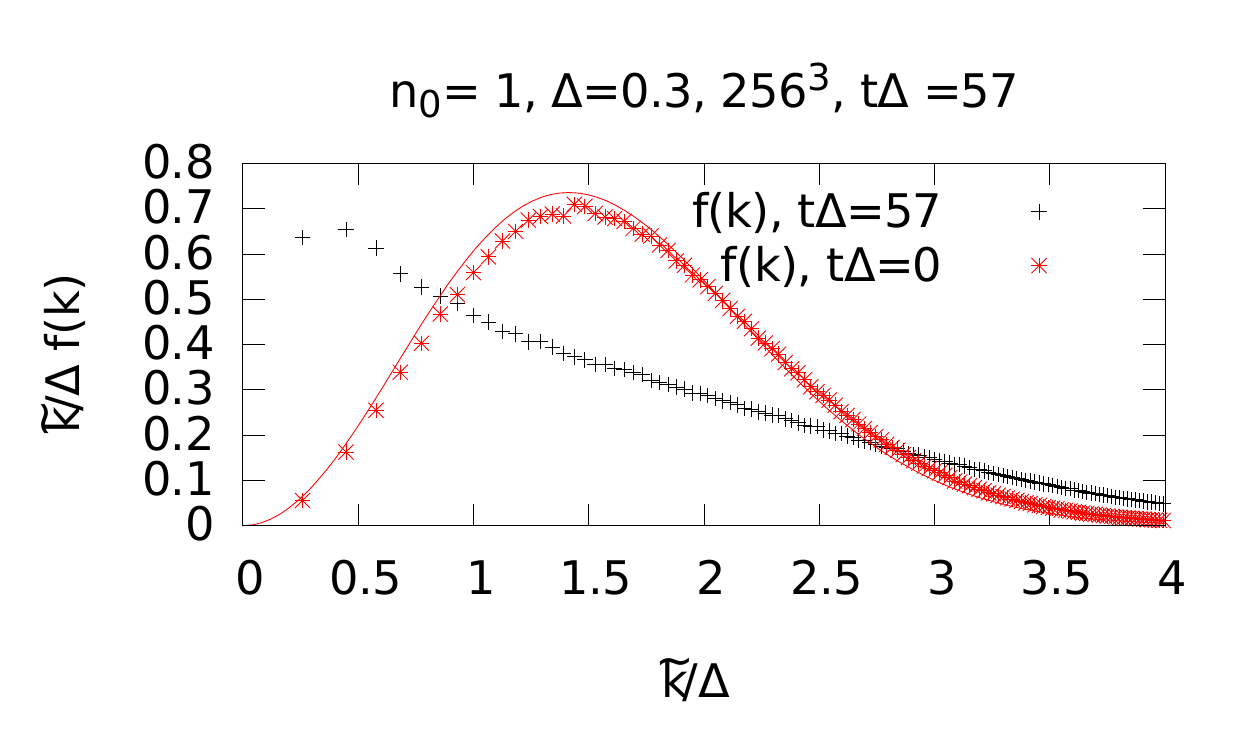}}
 \caption{Here we show the particle number at the initial time along with the analytical form of the initial condition and the at later time $t \Delta = 57.$}
\label{fig:fdifdef}
\end{figure}

The initial occupation number distribution is shown in \fig \ref{fig:fdifdef} along with the analytical curve. We also show the distribution after some time-evolution has taken place ($t \Delta = 57$). The reason for the slight deviation from the analytical curve might be that the initial links are obtained by exponentiating the gauge field, and the gauge field is then extracted by taking the trace. This procedure always introduces an error which is proportional to the lattice spacing. We have not, however, explicitly tested this by going to smaller lattice spacings.

\begin{figure}
\centerline{\includegraphics[width=0.7\textwidth]{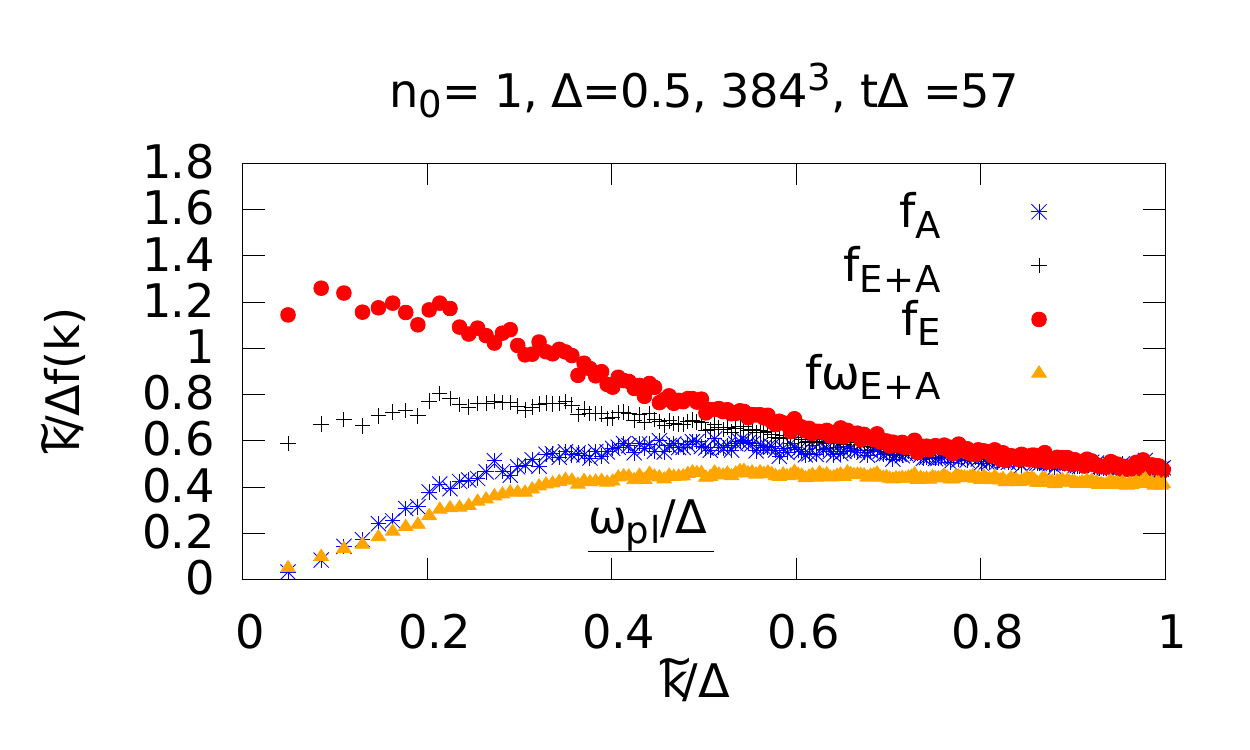}}
 \caption{Comparison of different definitions for the occupation number distribution. The extraction of $f_E,$ $f_A$ and $f_{E+A}$ are done using \eqs (\ref{eq:discF}), (\ref{eq:discFE}) and (\ref{eq:discFA}). The curve $f\omega_{E+A}$ has been obtained by first extracting the dispersion relation as explained in \se \ref{subsec:effdr}, and using the numerically extracted dispersion relation as an input for the occupation number extraction.  }
\label{fig:fmethods}
\end{figure}

Figure \ref{fig:fmethods} shows the occupation number extracted using different definitions. We observe that the deviations between the definitions shrink as we go to higher momenta. We have also marked the approximate location of the plasmon mass scale in the figure. Above the Debye scale, which perturbatively lies at $\sqrt{3} \omega_{pl}$ we would expect the differences between the different definitions to vanish, which indeed seems to take place.  

\subsection{Dependence on lattice cutoffs}
\begin{figure}
\centerline{\includegraphics[width=0.7\textwidth]{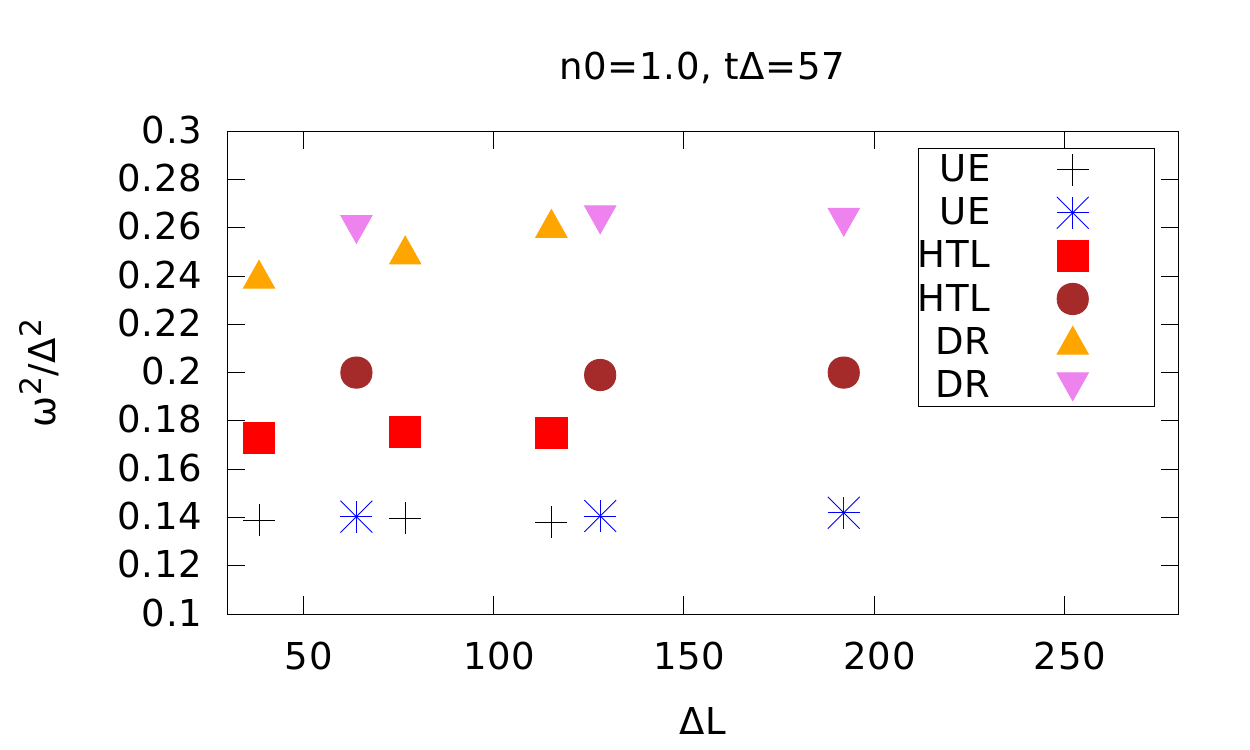}}
 \caption{Dependence of the plasmon mass on the  infrared cutoff. We have performed these computations for two separate UV cutoffs. For the upper datapoints  $a_s \Delta = 0.3$ and for the lower points $a_s \Delta = 0.5.$ The HTL and UE methods are not sensitive to the infrared cutoff. Using the DR method one can observe a slight decrease for smaller system sizes, however, one has to keep in mind that the statistical accuracy of the DR method is not as good as that of the other methods.  }
 \label{fig:ircutoff}
\end{figure}

The dependence on the IR cutoff $L \Delta$ is shown in \fig \ref{fig:ircutoff} for two different UV-cutoffs $a_s \Delta$. We find that HTL and UE methods are insensitive to the IR cutoff. For the DR method we observe a possible decreasing trend at small IR cutoffs. However, since the trend is not very drastic and the statistical accuracy of the DR method is worse than that of the other methods, we can not conclusively say that there is a cutoff dependence in the DR method.

\begin{figure}
\centerline{\includegraphics[width=0.7\textwidth]{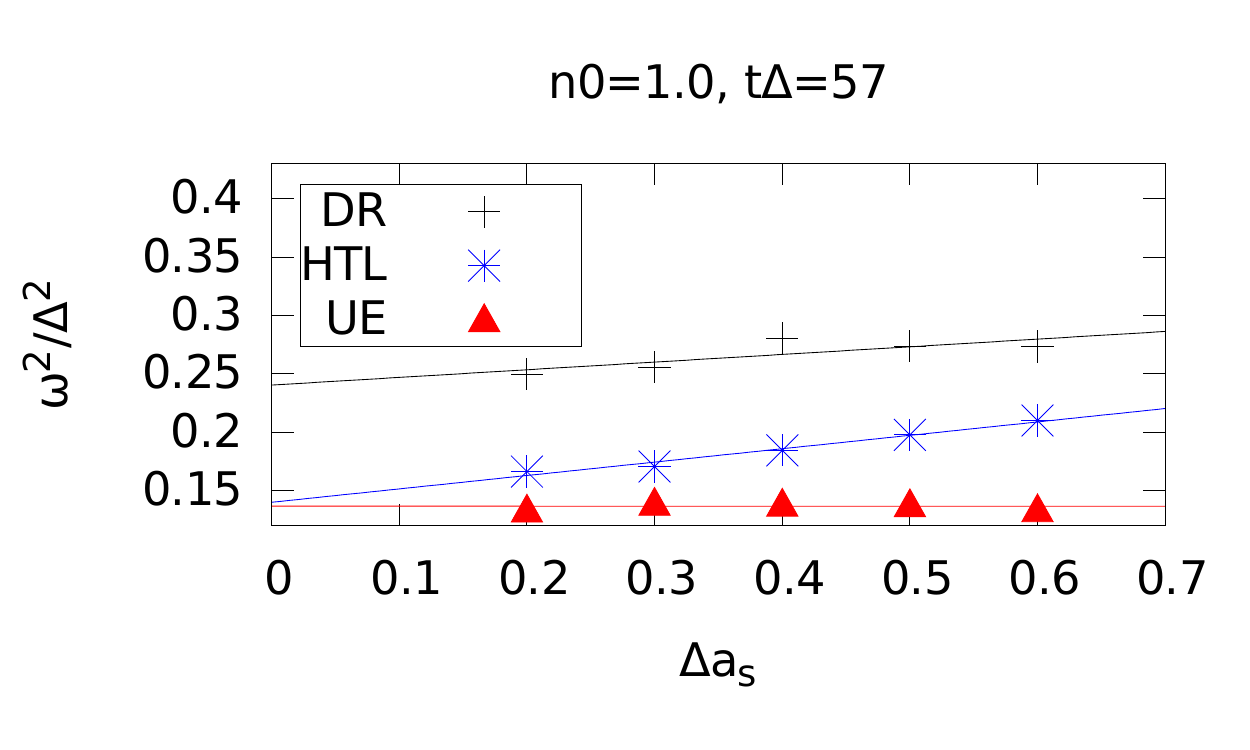}}
 \caption{Here we show the ultraviolet cutoff dependence of our results. The straight lines correspond to a linear extrapolation to $a_s = 0.$  }
 \label{fig:uvcutoff}
\end{figure}

The ultraviolet cutoff dependence is shown in \fig \ref{fig:uvcutoff}. We observe that while DR and UE methods seem to be insensitive to variations in UV cutoff, the HTL method seems to have a decreasing trend when we go to the continuum limit. It is also noteworthy that the continuum is on the left in \fig \ref{fig:uvcutoff}, and since we have to keep the physical lattice size fixed, we have used larger lattices (in terms of number of points) for the datapoints on the left. Thus we expect their statistical accuracy to improve while going to the left in this plot.

\begin{figure}
\centerline{\includegraphics[width=0.8\textwidth]{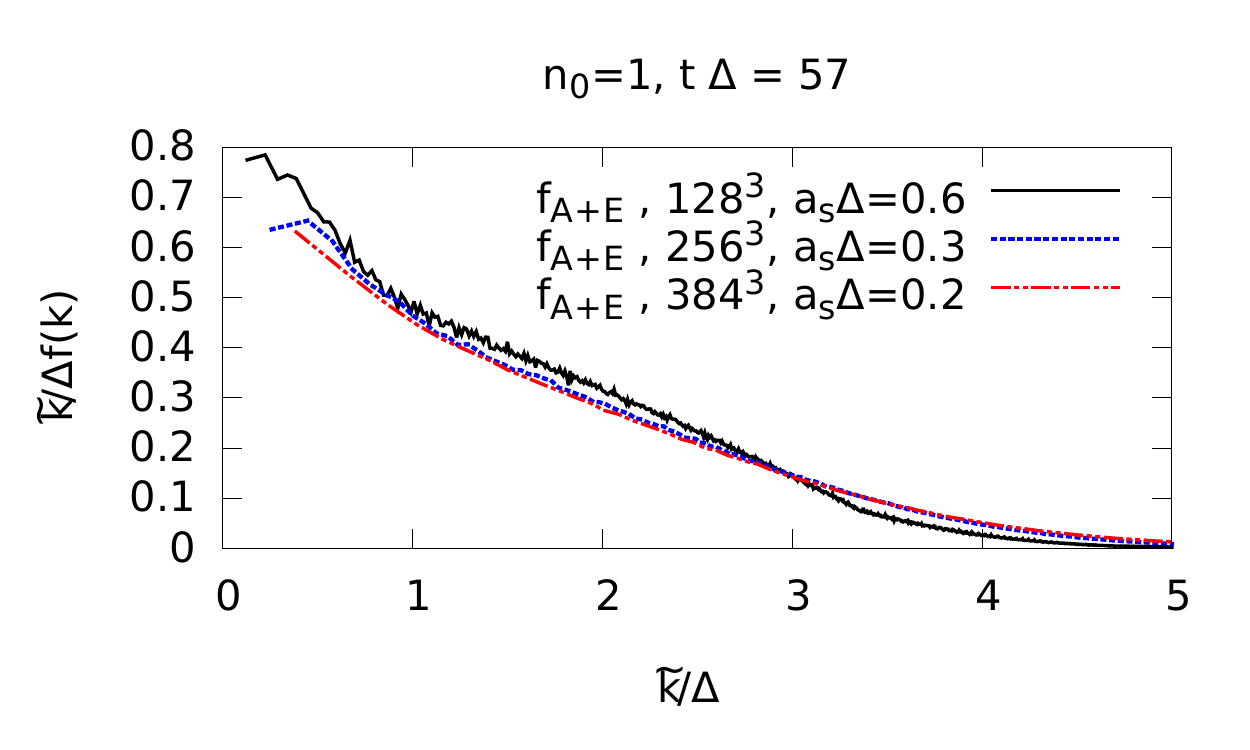}}
 \caption{ Ultraviolet cutoff dependence for the integrand of the HTL formula (\ref{eq:htlintegral}).   }
 \label{fig:htlintegrand}
\end{figure}

In order to better understand how the cutoff dependence arises, we have plotted the HTL integrand for various UV cutoffs in \fig \ref{fig:htlintegrand}. We observe that the choice of the ultraviolet cutoff also changes the behaviour of the quasiparticle distribution in the infrared region. It seems that when more phase space opens up in the ultraviolet region more energy gets transferred from the IR to the UV. 

\subsection{Dependence on time and occupation number}
\begin{figure}
\centerline{\includegraphics[width=0.7\textwidth]{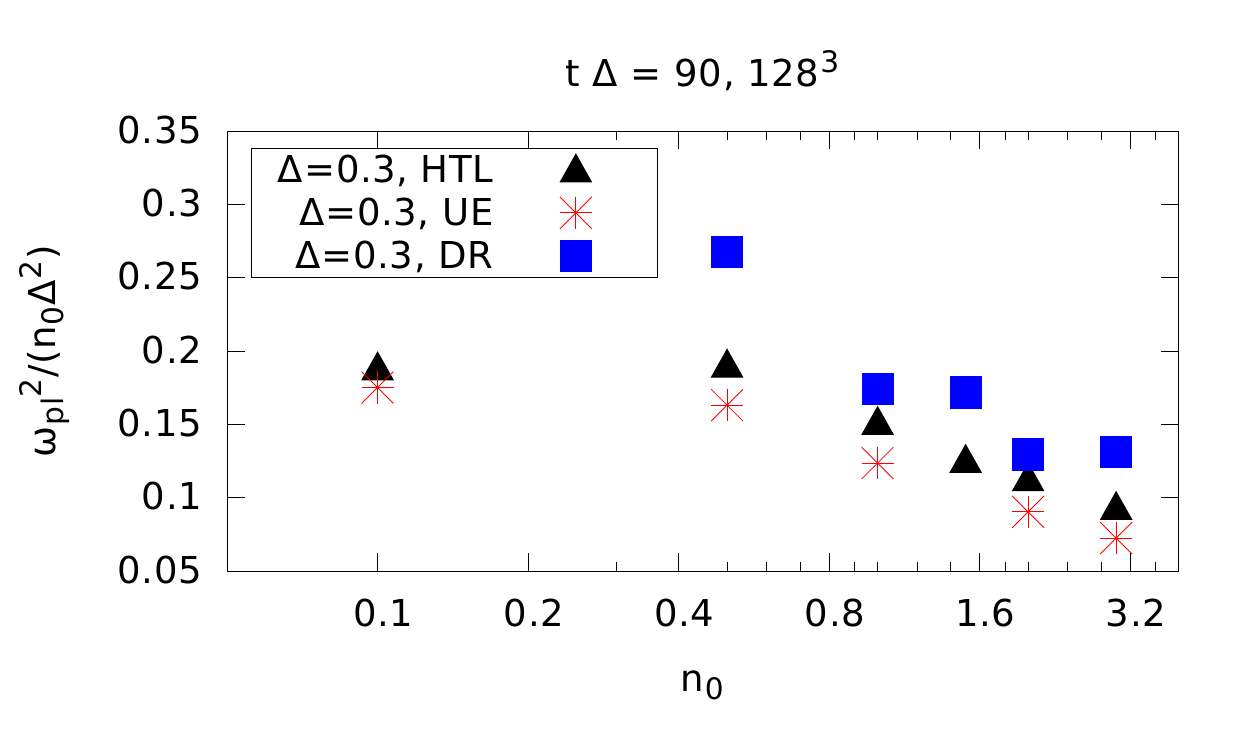}}
 \caption{Occupation number dependence of the plasmon mass scaled by the occupation number parameter $n_0$ for different methods of evaluating the plasmon mass scale. }
\label{fig:occupnumberdep}
\end{figure}

Next we study the dependence of the plasmon mass on the physical simulation parameters: time and occupation number. Figure \ref{fig:occupnumberdep} shows the dependence of the plasmon mass on the occupation number of the system at a fixed time. We observe that the relationship between the different methods depends only weakly on the occupation number (at smaller occupation numbers the system seems to be still in the initial transient regime). The relevant scales of the problem are the mass scale $\omega_{pl}$ and the hard scale $\Delta$. The separation between these two scales is set by the occupation number, since according to the HTL formula (\ref{eq:htlintegral}) we have $\omega_{pl} \sim n_0$. When we increase the occupation number, these scales approach each other, and also the ambiguity in the definition of the occupation number grows larger (since the definition of the occupation number becomes more difficult below the Debye scale). We observe that the decrease in the plasmon mass is less than linear in $n_0$, even though one would expect linear dependence based on \eq (\ref{eq:htlintegral}). Instead of nonlinear effects, we expect that the reason for this behavior is that the time dependence is also different for different $n_0$.

\begin{figure}
\centerline{\includegraphics[width=0.9\textwidth]{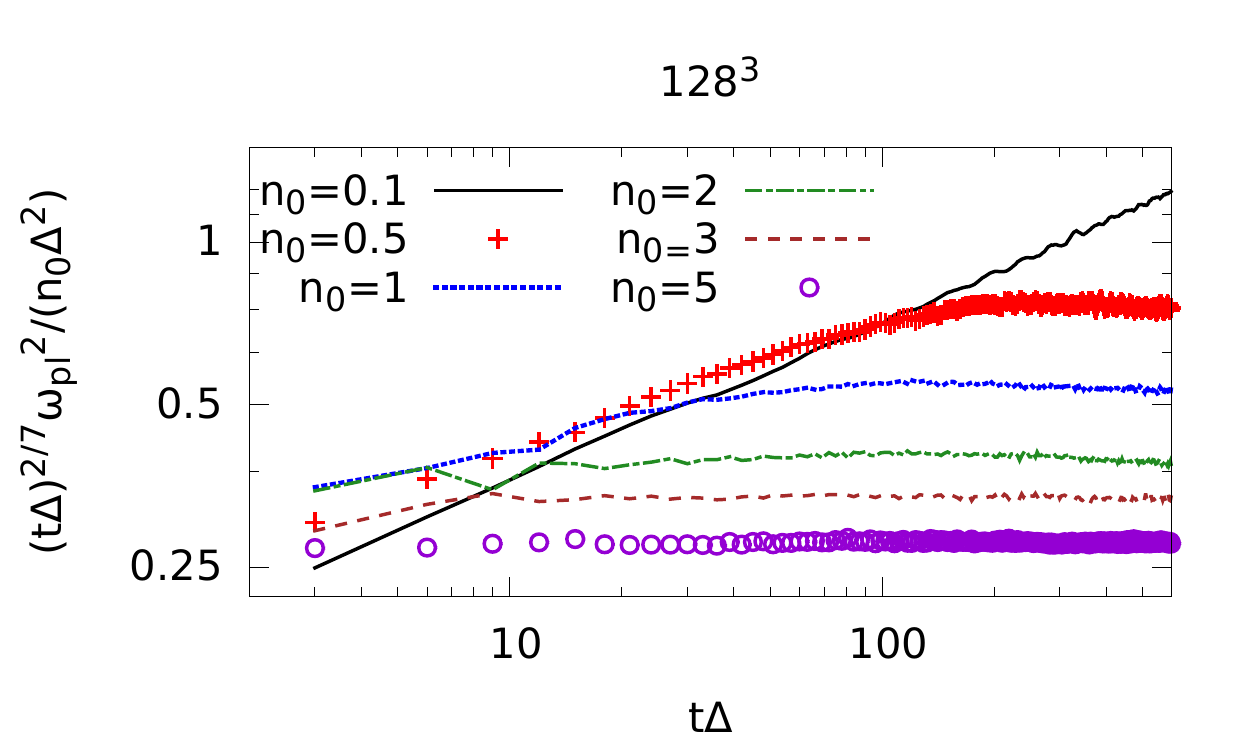} }
\centerline{\includegraphics[width=0.9\textwidth]{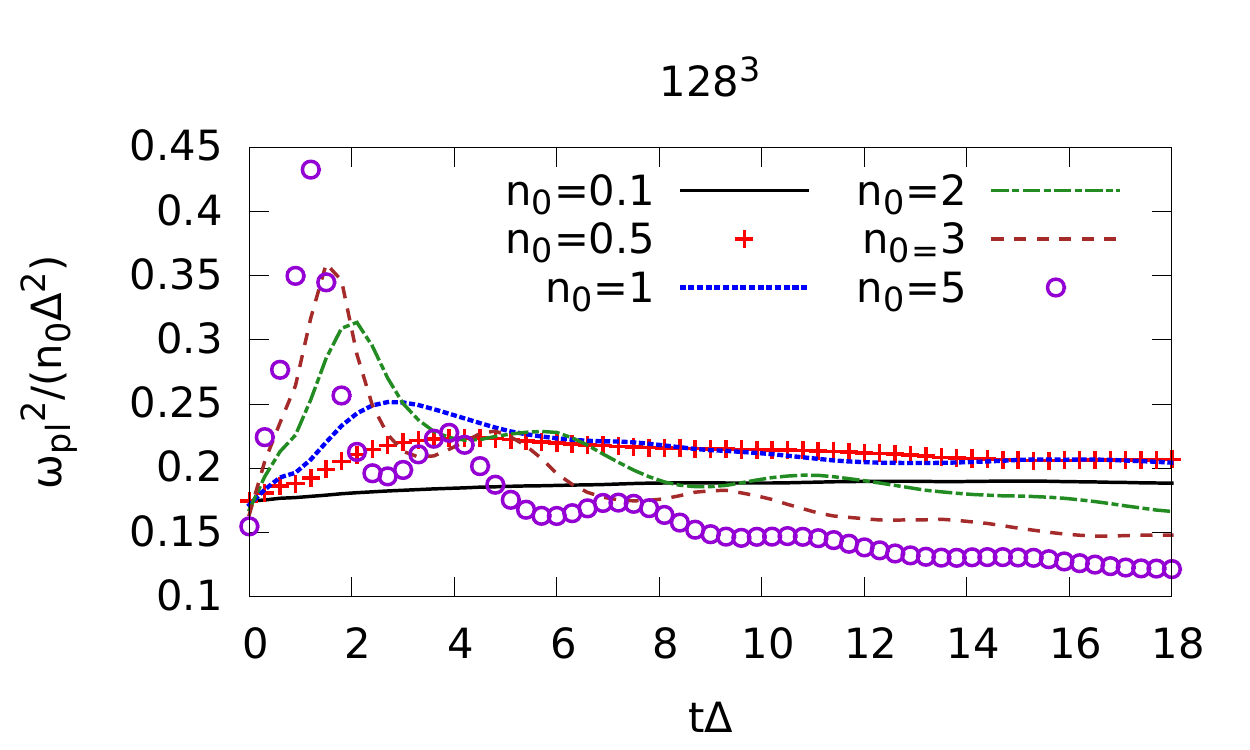}}
 \caption{Time dependence of the plasmon mass scale (scaled by the occupation number and momentum scale) using the HTL method. The upper figure concentrates on longer timescales, and the lower figure on the initial transient behaviour. We observe that simulations with larger occupation numbers settle faster into the asymptotic behavior. We find that the late time behavior is consistent with the $t^{\nicefrac{-2}{7}}$ power law. Here we use $\Delta = 0.3.$ }
 \label{fig:tdep}
\end{figure}

The time dependence of the plasmon mass scale is studied using the HTL method in \fig \ref{fig:tdep}. The upper plot shows the time-dependence on longer time scales and the lower plot shows the behavior during the initial transient time. The late time behavior seems to be consistent with the $t^{\nicefrac{-2}{7}}$ power law proposed in \re \cite{Kurkela:2012hp}. The duration of the initial transient behavior seems to be set by the occupation number, providing an explanation for the observations we made when interpreting \fig \ref{fig:occupnumberdep}. More highly occupied systems have spent a larger fraction of their lifetime in the scaling solution regime.

\begin{figure}
\centerline{\includegraphics[width=0.7\textwidth]{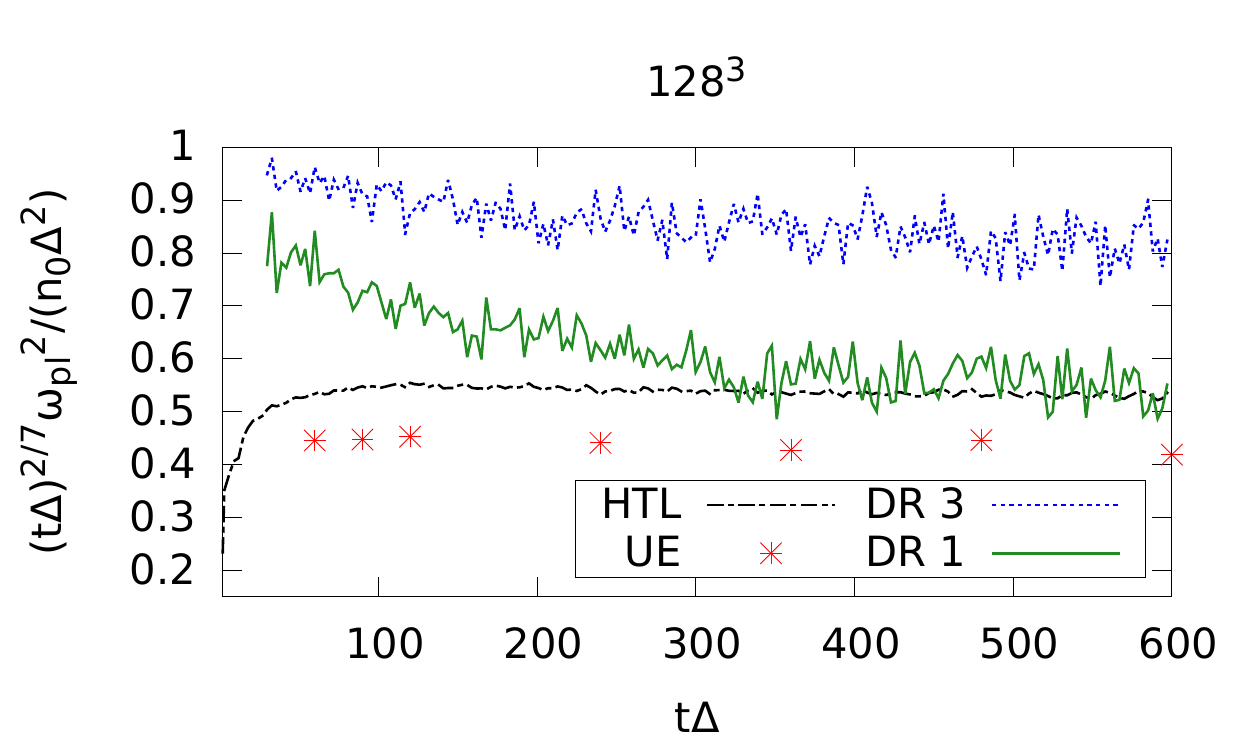}}
 \caption{The time dependence of the plasmon mass scale for all methods is shown here. The labels DR 1 and DR 3 refer to the cutoffs used in the fit (in $\nicefrac{k^2}{\Delta^2}$) when estimating the plasmon mass scale using the DR method. We observe that the behavior at late times is consistent with the $t^{\nicefrac{-2}{7}}$ power law for all methods (taking into account the large uncertainty in the DR method). The simulations parameters here are $\Delta = 0.3$ and $n_0 = 1$.} 
 \label{fig:tdep2}
\end{figure}

Figure \ref{fig:tdep2} shows the time dependence of the plasmon mass scale using all three methods.  We find that all methods agree on the asymptotic power law behavior, and that the UE and HTL methods are in agreement at late times. The DR method suffers from poor statistical accuracy and strong cutoff dependence (the two curves refer to two fits with different fit cutoffs), which prohibits us from drawing any firm conclusions about its behavior. The poor statistical accuracy is also visible, even though the results of the DR method have been averaged over 20 runs (while the others are results from a single run). At late times also the agreement of the DR with other methods seems to improve, when the separation between to two mass scales becomes clearer.

In the next section we will study the mass scale in two-dimensional systems. In there we will denote the momentum scale by $Q$ instead of $\Delta,$ which was used here, for reasons which will become apparent.

\subsection{Summary of the results in three dimensions} 
Our results indicate that the UE method is probably the most reliable method for the extraction of the plasmon mass scale. The reason for this is its insensitivity to cutoffs. The main drawback is its big computational cost. The UE method agrees with HTL when the ultraviolet cutoff is sufficiently small. This agreement also signifies that the kinetic theory description seems to be a valid way to  understand an overoccupied classical system of gauge fields. 

The DR method seems to give 50 \% larger values for the plasmon mass scale than the other methods. It also requires much more statistics than the other methods, and is sensitive to the cutoff used in the fit.

\section{Plasmon mass scale in two-dimensional system}
\subsection{Initial conditions}
Our next step is to proceed into two-dimensional system, which we implement by doing simulations on a three-dimensional lattice with $N_z = 1.$ As explained in the introduction, our main motivation is to mimic the boost invariant glasma fields created in an ultrarelativistic heavy-ion collision. Here we want to use a similar initial condition as in three dimensions in order to make the comparison as straightforward as possible. However, it turns out that gauge fixing makes this more complicated in 2 dimensions than in 3 dimensions.

The initial gauge fields are chosen to satisfy
\begin{equation}
g^2 f\left(k, t=0 \right) = n_0 \dfrac{\ktt}{Q} \exp{\left( \dfrac{-\ktt^2}{2 Q^2}\right)}
 \frac{(2\pi)\delta(k_z)}{a_s}.
\label{eq:initdist}
\end{equation}
Similarly as in the three-dimensional case, our initial condition contains only magnetic modes to guarantee that Gauss' law is satisfied at $t=0$. 

\begin{figure}[t]
\centering
\begin{minipage}{0.48\textwidth}
\includegraphics[width=1.0\textwidth]{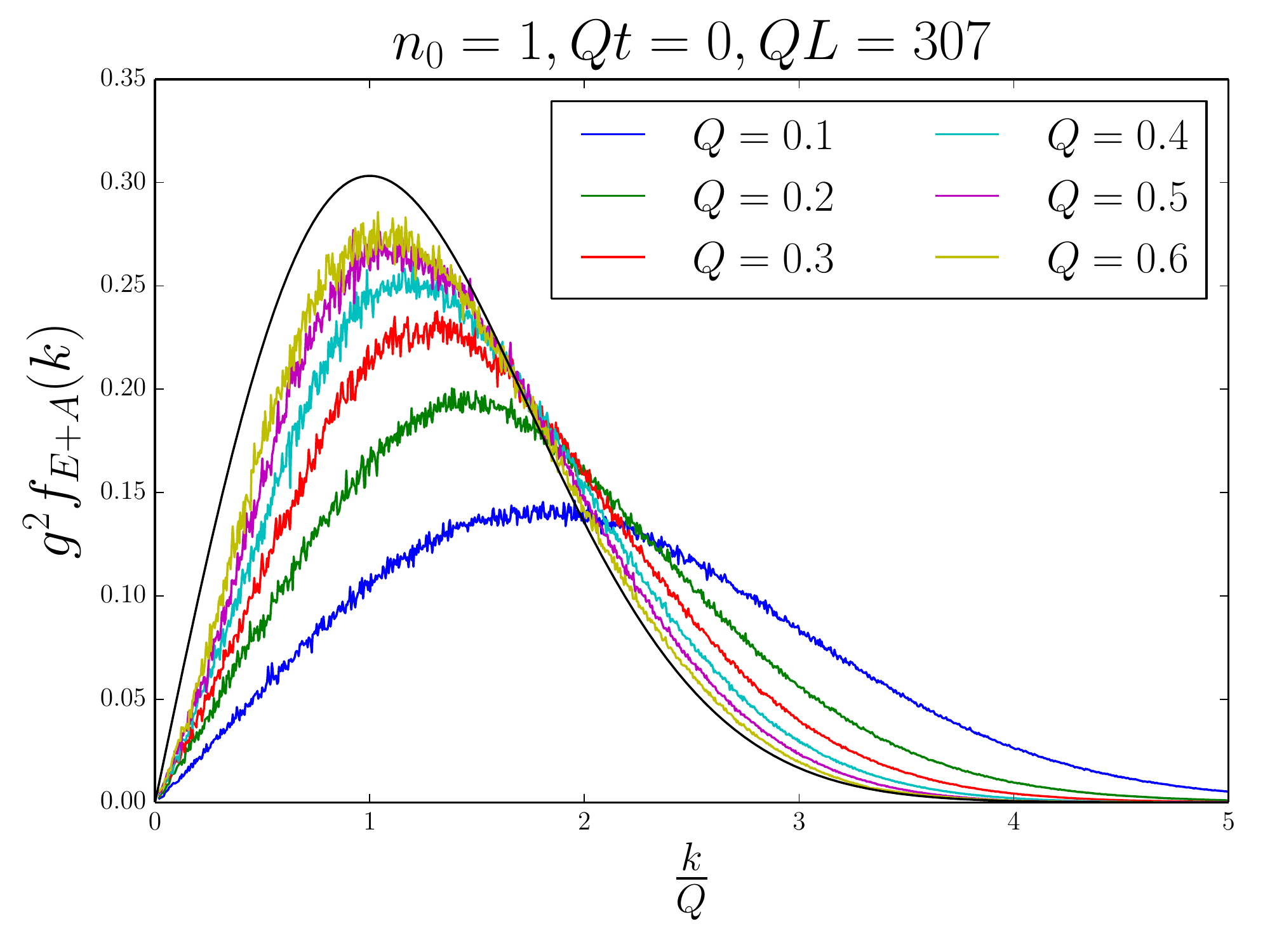}
\end{minipage}
\begin{minipage}{0.48\textwidth}
\includegraphics[width=1.0\textwidth]{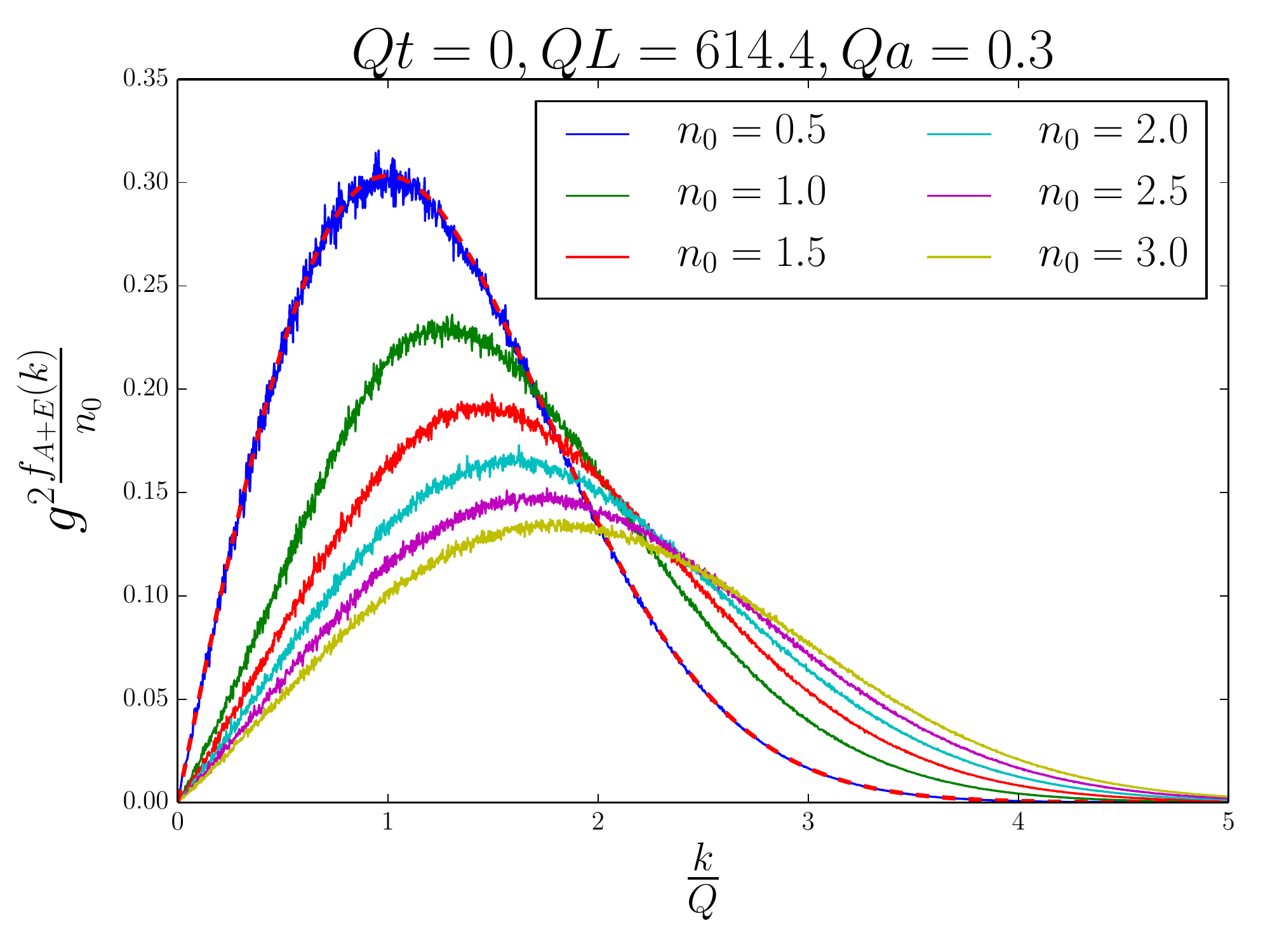}
\end{minipage}
 \caption{Left: The quasiparticle spectrum for several initial momentum scales. The analytical distribution \eq (\ref{eq:initdist}) is shown as solid black line. The deviations are the worst for the configurations with smallest $Q a_s$. Right: The effect of the occupation number on the quasiparticle distribution. We observe that more highly occupied systems are the most affected by the gauge fixing procedure. The analytical curve is shown as the red dashed line here. Averaged over 10 runs. Note that the y-axis in the right hand side plot is $g^2 \nicefrac{f}{n_0}$ contrary to what was shown in paper \cite{Lappi:2017ckt}, where the factor of $n_0$ was missing.}
 \label{fig:fvsn0}
\end{figure}

Figure \ref{fig:fvsn0}  demonstrates how the initial occupation number distribution behaves as a function of $Q$ and $n_0$. We observe that at larger $Q$ and $n_0$ the system greatly deviates from the analytical form. The reason for this is that the spectrum is a gauge fixed observable, and the gauge fixing has a large deforming effect on the spectrum.  This happens in spite of the fact that we project out the longitudinal component of the gauge field before the exponentiation. This happens because we construct the gauge links by exponentiating the gauge fields and use the antihermitean traceless part as its inverse operation. This procedure is sufficient when the field amplitudes are sufficiently small. However at larger amplitudes one starts to observe deviations. Thus the links may not satisfy the gauge condition to desired accuracy, forcing us to do some additional gauge fixing when the occupation number or the hard scale is large. This problem is much more severe in 2 dimensions than in 3 dimensions.  Because of this problem, it is not meaningful to compare our observables to these initial parameters (since even the initial deviation might be large). Thus we want to characterize the initial $Q$ and the occupation number with gauge invariant observables, which agree with the initial parameters in the weak field limit. 

\begin{figure}[t]
\begin{minipage}{0.5\textwidth}
\centerline{\includegraphics[width=\textwidth]
{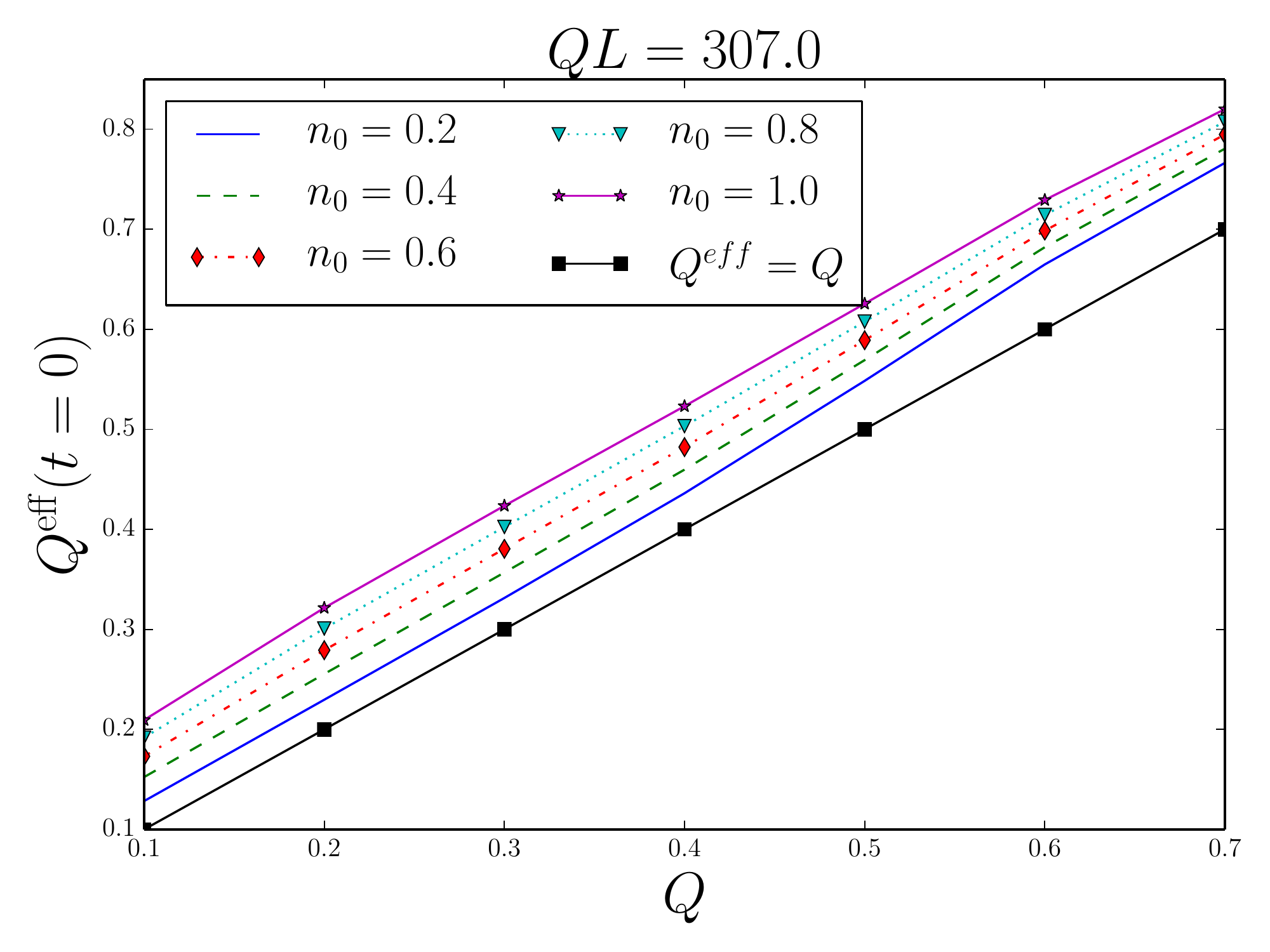}}
\end{minipage}
\begin{minipage}{0.5\textwidth}
\centerline{\includegraphics[width=\textwidth]
{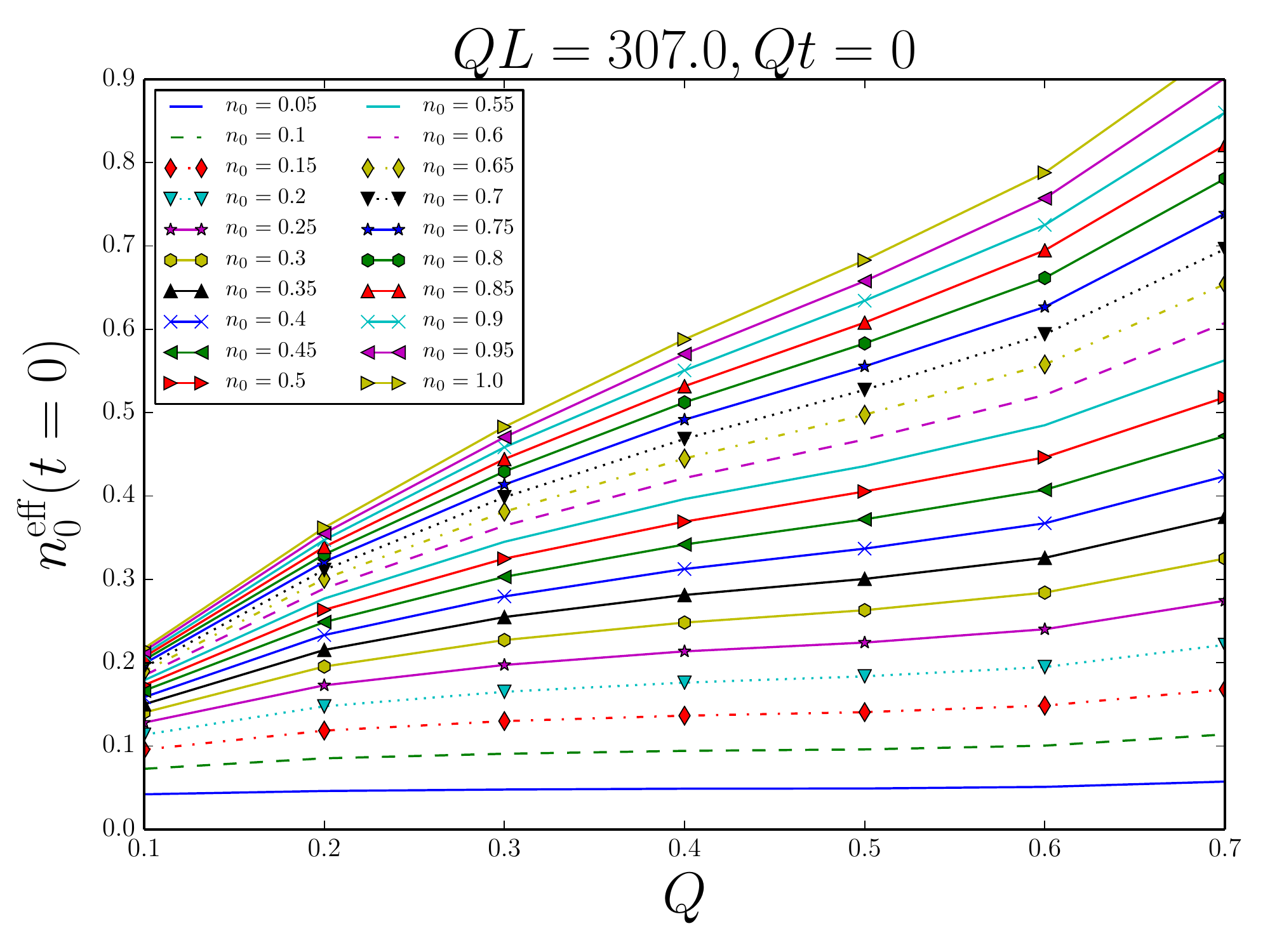}}
\end{minipage}
\caption{Right: The effective occupation number as a function of the initial scale for various $n_0$. For the smallest $n_0$ we find that $n_0$ corresponds to $n_0^{\mathrm{eff}}$ independently of $Q$. For large $n_0$ the choice of the initial scale also contributes. Left: The effective momentum scale as a function of the scale set by the initial condition for various $n_0$. The effective scale is linear in the initial scale, but the occupation number seems to introduce a constant shift to $\Qeff$. }
\label{fig:n0effvsQ}
\end{figure}

Let us start by defining the typical momentum of the chromomagnetic field squared (similarly as in \cite{Bodeker:2007fw,Kurkela:2012hp} )
\begin{equation}
\label{eq:Qeff}
p_{\mathrm{eff}}^2(t) = \dfrac{\left< \mathrm{Tr}\left(\boldsymbol{D} \times \boldsymbol{B} \right)^2 \right>}{\left< \mathrm{Tr} \left(\boldsymbol{B}^2\right) \right>}.
\end{equation} 
For our initial condition we estimate this perturbatively as 
\begin{equation}
p^2_{\mathrm{eff}}(t=0) \approx \dfrac{\int \mathrm{d}\ktt \ktt^4 f(\ktt)}{\int \mathrm{d}\ktt \ktt^2 f(\ktt)} = 4Q^2.
\end{equation}
Thus in the dilute limit we define the effective momentum scale in such a way that it matches the initial momentum scale $Q$
\begin{equation}
Q_{\mathrm{eff}}\left(t \right) = \dfrac{p_{\mathrm{eff}}\left(t \right)}{2}.
\end{equation}

\begin{figure}[t]
\centerline{\includegraphics[width=0.7\textwidth]{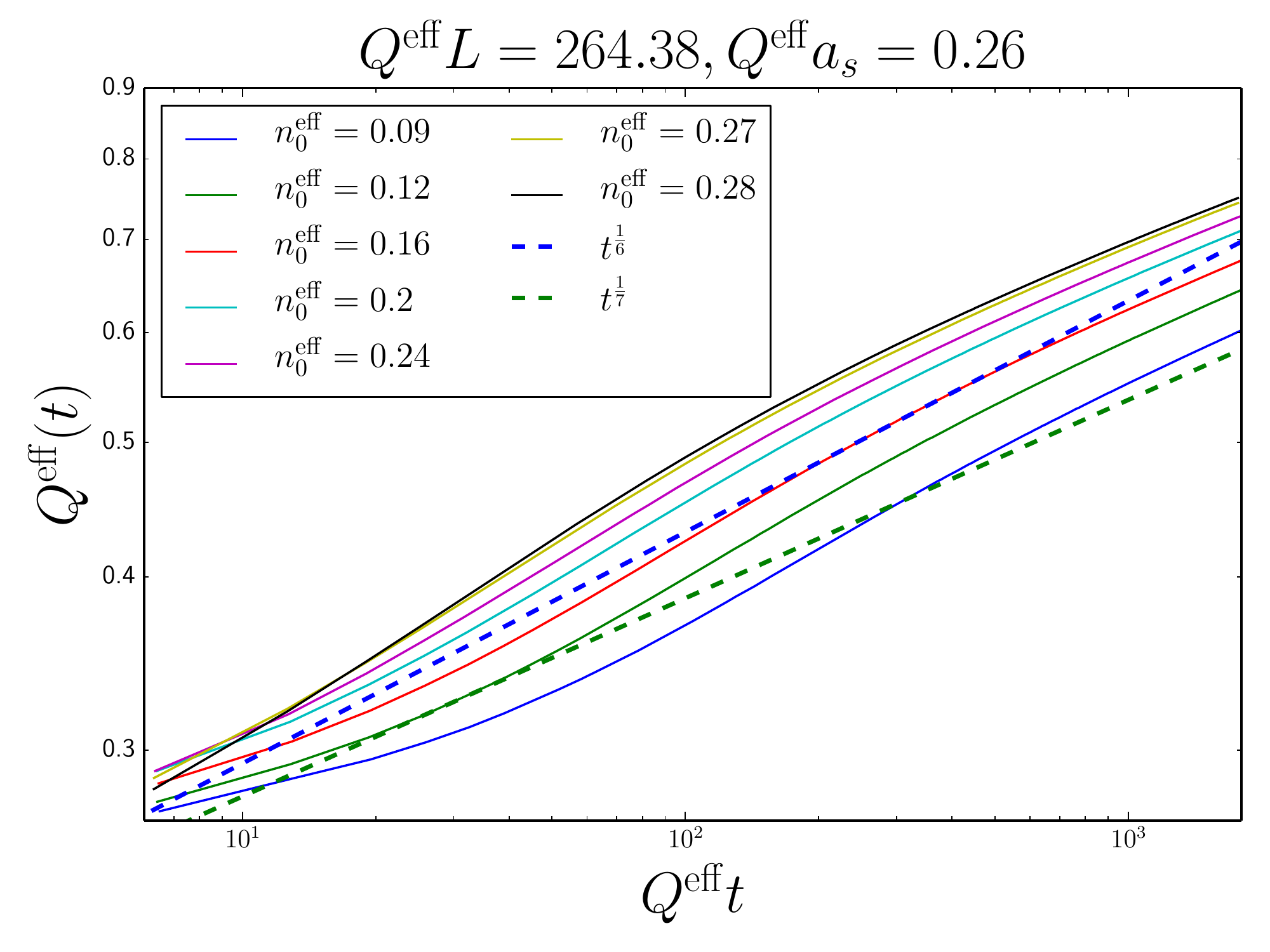}}
\caption{Time dependence of the effective momentum scale is shown for various $\neff$.  The effective momentum scale seems to exhibit power law behavior with the power lying between $t^{\nicefrac{1}{7}}$ and $t^{\nicefrac{1}{6}}$ power laws. Averaged over five runs.}
\label{fig:Qeffvst}
\end{figure}

In order to find a corresponding estimate for the occupation number, consider the energy density of the system. Using \eq (\ref{eq:edensity}) we get 
\begin{equation}
\epsilon \approx  n_0 Q^3 \dfrac{(\nc^2-1)}{\pi}\frac{1}{a_s g^2}.
\end{equation}
Here the momentum scale can be replaced by the gauge invariant momentum scale, and the equation can be solved for $n_0$.
Next define  $\epsilon^{2d}\equiv a_s\epsilon.$ We get
\begin{equation}
n_0^{\mathrm{eff}} = \dfrac{\pi g^2}{(\nc^2-1)} \dfrac{ \epsilon^{2d}}{ Q_{\mathrm{eff}}^3}.
\end{equation}

The left and right panels of figure \ref{fig:n0effvsQ}  demonstrate the behaviour of these effective observables as a function of the initial $Q$. We observe that the $\Qeff$ behaves linearly in $Q,$ but we also see some non-linearities, since the observed $\Qeff$ also increases when the initial occupation number increases. A similar effect was also observed in \fig \ref{fig:fvsn0}, where the peak of the occupation number distribution was shifted towards larger $k$, corresponding to an increase of $\Qeff$. We also see that the effective occupation number is sensitive to the initial momentum scale. The left panel of \fig \ref{fig:n0effvsQ} exhibits this effect. Only the smallest occupation numbers are insensitive to the variations in the initial momentum scales. In practice we will use \fig \ref{fig:n0effvsQ} to bridge the gap between the initial simulation parameters and the effective observables. %In practice, one wants to perform simulations by varying one parameter and keeping the other fixed.  Figure \ref{fig:n0effvsQ}  will allow us to pick suitable (effective) initial parameters for our simulations.

The measured scales are functions of time. Since our simulations are performed in a fixed box, the energy density is a constant quantity. Thus the only parameter with time-evolution is $\Qeff$. Its time-evolution is shown in \fig \ref{fig:Qeffvst}. The time-evolution is consistent with a $Q_{\mathrm{eff}}(t) \sim t^{\nicefrac{1}{7}}-t^{\nicefrac{1}{6}}$ power law. Consequently $n_0^{\mathrm{eff}}(t) \sim  t^{\nicefrac{-3}{7}}-t^{\nicefrac{-1}{2}}$. For the rest of this section we will use the initial scales as the reference scales with the notation $Q_{\mathrm{eff}}(t=0)=Q_{\mathrm{eff}}$ and $n_0^{\mathrm{eff}}(t=0) = n_0^{\mathrm{eff}}$.

\begin{figure}[t]
\begin{minipage}{0.5\textwidth}
\centerline{\includegraphics[width=\textwidth]{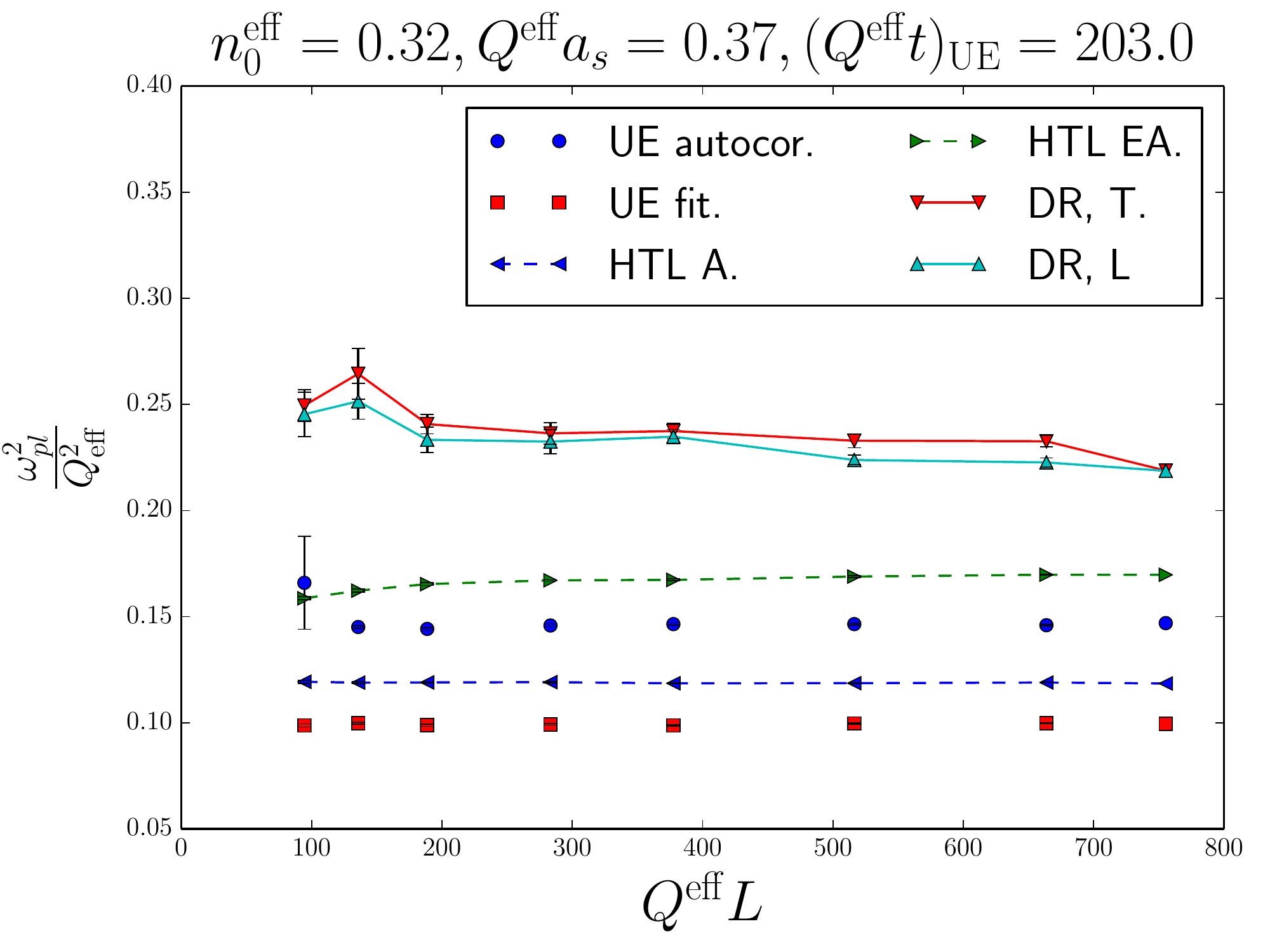}}
\end{minipage}
\begin{minipage}{0.5\textwidth}
\centerline{\includegraphics[width=\textwidth]{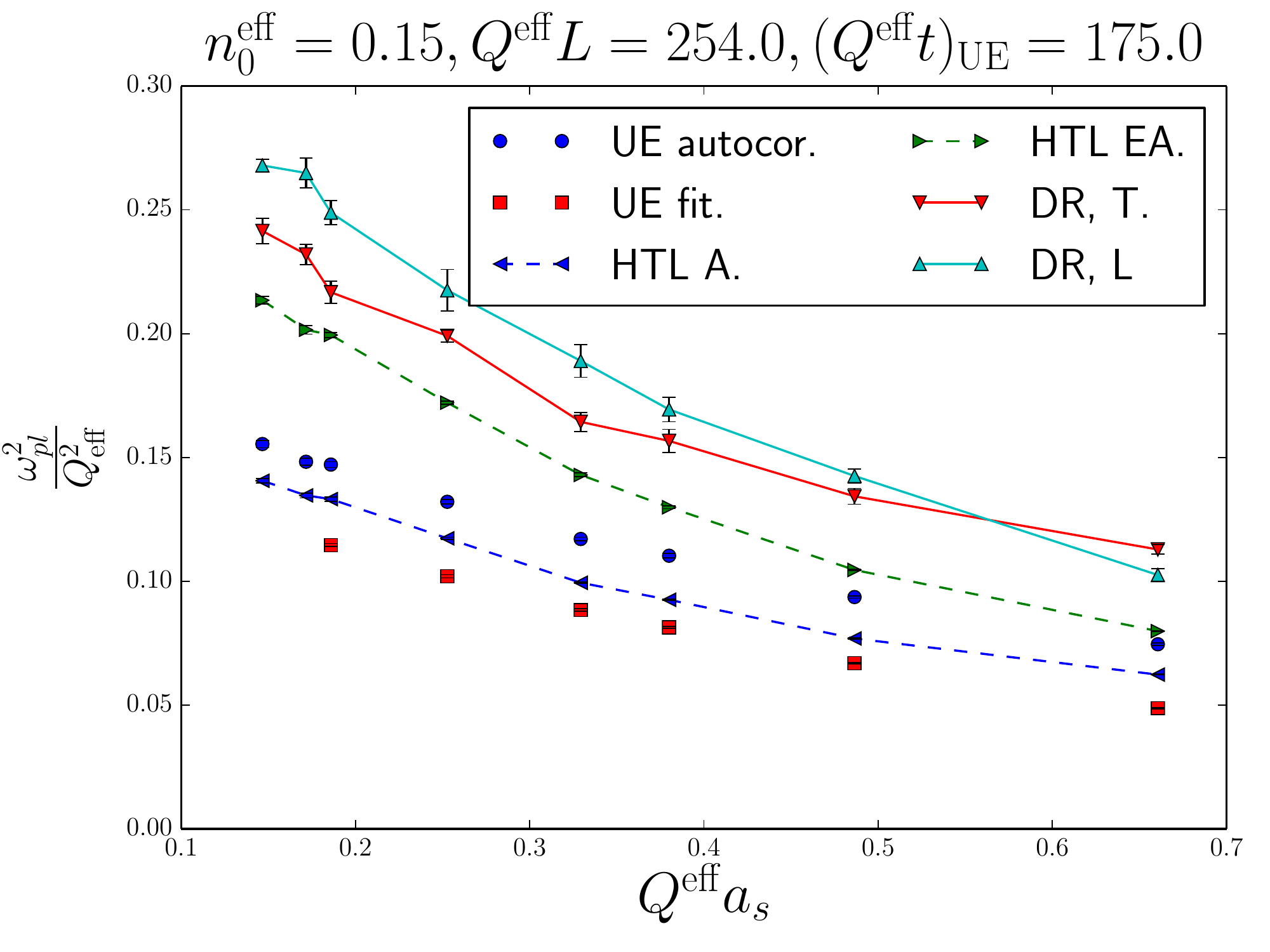}}
 \end{minipage}
  \caption{The right figure shows the dependence on the UV-cutoff. We observe that results obtained using all methods increase when we go towards the continuum. The  lattice sizes we used were $376^2, 514^2, 646^2, 763^2, 1008^2, 1326^2, 1575^2, 1800^2$ with averages taken over 10, 9, 8, 7, 6, 5, 4, 3 simulations. 
  The left figure shows the infrared cutoff dependence for various methods. We find no infrared cutoff dependence.  Here $n_0^{\mathrm{eff}}$ varies between $n_0^{\mathrm{eff}} = 0.3240-0.3246,$ and  $Q^{\mathrm{eff}} = 0.3686-0.369,$ and the lattice sizes used were 256, 368, 512, 768, 1024, 1400, 1800, 2048, and results were averaged over 20, 15, 15, 9, 10, 8, 6, 5 configurations. The title of the plot shows the average values of  $Q^{\mathrm{eff}}$ and $n_0^{\mathrm{eff}}$. 
 } 
 \label{fig:2dcutoff}
\end{figure}

\subsection{Results}
First we want to check how our results depend on the lattice cutoffs, which are given by $\Qeff a_s$ and $\Qeff L$. Some additional uncertainty will arise from the fact, that we can not directly adjust $\Qeff$ and $\neff$. Instead, using left and right plot of \fig \ref{fig:n0effvsQ} we must choose such $n_0$ and $Q$ that we stay as close to fixed $\Qeff$ and $\neff$ as possible. The error bars featured on the plots are statistical errors which are given by the standard error of the mean. They will be insignificant for the UE and HTL measurements. The DR method is statistically less accurate than the other methods.

The cutoff effects are shown in the \fig \ref{fig:2dcutoff}. We observe that all measurements are insensitive to the infrared cutoff. On the contrary, all observables feature a non-negligible ultraviolet cutoff dependence. This result is completely different from the three-dimensional one, but the results do not seem to be UV-divergent.

\begin{figure}[t]
\begin{minipage}{0.5\textwidth}
\centerline{\includegraphics[width=\textwidth]
{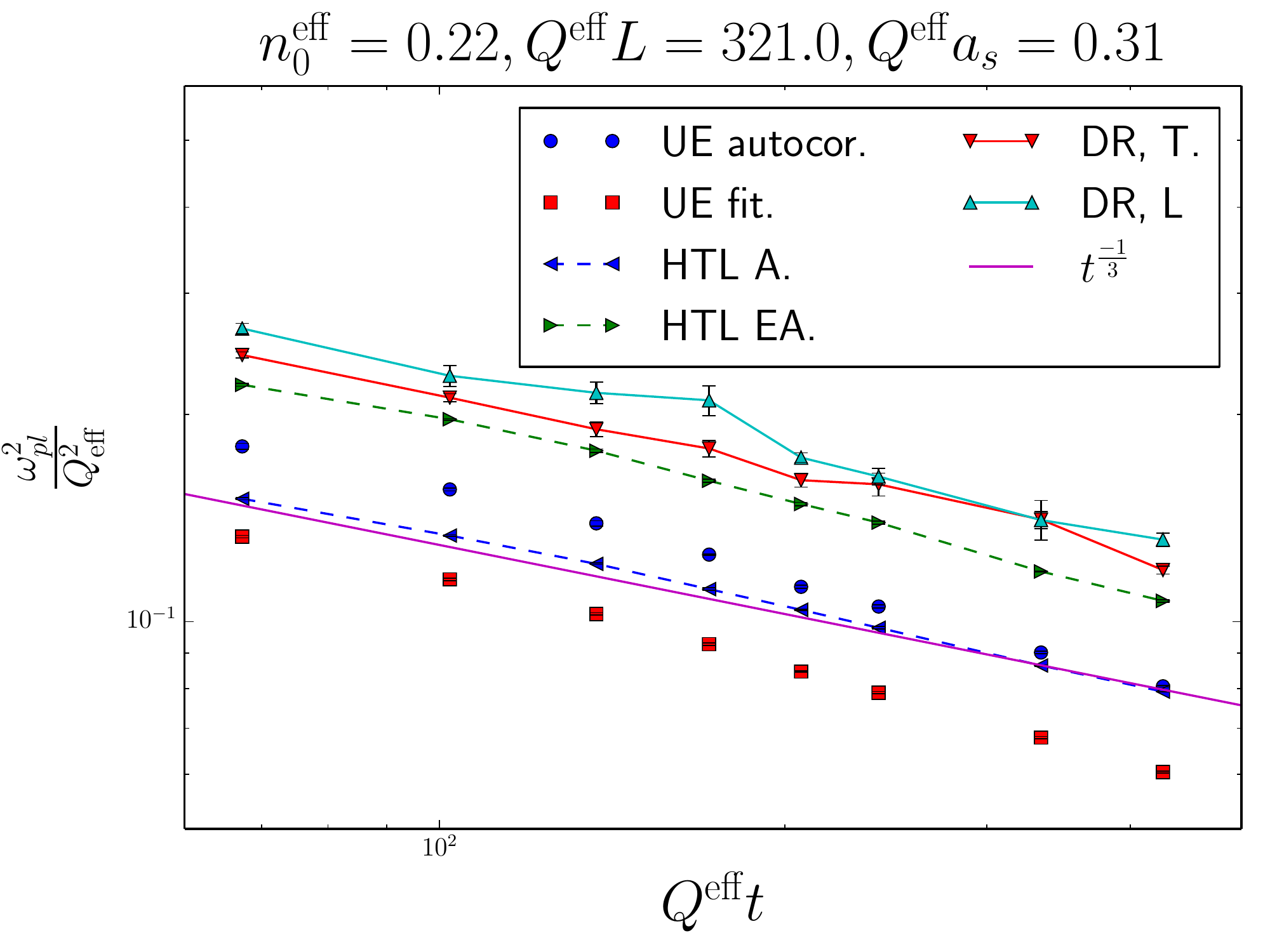}}
\end{minipage}
\begin{minipage}{0.5\textwidth}
\centerline{\includegraphics[width=\textwidth]
{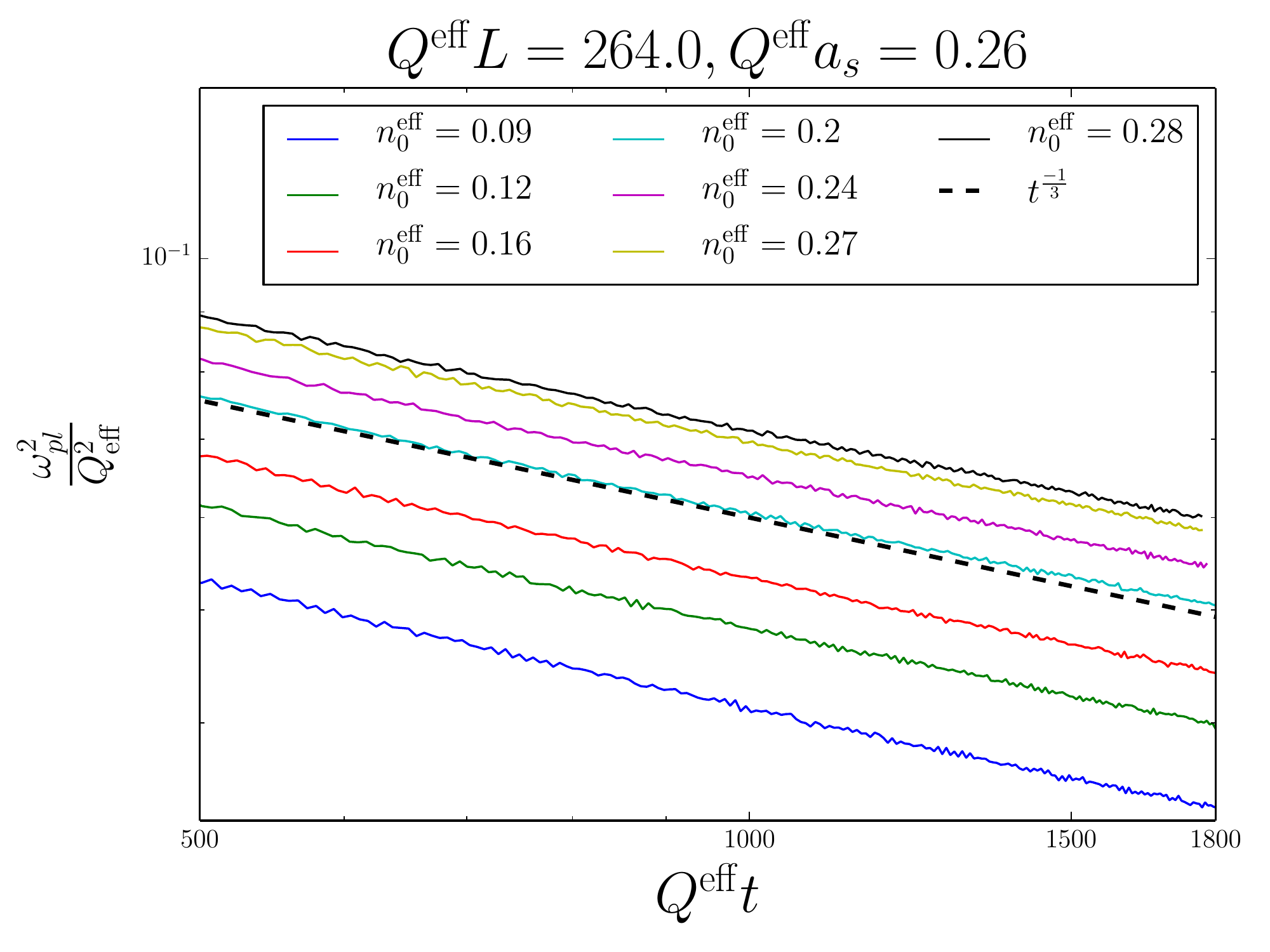}}
\end{minipage}
\caption{On the left we show the measured plasmon mass scale as a function of time for all methods. We also show a curve which corresponds to $t^{\nicefrac{-1}{3}}$ power law. The HTL-A method almost agrees with the power law, but the value given by the UE method seems to decrease faster than the power law. In the right hand side figure we focus on the late time behavior of the plasmon mass scale using exclusively the HTL-A method. At late times the time dependence of the mass scale is consistent with the power law.  The results are  averaged over five runs.}
\label{fig:HTLAvst}
\end{figure}

The time dependence of the plasmon mass scale is shown in \fig \ref{fig:HTLAvst}. The plot on the left shows the time-dependence using all methods, and the plot on the right shows the time-dependence at later times using only the HTL-A method. We observe that the difference between the DR and the other methods persists also at later times. At late times the plasmon mass scale agrees with $t^{\nicefrac{-1}{3}}$ powerlaw.

\begin{figure}[]
\centerline{\includegraphics[width=0.6\textwidth]{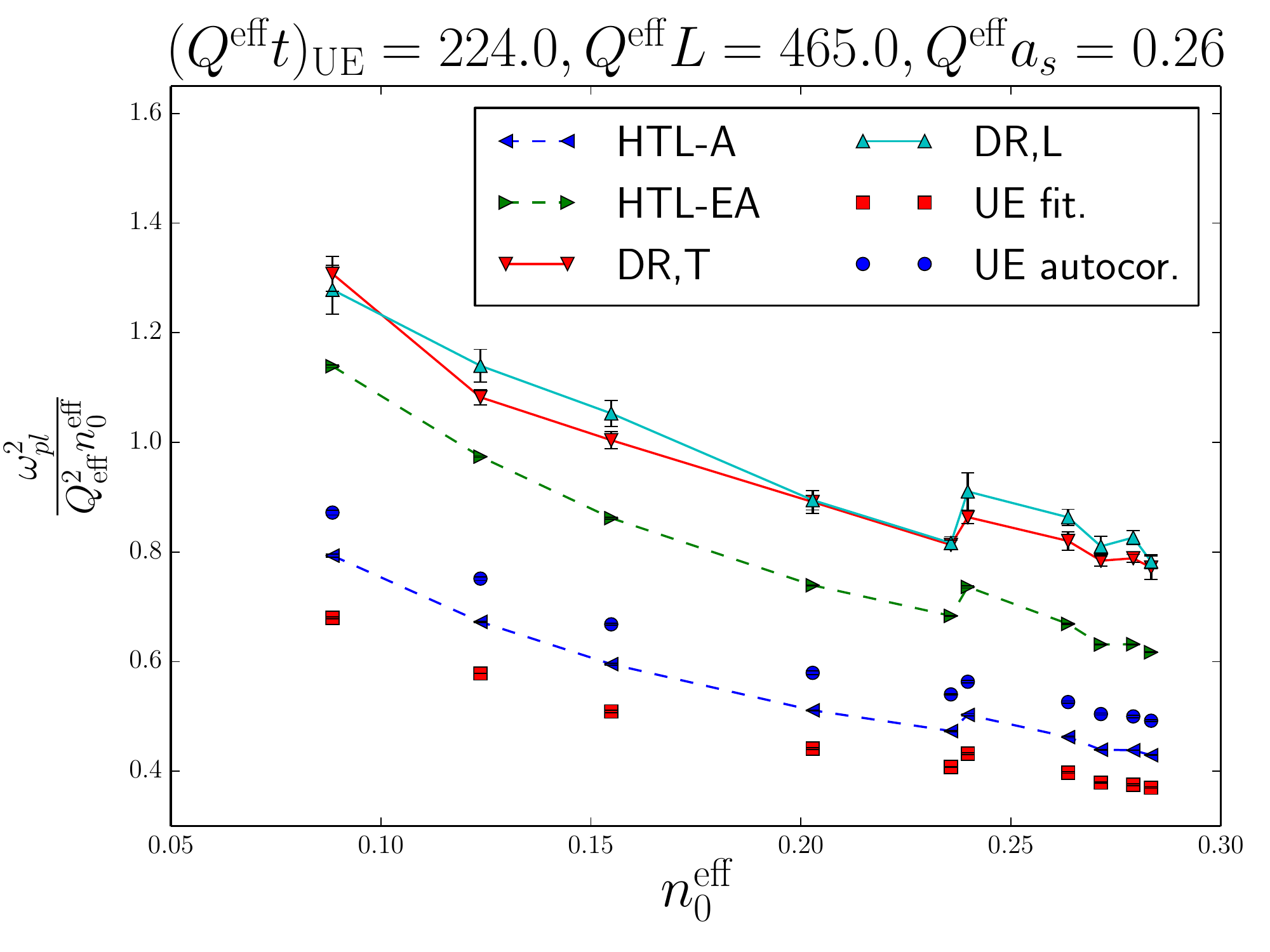}}
 \caption{The dependence of the plasmon mass scale on the occupation number $\neff$ for all methods. The trend turns out to be similar as in three dimensions, the scale decreases faster for more dense systems at a fixed time. Here $Q^{\mathrm{eff}}L=432-480.$ and $Q^{\mathrm{eff}} = 0.240-0.267.$ $(Q^{\mathrm{eff}}t)_\mathrm{UE} = 205-227.$ The bumb at $n_0 \approx 0.24$ is caused by a deviation in these parameters.  These results are averaged over five configurations.}
\label{fig:occupnumberdep2d}
\end{figure}

The dependence on the occupation number is shown in \fig \ref{fig:occupnumberdep2d}. The results are qualitatively similar to the three-dimensional system. The plasmon mass scale decreases (at fixed time) as a function of increasing occupation number. The bump in \fig \ref{fig:occupnumberdep2d} is caused by a slight deviation in the estimation of $n_0$ and $Q$ from the desired $\neff$ and $\Qeff$. We presume that this behavior is explained similarly as in three spatial dimensions, namely that the more dense systems enter the time scaling regime more rapidly, and thus more dense systems have spent more time in this regime. 

The UE and HTL-A method seem to be in rough agreement, and they can be used almost interchangeably to measure the plasmon mass scale. However, since the computational cost of the UE method is overwhelmingly larger than that of the HTL method, we consider the HTL-A method to be the best method to measure the plasmon mass scale in this two-dimensional system. However, one should bear in mind that in two dimensions the way one defines the occupation number distribution has a more dramatic impact on the results than in three-dimensional systems. The reason for this is that the infrared contribution to the HTL integral is larger than in three dimensions. The DR method works similarly as in three dimensions: it agrees with the other methods within a factor of two.

%\pagenumbering{arabic}
\chapter{Time-evolution of linearized gauge field fluctuations}
\label{flucts}
Next we will consider linearized perturbations of classical Yang-Mills fields. We want to explicitly exclude interactions between the fluctuations and eliminate the backreaction to the classical fields. We have several possible applications and physical motivations for this approach. The first one is to study unstable quantum fluctuations which break the boost invariance of the glasma fields shortly after a heavy-ion collision. There have been attempts to study their contribution to the thermalization process, but these have suffered from issues concerning the numerical treatment of the fluctuations due to their backreaction to the classical fields. The second application is to study linear response of classical Yang-Mills theory, which enables us to study the dispersion relation and the spectral function of the classical theory. The linear response analysis of a three-dimensional isotropic system will be discussed in \ch \ref{response}.

The major drawback of the linearized fluctuation approach is the fact that the presence of the background field will violate several conservation laws. The background field is a function of time and position, thus the time translational invariance of the fluctuations is violated and energy conservation is lost. Similarly the breaking of translational and rotational invariance leads to losing conservation of linear and angular momentum. All of these are of course recovered when we take the background field to zero. 

To our knowledge, no one has formulated the time-evolution equations of the linearized classical Yang-Mills fields on the lattice, even though we are expecting this to have several interesting applications in the future. %The potential reasons for this are problems concerning Gauss' law (which we will be dealing with in detail). 
The potential reason is the nonconservation of Gauss' law, which we will be dealing with in detail.

%We have already studied these aspects of the classical theory in fixed three-dimensional box, as a warm up for more phenomenologically interesting scenarios. We will discuss the results in detail in chapter \ref{response}. 

We will start by deriving the equations of motion for the linearized fluctuations in the continuum. Then we will proceed to the discretized case. At the end of this section we will discuss why the naive approach fails to conserve Gauss' law and then we derive the equations of motion in the Lagrangian formalism. The time-evolution equations of the linearized fluctuations were originally presented in the Hamiltonian formalism in paper \cite{Kurkela:2016mhu}.

\section{Equations of motion for the fluctuations}
We start from the Hamiltonian equations of motion presented in subsection \ref{sec:lathamilton}. The equations of motion of the fluctuations are obtained by introducing a linearized perturbation to the equations of motion of the background. In practice this is done by a direct substitution 
\begin{equation}
 (E^i,A_i) \to (E^i  + e^i,  A_i + a_i),
 \end{equation}
 where we refer to the fluctuations with the lower case letters, and to the classical background with the upper case letters. The equations we obtain are
\begin{eqnarray}
 \dot{a}_i &= &e^i  \label{eq:contaeom}
\\
\dot{e}^i &=& \left[D_j,\left[D_j,a_i\right]\right] - \left[D_j,\left[D_i,a_j\right]\right] + i g \left[a_j,F_{ji}\right],
\label{eq:conteeom}
\end{eqnarray}
and a counterpart of Gauss' law
\begin{equation}
\label{eq:contgauss}
c(\xx,t) =\left[ D_i,e^i \right]  + i g \left[a_i,E^i\right] = 0.
\end{equation}

Next we wish to write down the equations of motion of the linearized fluctuations corresponding to \eqs (\ref{eq:conteeom}) and (\ref{eq:contaeom}) on the lattice. In principle there are many ways one can carry out this task. We have the following requirements for the discretized equations:
\begin{enumerate}
\item \label{it:contlim} Correct continuum limit, corresponding to reduction to the equations of motion \nr{eq:contaeom},~\nr{eq:conteeom} in the limit $a_s \to 0, \  \ud t \to 0$.
\item \label{it:gt} The equations of motion should be gauge invariant.
\item \label{it:linearity} Linearity in the fluctuation fields.
\item \label{it:gauss} Gauss' law should be exactly conserved. The lattice version of  Gauss's law has to reduce to \nr{eq:contgauss} in the limit $a_s \to 0, \ \ud t\to 0$ at every time step.
\item \label{it:treversal}Time reversal invariance (under $\ud t \to -\ud t$).
\end{enumerate}
In lattice field theory gauge invariance is of uttermost importance, and that is why we start from requirement \ref{it:gt}. We require the gauge transformation properties of the linearized fluctuation to be similar to the electric field
\begin{eqnarray}
 a_i(\xx) &\to& V(\xx) a_i(\xx) V^\dag(\xx)
\\
 e^i(\xx) &\to& V(\xx) e^i(\xx) V^\dag(\xx).
\end{eqnarray}
It immediately follows that the fluctuation of the gauge field corresponds to the variation of the link matrix from the left
\begin{equation}
\label{eq:fluctintroduction}
 U_i(\xx)_\textup{bkg + fluct} = e^{i a_i(\xx)}U_i(\xx) \approx
U_i(\xx) + i a_i(\xx)U_i(\xx).
\end{equation}
Here the factors of $g$ and $a_s$ have been absorbed into the definition of the fluctuation of the gauge field on the lattice. Thus the connection between the continuum and lattice variables is the same as for the background fields in the Hamiltonian formalism, i.e. $a_i^\textnormal{lat} = a_s g a_i^\textnormal{cont}$ and $e_i^\textnormal{lat} = a_s g e_i^\textnormal{cont}$  %assuming $E^i_{\mathrm{bkg} + \mathrm{fluct}} =E^i+e^i$. 

Inserting the fluctuations of the links \equref{eq:fluctintroduction} into \eq (\ref{eq:clestep}) gives the equation of motion for the fluctuation of the electric field
\begin{align} \label{eq:estep}
a^2 e^i(t+\ud t) &= a^2 e^i(t) - \ud t \sum_{j\neq i}\Bigg[ i \Big(
 a_i(\xx)U_{ij}(\xx) 
+ a_j(\xx+\boldsymbol{\I} \to \xx)U_{ij}(\xx) \\
 & - U_{ij}(\xx)a_i(\xx+\boldsymbol{\J} \to \xx) 
- U_{ij}(\xx)a_j(\xx)+ a_i(\xx)W_{ij}(\xx)  \nonumber
\\
& 
 - a_j(\xx+\boldsymbol{\I}-\boldsymbol{\J} \to \xx+\boldsymbol{\I} \to \xx)W_{ij}(\xx) 
 - W_{ij}(\xx) a_i(\xx-\boldsymbol{\J} \to \xx) \nonumber \\
& +W_{ij}(\xx)a_j(\xx-\boldsymbol{\J} \to \xx)
 \Big) \Bigg]_\ah, \nonumber
\end{align}
Here the parallel transported links are denoted by 
\begin{equation}
a_j(\xx+\I \to \xx)
 \equiv
U_i(\xx)a_j(\xx+\I)U^\dag_i(\xx).
\end{equation}
Fields parallel transported over two links are denoted similarly
\begin{equation}
a_j(\xx+\I-\J \to \xx+\I  \to \xx) 
\equiv U_i(\xx)a_j(\xx+\I-\J \to \xx+\I)U^\dag_i(\xx)\nonumber.
\end{equation}

Similarly we can derive Gauss' law by performing the same substitutions to Gauss' law of the background field (\ref{eq:gauss}). We get
\begin{equation} \label{eq:fluctgauss}
 c(\xx,t)
= \sum_i \frac{1}{a^2} \Big\{ e^i(\xx)- e^i(\xx-\I \to \xx) + i [a_i,E^i] \left(\xx -  \I \to \xx \right) \Big\} .
\end{equation}
For a complete set of equations we still need a time-evolution equation for the fluctuation of the gauge potential $a_i \vas \xx \oik.$ The most obvious choice would be to perform a similar substitution procedure as we did for the electric field on \equref{eq:cllinkstep}.
The resulting equation of motion is 
\begin{equation} \label{eq:naiveequ}
a_i\left(x+\ud t\right) = a_i\left(x\right) + \mathrm{d}_t \left( i \left[ E^i\left(x\right) ,a_i\left(x\right) \right] + e^i\left(x\right)\right),
\end{equation}
where the notation $x + \ud t$ refers to the gauge field at position $x$ at the next time step.
This approach, however, turns out to break Gauss' law, corresponding to unphysical charge creation. We will refer to this method as the naive approach.

Instead, a better way is to start by imposing that Gauss' law is conserved
\begin{align}
c(\xx,t)  = c(\xx,t+\ud t),
\end{align}
and use this to construct the proper timestep for the fluctuation of the gauge field. When one imposes this condition, it turns out to be equivalent to the following condition
\begin{equation}\label{eq:genstep}
\left[ E^i,a_i(t+\ud t) \right] = -i \left(\Box_{0i}e^i \Box^\dagger_{0i} - e^i\right)
+
\left[E^i,\Box_{0i} a_i(t)\Box^\dagger_{0i}\right].
\end{equation}
Here we use the notation $\Box_{0i}=e^{i E^i\ud t}$ to refer to the ``timelike plaquette''. This equation can be solved for $a_i\left(x+\ud t\right)$ only partially. In general we can divide $a_i$ into two parts in color space, one component that is parallel to $E^i$ in color space, and the other that is perpendicular. Equation (\ref{eq:genstep}) does not put any constraints on the parallel component, but it can be used to solve the perpendicular component. The perpendicular and parallel components are obtained as follows
\begin{eqnarray}
 f^\parallel  &=& \frac{\tr \left[f E^i\right]}{\tr \left[ E^i E^i\right]}E^i,
\\
 f^\perp  &=& f - \frac{\tr \left[f E^i\right]}{\tr \left[ E^i E^i\right]}E^i.
\end{eqnarray}
Since we know how to solve \equref{eq:genstep} for the perpendicular component, we will first briefly address how to deal with the parallel component. The equation of motion for the parallel component is obtained from \equref{eq:naiveequ}, and it is given by 
\begin{equation}
a^\parallel_i(t+\ud t)=a^\parallel_i(t) + \ud t e^{i \parallel }(t+\ud t/2).
\end{equation}
This trivially satisfies \equref{eq:genstep}.

Then let us move on to solve \equref{eq:genstep}. For general $N$ it is the best to solve \equref{eq:genstep} in the adjoint representation. The solution takes a particularily simple form for $N=2$ which we will also feature along with the general solution. Multiplying an element of $\mathfrak{su}(n)$ from left by an element of $SU(N)$ and from right by its hermitean conjugate as in $\Box_{0i} a_i \Box_{0i}^\dagger$ corresponds to multiplying by a matrix $\left(\widetilde{\Box}_{0i}\right)^{ab} = 2 \tr \left[t_a \Box_{0i} t_b \Box_{0i}^\dag \right]$ in the adjoint representation. The elements of the algebra are given by the $N^2-1$ component vectors $\underline{a}_i$ and $\underline{e}^i,$ which have the components $\left(a_i\right)^a$ and $(e^i)^a.$ The commutator of two elements of the algebra corresponds to a multiplication by a matrix in the adjoint representation $\left(\widetilde{E}^i\right)^{ab} = E^i_c \left(T^c\right)^{ab}
= -if_{cab}E^i_c$. In the adjoint representation we can write \equref{eq:genstep} as 
\begin{equation}
 \widetilde{E}^i \underline{a}_i(t+\ud t) = -
i \left( \widetilde{\Box}_{0i} - \mathbb{1} \right) \underline{e}^i
+ \widetilde{E}  \widetilde{\Box}_{0i}\underline{a}_i(t).
\end{equation}
In the most general case the matrix inversion of $\widetilde{E}^i$ can not be done, since it corresponds to the commutator, and projects the parallel components to zero. This also means that the timelike plaquette is equivalent to the identity matrix for the parallel components. When we consider only perpendicular directions, the matrix $\widetilde{E}^i$ can be inverted, and the timestep can be written as 
\begin{equation}\label{eq:finalastep}
 \underline{a}_i(t+\ud t) = \left(\widetilde{E}^i\right)^{-1}_{\perp}\left[
-i \left( \widetilde{\Box}_{0i} - \mathbb{1} \right) \underline{e}^{i \perp}
+ \widetilde{E}^i  \widetilde{\Box}_{0i}\underline{a}_i^\perp(t)
\right]
+ 
\underline{e}^{i \parallel} \ud t
+
\underline{a}_i^\parallel(t),
\end{equation}
where the notation $ \left(\widetilde{E}^i\right)_\perp$ refers to the projection to the subspace of the perpendicular components where the matrix is invertible. 

Next we can check if the equation of motion satisfies the requirement \ref{it:contlim}, which imposes that the equation must have the correct continuum limit. In the limit of small $\mathrm{d}t$ we have $\widetilde{\Box}_{0i} \approx \mathbb{1} + i \widetilde{E}^i \ud t.$ Plugging this into \equref{eq:finalastep} yields
\begin{equation}
\label{eq:smalldtlim}
\underline{a}_i\left(t + \ud t \right) \approx \underline{a}_i \left( t \right) + \underline{e}^i \ud t + i \ud t \widetilde{E}^i a_i \left(t \right),
\end{equation}\footnote{In paper \cite{Kurkela:2016mhu} the term involving the background electric field was missing in the corresponding equation.}
which corresponds to the naive discretization given by \equref{eq:naiveequ}. On the first sight it seems that \equref{eq:smalldtlim} has an extra term which could cause concern in the continuum limit. However, since the lattice definitions of the electric field and the gauge field fluctuation both include one factor of lattice spacing, the extra term (involving the commutator) is actually of higher order in lattice spacing. Thus we can conclude that the requirement \ref{it:contlim} is satisfied. It is also straightforward to check that the requirement \ref{it:treversal} of time reversal invariance is satisfied.
 
Finally we will show the simplified results for the $SU(2)$ gauge group. In this case one can simply multiply \equref{eq:finalastep} by $\widetilde{E}^i$ from the left to obtain the solution. There one must make use of an identity for $f^{abc}f^{ade},$ which simplifies  to $\epsilon^{ijk}\epsilon^{ilm} = \delta^{jl}\delta^{km}-
\delta^{jm}\delta^{kl}$ in the case of $SU(2)$. Thus $\widetilde{E}^i\widetilde{E}^i \underline{a}_i^\perp = 2 \mathrm{Tr} \vas E^i E^i \oik \underline{a}_i^\perp,$ and \equref{eq:finalastep} becomes
\begin{align}\label{eq:su2adj}
 \underline{a}_i(t+\ud t) &= \dfrac{1}{2  \mathrm{Tr}\left(E^i E^i\right)}  \widetilde{E}^i \Bigg\{ \left[
-i \left( \widetilde{\Box}_{0i} - \mathbb{1} \right) \underline{e}^{i \perp}
+ \widetilde{E}^i  \widetilde{\Box}_{0i}\underline{a}_i^\perp(t)
\right] \nonumber
\\
&+ 
\underline{e}^{i \perp} \ud t
+
\underline{a}_i^\perp(t) \Bigg\}.
\end{align}
In the fundamental representation of $SU(2)$ the update can be written as 
\begin{align}\label{eq:su2fund}
a_i(t+\ud t) &= \frac{i}{2\tr \left[E^i E^i\right]} 
\Bigg[E^i,
-i \left(\Box_{0i}e^{i \perp} \Box^\dag_{0i} - e^{i \perp}\right)
+
\left[E^i,\Box_{0i} a^\perp_i(t)\Box^\dag_{0i}\right]
\Bigg] \nonumber \\
& + \ud t e^{i \parallel} + a^\parallel_i(t),
\end{align}
and can be simplified further to 
\begin{align}
a_{ i}\left(x+ \ud t\right)  & = \dfrac{1}{2 \mathrm{Tr}\left(E^i E^i\right)} \Bigg[- i\commutator{E_i}{\Box_{0i} e_{\bot,i} \Box_{0i}^\dagger }   \\ & + i\commutator{E_i}{ e_{\bot,i}} \Bigg] + \left( \Box_{0i} a_{\perp,i}\left(t\right) \Box_{0i}^\dagger\right) + \ud t e^{i \parallel} + a^\parallel_i(t). \nonumber \label{eq:datimeevol}
\end{align}

\begin{figure}[t]
\centerline{\includegraphics[width=0.8\textwidth]{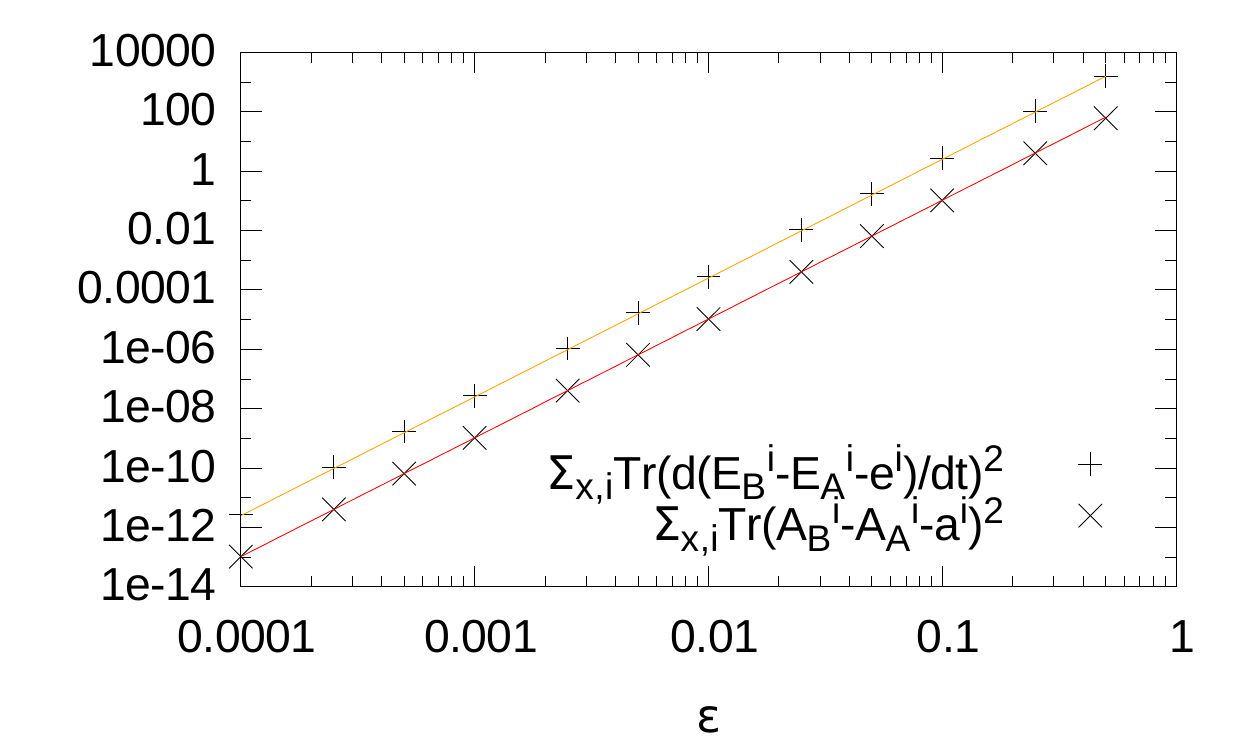}}
\caption{Here we test the linearization with the decomposition to the background field and the fluctuations using quantities $\delta_A$ and $\delta_{\dot{E}}$ as shown by \equref{eq:da} and \equref{eq:de}. The solid lines correspond to fits of the form $a \epsilon^4,$ demonstrating the desired power law. }
\label{fig:a_t_comparison}
\end{figure}

\begin{figure}[]
\centerline{\includegraphics[width=0.8\textwidth]{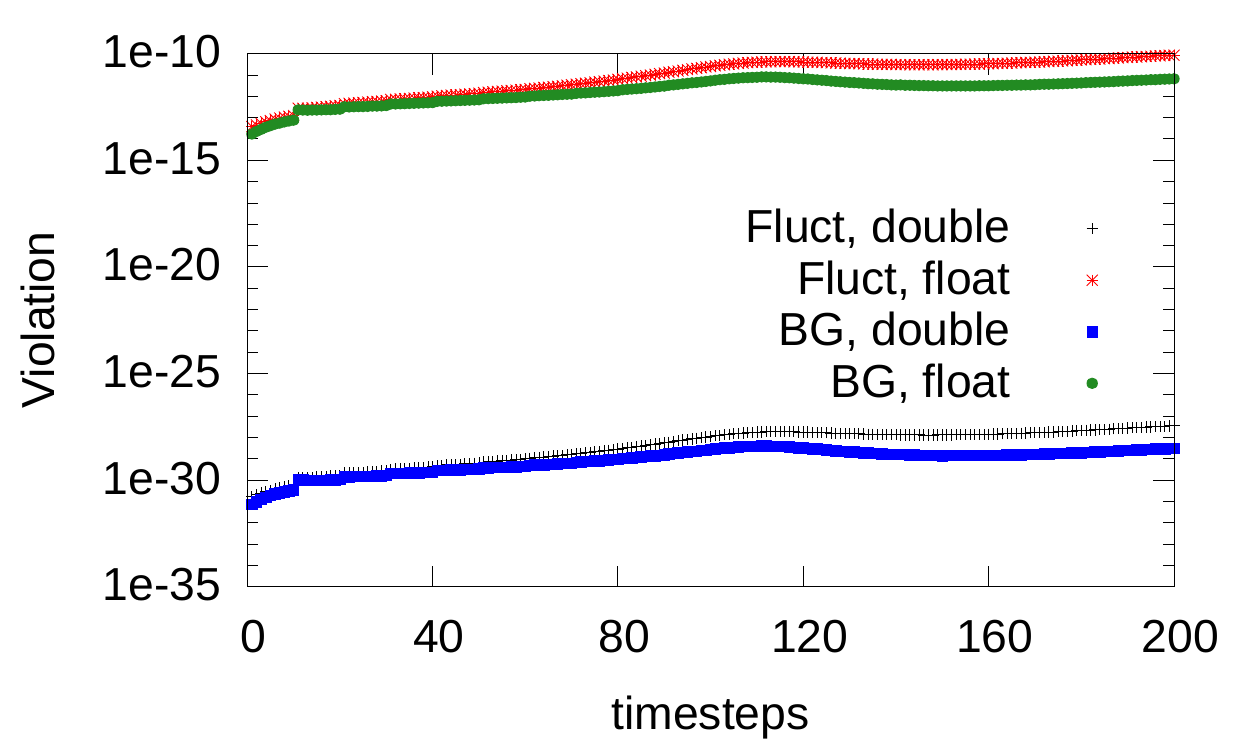}}
\caption{Violation of Gauss' law for the background fields and for the fluctuations measured using \equref{eq:bggaussviol} and \equref{eq:flucgaussviol} with single and double precision numbers. In order to also test for gauge invariance we have performed a random gauge transformation on every time-step and fixed Coulomb gauge on every tenth timestep. This shows as a slight discontinuity on every tenth timestep.  The parameters used here are $\epsilon = 0.1$ and $\ud t = 0.01.$}
\label{fig:gausslaw}
\end{figure}

\section{Numerical tests}
In order to test the remaining requirements \ref{it:linearity} and \ref{it:gauss} we have performed a couple of numerical tests. The tests are performed with a numerical implementation for $SU(2).$ The initial condition for the test is constructed by taking random gauge fields with all components uniformly distributed between $[0,0.9]$. The link matrices are then constructed by the usual exponentiation procedure. The electric fields are set to zero to guarantee that Gauss' law is satisfied at $t=0$.  The gauge field fluctuations $a_i$ are taken from the same distribution as the background gauge fields. In order to establish a clear scale, which we can later change, the fluctuations are multiplied by a small parameter $\epsilon,$ which ranges between 0.5 and 0.0001. Next we evolve two systems separately in time. The first system consists of the background fields $A_i,$ $E^i$ and the linearized fluctuations $e^i$ and $a_i$ evolved with their own equations of motion.  The second system has only the background fields, which also include the fluctuation. We refer to these background fields as $\hat E^i(t=0) = E^i(t=0) + e^i(t=0)$ and
$\hat A^i(t=0) = A^i(t=0) + a^i(t=0)$. 

Since the equations have been linearized, the following quantity should be proportional to $\epsilon^4.$
\begin{equation}
\label{eq:de}
\delta_E = \sum \limits_{x,i}\tr(\hat E^i-E^i-e^i)^2.
\end{equation}
We can construct an observable with similar scaling behavior for the gauge fields
 \begin{align} 
\label{eq:da}
 \delta_A & = \dfrac{1}{2} \sum \limits_{x,i,a}\left(2 {\rm Im}\mathrm{Tr}\left( t^a \hat U_{i}U_{i}^\dagger\right) - a_i^a\right)^2\\
  &\approx\sum \limits_{x,i} \tr(\hat A^i-A^i-a^i)^2.
 \end{align} 
For numerical convenience we use the analogous quantity involving the time derivatives of the electric fields instead of \equref{eq:de}. Figure \ref{fig:a_t_comparison} demonstrates that the correct scaling behavior is indeed established, and thus the linearization works as desired. For the naive timestep \equref{eq:naiveequ} the correct scaling in $\epsilon$ is observed only for smaller values of $\ud t$ because the naive timestep has violations of linearity of the order $\delta_A \sim \epsilon^2 \ud t^4.$ We have observed this numerically by checking that, for larger values of $\ud t$ the quantities $\delta_E$ and $\delta_A$ scale as $\epsilon^2$ for the naive timestep. Thus, the naive timestep does not even fully satisfy our requirement of the correct linearized behavior. 

Next we want to construct a way to measure the violation of Gauss' law as a function of time. We measure the violation with a sum of squares vs. square of sums type of measurement, which tells us how well the violation of Gauss' law cancel each other between the different directions. For the background field this observable is
 \begin{equation} \label{eq:bggaussviol}
\frac{2\sum \limits_x \mathrm{Tr}\left(\sum \limits_{i}\left[E^i(\xx) - E^i(\xx-\I \to \xx )\right]\right)^2}{2 \sum \limits_{x,i} \mathrm{Tr}\left[E^i(\xx) - E^i(\xx-\I \to \xx )\right]^2}.
\end{equation}
The numerator is the sum of Gauss' law violations over the entire lattice, and the denominator is the sum of squares. In principle the numerator should be exactly zero, but in practice a nonzero error is always observed due to limited machine precision. In spite of this, this quantity should always remain extremely small.
The corresponding observable for the fluctuations is
\begin{equation} \label{eq:flucgaussviol}
 \frac{2 \sum \limits_x \mathrm{Tr}\left(\sum \limits_i \left[
e^i(\xx)- e^i(\xx-\I \to \xx)
 + i [a_i,E^i](\xx-\I \to \xx)
\right]\right)^2 
}{
2 \sum \limits_{x,i} \mathrm{Tr} \left[
e^i(\xx)-  e^i(\xx-\I \to \xx) 
 + i [a_i,E^i]\left(\xx-\I \to \xx \right)
\right]^2}.
\end{equation} 
Figure \ref{fig:gausslaw} demonstrates the time-evolution of these quantities for both the background field and fluctuations. We observe that when we compute these quantities in single precision the error is several orders of magnitude larger than in double precision. This indicates that the error is really dominated by limited machine precision, since if we had an additional source of error, it would very quickly overwhelm the error arising from the limited machine precision. 

Thus we have managed to construct a time-evolution scheme which satisfies our initial requirements. In the next chapter we will utilize this formalism by performing linear response analysis of a classical Yang-Mills system in a three-dimensional fixed box. 

\section{Time-evolution in the Lagrangian formalism}
The naive time-evolution equation for the gauge field fluctuation turned out to violate the Gauss' law when time was discretized. In the Hamiltonian formalism time is kept as a continuous variable, which must be discretized for practical computations. Furthermore, Gauss' law is not an integral part of Hamiltonian formalism as we saw in the introduction. Instead, it must be imposed by hand. In the Lagrangian formalism Gauss' law is one of the equations of motion and its conservation should be guaranteed automatically. Thus we want to compute the time-evolution equation of the gauge field in the Lagrangian formalism. For compatibility with the rest of this section we will stick to same lattice fields, i.e. $E^i_L = a_s g E_i^{cont}$ and similarly for the gauge fields and fluctuations. The time-evolution equations for the electric fields and their fluctuations will stay intact when we go to the Lagrangian formalism. For the background gauge field we use the procedure described in \se \ref{sec:lagrangianlattice}.

We start from the definition of the electric field \equref{eq:elatdef}, using the same lattice units as we have done throughout this section
\begin{equation}
\label{eq:elatdefhamilton}
E_i^a(x) = \dfrac{2}{a_t } \mathfrak{Im}\mathrm{Tr}(t^a U_{i0}\left(x\right)).
\end{equation}
We introduce the fluctuations as we did in \equref{eq:elatdef}.
\begin{equation}
e_i^a\left(x\right) = \dfrac{2}{a_t}\mathfrak{Re}\mathrm{Tr}\left(t^a a_i\left(x\right) U_{i0}\left(x\right)\right) - \dfrac{2}{a_t}\mathfrak{Re}\mathrm{Tr}\left(t^a  U_{i0}\left(x\right)a_i\left(x+\hat{t}\right) \right).
\end{equation}
The next step is to decompose the temporal plaquette using \equref{eq:linkevolution}, but absorbing the factors of $a_s$ and $g$ into the definition of the electric field. We evaluate the traces using  \equref{eq:2trace} and \equref{eq:3trace}. Simplifying the result yields
\begin{align}
\label{eq:efluctlagrangianform}
e_i^a\left(x\right) &= \dfrac{a_i^a\left(x\right) - a_i^a\left(x+\hat{t}\right)}{a_t} \sqrt{1-\left(\dfrac{a_t}{2} E_i^b\left(x\right) \right)^2} \nonumber \\ &+ \dfrac{1}{2}a_t f^{abc} E_i^b\left(x\right) \left(a_i^c\left(x\right) + a_i^c\left(x + \hat{t}\right) \right). 
\end{align}
This holds for both $N=2$ and $N=3$ since the symmetric structure constants are real for both groups. Here one can observe that this equation behaves correctly in the limit of continuous time, it reduces to $e_i^a = -\partial_t a_i.$ We will present a solution to this equation for $N=2$ which is the relevant group for numerical purposes. There the structure constants are given by Levi-Civita symbols, and the terms involving the structure constants become analogous to cross products of $E$ and $a$ in color space. 

In order to solve \equref{eq:efluctlagrangianform} we wish to write it in matrix form in color space. The cross product matrix in the adjoint representation is given by 
\begin{equation}
E_i = \begin{pmatrix}
0 & -E_i^3 & E_i^2 \\
E_i^3 & 0 & -E_i^1 \\
-E_i^2 & E_i^1 & 0 
\end{pmatrix}.
\end{equation}
Next we move all terms involving $a_i^a\left(x + \hat{t}\right)$ to the left hand side of the equation \equref{eq:efluctlagrangianform} and write it in matrix form
\begin{equation}
\label{eq:matrixaupdate}
A a_i\left(x + \hat{t}\right) = B a_i\left(x\right) - e_i\left(x\right).
\end{equation}
The matrix $A$  is given by
\begin{equation}
A = \begin{pmatrix}
\dfrac{1}{a_t}\sqrt{1-\left(\dfrac{a_t}{2} E_i^b\right)^2} & \nicehalf E_i^3 & -\nicehalf E_i \\
-\nicehalf E_i^3 & \dfrac{1}{a_t}\sqrt{1-\left(\dfrac{a_t}{2} E_i^b\right)^2} & \nicehalf E_i^2 \\
\nicehalf E_i^2 & -\nicehalf E_i^1 & \dfrac{1}{a_t}\sqrt{1-\left(\dfrac{a_t}{2} E_i^b\right)^2} 
\end{pmatrix},
\end{equation}
where we have omitted the position arguments of the electric fields for brevity. The matrix $B$ is given by
\begin{equation}
B = \begin{pmatrix}
\dfrac{1}{a_t}\sqrt{1-\left(\dfrac{a_t}{2} E_i^b\right)^2} & -\nicehalf E_i^2 & \nicehalf E_i^3 \\
\nicehalf E_i^3 & \dfrac{1}{a_t}\sqrt{1-\left(\dfrac{a_t}{2} E_i^b\right)^2} & - \nicehalf E_i^1 \\
- \nicehalf E_i^2 & \nicehalf E_i^1 & \dfrac{1}{a_t}\sqrt{1-\left(\dfrac{a_t}{2} E_i^b\right)^2}
\end{pmatrix}
\end{equation}
Using these matrices we can solve \equref{eq:matrixaupdate}
\begin{equation}
a_i\left(x + \hat{t}\right) = A^{-1}\left( B a_i\left(x\right) - e_i\left(x\right) \right),
\end{equation}
provided that the matrix $A$ is invertible. The determinant of $A$ is 
\begin{equation}
\det\left(A\right) = \frac{\sqrt{4-a_t^2 \left(\left(E_i^b\right)^2\right)}}{2a_t^3}.
\end{equation}
This means that we may also encounter a specific combination of background electric field and time-step, which exactly balance each other out in such a way that the matrix $A$ becomes non-invertible. In this case we can only solve equation \equref{eq:efluctlagrangianform} for the perpendicular component of $a_i\left(x + \hat{t}\right),$ and we have to supplement it with the parallel component arising from the naive equation of motion \equref{eq:naiveequ} similarly as we did in the Hamiltonian formalism.

The way we obtained the solution here resembles the procedure we had to carry out in the Hamiltonian formalism. There we ended up dividing the field in parallel and perpendicular components. Here we took a more straightforward approach and wrote the equations in the form of matrices. However, both involved inverting commutators (or cross products) in color space.

\section{Discussion concerning the numerical treatment of fluctuations}
Next we want to discuss the potential reason for the violation of Gauss' law which we observed when we discretized the equations of motion of the fluctuations on the lattice. The equations of motion of the fluctuations did not break Gauss' law when time was kept as a continuous variable. However, when we discretized the time, we observed that Gauss' law was violated. When we realized this violation, we derived equations of motion starting from the demand that Gauss' law has to be conserved, and obtained modified equations of motion which satisfied our demands. However, this is not the only way to derive the equations of motion. We also presented a derivation in the Lagrangian formalism. There Gauss' law is one of the equations of motion from the very beginning, and that is why we have a reason to expect that it should be conserved without any further modifications. Thus the potential reason for the nonconservation of Gauss' law is the fact that in the Hamiltonian formalism Gauss' law is not an integral part of the equations of motion in the same way as it is in the Lagrangian formalism.

We would like to emphasize that this does not mean that the Hamiltonian approach we derived is invalid. As we saw in the case of the background field, it is possible to construct several equations of motion which satisfy Gauss' law, like the Lagrangian and Hamiltonian equations of motion did.

%The surest way to construct equations of motion which would certainly conserve Gauss' law would be to construct an action, from which the equation of motion of the fluctuations would be derived. In this case Gauss' law would be given by the zeroth component of the Euler-Lagrange equations of motion. This computation would in principle be straigthforward but tedious. One has to introduce the fluctuations on every link, and one has to expand to second order in fluctuation (first order would correspond to the equation of motion of the background field). 

%\pagenumbering{arabic}
\chapter{Spectral properties of classical Yang-Mills systems}
\label{response}
In this section we  apply the linearized fluctuation formalism to linear response measurements. We measure spectral properties of classical Yang-Mills theory in a fixed box focusing on observables which are essential for addressing the existence of quasiparticles and their properties. These observables include the spectral function, the dispersion relation, the damping rate and the plasmon mass. We  also compare these to the HTL predictions. These results were presented for the first time in paper \cite{Boguslavski:2018beu}.

We will start by going through our initial conditions and reference scales. Then we will briefly introduce the spectral and statistical correlation functions and explain how the retarded propagator and spectral function can be extracted using linear response theory. Then we will go through our numerical results.

\section{Initial conditions and scales}
We divide the fields into transverse and longitudinal components, and we initialize only the transverse components of the gauge fields and the electric fields according to the prescription
\begin{align}
\label{eq_fields_init}
 A_j^a(t=0,\mbf p) &= \sqrt{\frac{f(t=0, p)}{p}} \sum_{\lambda = 1,2} c^{(\lambda)}_a(\mbf p)\, v_j^{(\lambda)}(\mbf p) \nonumber \\
 E^j_a(t=0,\mbf p) &= \sqrt{p\,f(t=0, p)} \sum_{\lambda = 1,2} \tilde{c}^{(\lambda)}_a(\mbf p)\, v_j^{(\lambda)}(\mbf p),
\end{align}
where  $v$ refers to the orthonormal polarization vectors with $v^1$ and $v^2$ being transversely polarized and $v^3$ being the longitudinal polarization vector. The complex gaussian random numbers $c_{a}^{\left( \lambda \right)}$ satisfy 
\begin{align}
\label{eq_random_nums}
 \left\langle \left(c^{(\lambda)}_{a}(\mbf p)\right)^* c^{(\lambda')}_{a'}(\mbf p') \right\rangle_{\cl} = V \delta_{\mbf p, \mbf p'} \delta_{a,a'} \delta_{\lambda,\lambda'}.
\end{align}
A similar relation holds for $\tilde{c}$, while $\langle \tilde{c}^*  c \rangle_\cl = 0$. Here $\langle \cdot \rangle_{\cl}$ refers to a classical average over the random number distribution.

The initial fields correspond to the following occupation number distribution
\begin{align}
\label{eq_init_cond}
 f(t=0, p) = \frac{n_0}{g^2}\, \frac{\pInit}{p}\, e^{-\frac{p^2}{2\pInit^2}}.
\end{align}
Here the momentum scale $p_0$ plays analogous role to $\Delta,$ which we encountered in \se \ref{plasmon}. Similarly $n_0$ is a dimensionless number which determines the occupation number of our system. Since we initialize both the electric field and the gauge field at $t=0$ Gauss' law \equref{eq:gauss} is not satisfied. Thus we remove the unphysical charges by using the same algorithm \cite{Moore:1996qs} which was used when we introduced the uniform electric field. This initial condition guarantees high occupation numbers in the weak coupling $g \ll 1$ regime. The main difference between this initial condition and the previous initial conditions \equref{eq:initdist3d} is that at small $p$ this has larger infrared occupation number, and in this sense it is closer to the non-thermal attractor solution already at earlier times.

Next we define a characteristic energy scale in a similar fashion as we did in case of two-dimensional plasmon mass scale measurements
\begin{align}
 \Q = \sqrt[4]{5\,n_0}\; \pInit \propto \sqrt[4]{g^2 \epsilon}.
\end{align}
We will express all dimensionful quantities in units of $\Q$. 

Throughout this section we use a more elaborated procedure for the extraction of the occupation number
\begin{align}
\label{eq_distr_def}
 f_{\mrm{EE}}(t,p) &= \frac{1}{2 \left(N^2-1\right) N_\lambda V} \frac{\left\langle E^j_{a}(\mbf p) \left(E^i_{a}(\mbf p)\right)^* \right\rangle_\cl}{\sqrt{m_\HTL^2 + p^2}},
\end{align}
which also takes the quasiparticle mass $m$ into account.
The $m$ is computed using an iterative self-consistent prescription: %Since we consider isotropic systems, the average is taken over many modes with same magnitude $p$ of the momentum. 
we compute
\begin{align} 
\label{eq_mass_formula}
 m^2 = 2 N_c \int \frac{\mrm d^3 p}{(2\pi)^3}\,\frac{g^2 f(t,p)}{\sqrt{m^2 + p^2}}.
\end{align}
For the first iteration we use $m=0$. Once we obtain an estimate for the mass $m$, we can repeat the extraction of the quasiparticle spectrum with the new mass, followed by repeated estimation of $m.$ In practice we find that the value becomes stable after approximately the first four iterations.

\subsection{Spectral and statistical correlation functions}
Previously we have been studying equal time correlation functions. However they are only a tiny subset of the correlation functions we can compute. Unequal time correlation functions contain a lot of information concerning the existence of quasiparticles and their possible excitations. Thus we wish to study them in more detail.

The first correlation function we will introduce is the statistical correlation function. In general this is defined as the expectation value of the anticommutator of two field operators 
\begin{align}
 \label{eq_stat_fct}
 F (x, x') &= \frac{1}{2 N_\lambda \left(N^2-1\right)}\,\left\langle \left\{ \hat{A}_k^b(x), \hat{A}_k^b(x') \right\} \right\rangle, \nonumber \\
 \ddot{F} (x, x') &= \frac{1}{2 N_\lambda \left(N^2-1\right)}\,\left\langle \left\{ \hat{E}^k_b(x), \hat{E}^k_b(x') \right\} \right\rangle.
\end{align}
For a spatially homogenous system $F$ depends only on the relative coordinate $\boldsymbol{x}-\boldsymbol{x}^\prime,$ and can thus be Fourier transformed  with respect to the relative coordinate. In Fourier space $F$ depends only on $t, t^\prime$ and $p,$ since we consider isotropic systems.
For classical statistical simulations we can estimate this as %\textbf{Tämä pitää nyt varmaankin perustella jotenkin. Paperissa selitys Heisenberg kenttäoperaattoreista, enkä ymmärrä, mutta selitys ehkä täällä https://arxiv.org/pdf/cond-mat/0703163.pdf ja https://arxiv.org/pdf/hep-ph/0107129.pdf.}
\begin{align}
 \label{eq_stat_fct_class}
 F (t,t', p) &= \frac{1}{ N_\lambda \left(N^2-1\right)V} \,\left\langle \left(A_k^b(t, \mbf p)\right)^*\,A_k^b(t', \mbf p) \right\rangle_\cl \\
  \ddot{F} (t,t', p) &= \frac{1}{ N_\lambda \left(N^2-1\right)V} \,\left\langle \left(E^k_b(t, \mbf p)\right)^*\,E^k_b(t', \mbf p) \right\rangle_\cl,
\end{align}
where the subscript $\langle \rangle_\cl$ stands for the average over classical fields.  When one restricts the statistical function to equal times, one makes the straightforward observation that the statistical function is also connected to the distribution function \equref{eq:discFA}.

The spectral function is defined as commutator of two unequal time fields
\begin{align}
 \label{eq_spectral_fct}
 \rho (x, x') &= i\left\langle \left[ \hat{A}_k^b(x), \hat{A}_k^b(x') \right] \right\rangle  \nonumber \\
 \dot{\rho} (x, x') &= i\left\langle \left[ \hat{E}^k_b(x), \hat{A}_k^b(x') \right] \right\rangle. 
\end{align}
The spectral function can be Fourier transformed to momentum space just like the statistical function was transformed. The canonical commutation relations impose the following  equal time correlations 
\begin{align}
 \label{eq_rho_init_cond}
 \lim_{t \rightarrow t'} \rho_T(t, t', p) &= 0 \nonumber \\
 \lim_{t \rightarrow t'} \dot{\rho}_T(t, t', p) &= 1.
\end{align}
The normalization of the spectral function is given by the sum rules
\begin{align}
 \label{eq_rho_sum_rules}
 \dot{\rho}_T^\HTL (\Delta t = 0,p) &= 2\int_{0}^\infty \frac{\ud \omega}{2\pi}\,\dot{\rho}_T^\HTL (\omega,p) = 1 \nonumber \\
 \dot{\rho}_L^\HTL (\Delta t = 0,p) &= 2\int_0^{\infty} \frac{\ud \omega}{2\pi}\, \dot{\rho}_L^\HTL (\omega, p) = \frac{2m^2}{2m^2 + p^2}.
\end{align}
The relation \equref{eq_rho_sum_rules} is also satisfied by  $\ddot{F}_T(t,t',p)  \ddot{F}_T(t,t,p).$
%the transverse sum rule follows from the canonical commutation relations and the longitudinal one from the fact that $\int_{-\infty}^\infty \ud \omega/(2\pi)\, \tilde{\rho}_L^\HTL (\omega, p)/\omega = 2m^2 / (p^2(2m^2 + p^2))$. 
However, it is difficult to compute the spectral function directly in the classical theory, since in the classical theory the commutator becomes a Poisson-bracket. Thus we want to find an alternative way to estimate it. 

The spectral function is connected to the retarded propagator
\begin{align}
 \label{eq_GR_rho_relation}
 G_{R}(t,t',p) = \theta(t - t')\, \rho(t,t',p). 
\end{align}
The retarded propagator, on the other hand, can be extracted using linear response analysis.
In \se \ref{subsec:linresponse} we will show how we can carry out the extraction using the formalism developed in \ch \ref{flucts}.

We define the Fourier transforms of the statistical and spectral functions  as 
\begin{align}
 \label{eq_F_rho_w_def}
 F(\tcent,\omega, p) =\, & 2\int_{0}^{\infty} \ud \Delta t \, \cos( \omega\, \Delta t)\, F(\tcent +\Delta t/2, \tcent-\Delta t/2, p), \nonumber \\
 \rho(\tcent,\omega, p) =\, & 2\int_{0}^{\infty} \ud \Delta t \,\sin( \omega\, \Delta t)\, \rho(\tcent +\Delta t/2, \tcent-\Delta t/2, p),
\end{align}
where we have used the notation $\tcent = (t + t')/2$ and $\Delta t = t - t'.$
For numerical convenience we approximate these by
\begin{align}
\label{eq_F_rho_w_comp}
 F(\tcent,\omega, p) \approx\, & 2\int_{0}^{\dtmax} \ud \Delta t \, \cos( \omega\, \Delta t)\, F(\tcent +\Delta t, \tcent, p), \nonumber \\
 \rho(\tcent,\omega, p) \approx\, & 2\int_{0}^{\dtmax} \ud \Delta t \,\sin( \omega\, \Delta t)\, \rho(\tcent +\Delta t, \tcent, p).
\end{align}
Here we estimated $\tcent \approx t$ and $\Delta t_{max} \ll t.$ The dominant contributions to the integral are typically given by small values of $\Delta t.$ The reason is that the correlation functions are approximately damped oscillators.  Thus we can approximate the upper limit of the integral by $\Delta t_{\mathrm{max}}$. 

The gauge fixing procedure for the unequal time correlation functions is different from that of equal time correlation functions. For equal time correlation functions we imposed Coulomb gauge condition at all readout times. For unequal time correlation functions this introduces gauge artifacts to the spectral function, since the fields do not belong to the same gauge trajectory anymore. Thus, in the case of unequal time correlation functions, the Coulomb gauge is fixed only on the time step when the linearized fluctuation is introduced. After this timestep no further gauge fixing will be done.

%%%%%%%%%%%%%%%%%%%%%%%%%
%\begin{figure}[t]
%	\centering
%	\includegraphics[width=0.8\textwidth]{Pictures/Mode_rho_F_dt_multi.pdf}
%	\caption{Statistical and spectral functions in the time domain for different momenta. In order to normalize $\ddot{F}_T$ we divide it by the equal-time correlation function $\ddot{F}_T(t,t,p)$.}
%	\label{fig_rho_F_dt_multi}
%\end{figure}
%%%%%%%%%%%%%%%%%%%%%%%%%
\subsection{Extraction of the retarded propagator using linear response theory}
\label{subsec:linresponse}
Next we consider a perturbation $a_i^b$ corresponding to the linearized fluctuations, which appeared in the previous chapter. The perturbation is seeded by a source $j^k_b(x) = j^k_b(t, \mbf x)$ and given by
\begin{align}
\label{eq_lin_resp_basic}
 \langle \hat{a}_i^b(x)\rangle = \int \mrm{d}^4 x' G_{R,ik}^{~~bc}(x, x')\, j^k_c(x').
\end{align}
The propagator $G_{R,ik}^{~~bc}(x, x')$ is defined as 
\begin{align}
 G_{R,ik}^{~~bc}(x, x') = i\theta(t-t')\, \left\langle \left[ \hat{A}_i^b(x), \hat{A}_k^c(x') \right] \right\rangle. 
\end{align}
For a spatially homogenous system the propagator depends only on the relative coordinates. In Fourier space we can write \equref{eq_lin_resp_basic} as
\begin{align}
 \langle \hat{a}_i^b(t, \mbf p)\rangle = \int \ud t'\, G_{R,ik}^{~~bc}(t, t', \mbf p)\, j^k_c(t',\mbf p).
\end{align}
Because of the integral, the extraction of $G_R$ for a general source is complicated. However, we are free to choose the source of the perturbation the way we desire. We can get rid of the integral by using an instantaneous perturbation of the form
\begin{align}
\label{eq_source_init_cont}
 j^k_c(t',\mbf p) = j^k_{0,c}(\mbf p)\, \delta\left( t' - \tpert \right).
\end{align}
This leads to 
\begin{align}
 \langle \hat{a}_i^b(t, \mbf p)\rangle= G_{R,ik}^{~~bc}(t, \tpert, \mbf p)\, j^k_{0,c}(\mbf p).
\end{align}
We can not solve this equation directly by dividing by the source, since we are summing over multiple color components. Instead, we must rely on correlations to deduce the propagator. Use of correlations also permits us to initialize multiple momentum modes at the same time. 

We choose the initial source to satisfy
\begin{align}
\label{eq_source_relations}
 \left\langle \left(j^k_{0,b}(\mbf p)\right)^* j^{k'}_{0,b'}(\mbf p')\right\rangle_\uj = \delta_{b,b'} V \delta_{\mbf p, \mbf p'} \sum_\lambda \left(v_k^{(\lambda)}(\mbf p)\right)^* v_{k'}^{(\lambda)}(\mbf p),
\end{align}
by choosing 
\begin{align}
 \label{eq_j_init}
 j^k_{0,b}(\mbf p) = \sum_{\lambda} c^{(\lambda)}_b(\mbf p)\, v_k^{(\lambda)}(\mbf p),
\end{align}
where the $c^{(\lambda)}_b(\mbf p)$ are random phase factors satisfying \equref{eq_random_nums} and $\left\langle \cdot \right\rangle_\uj$ stands for an average over sources. If not stated otherwise, we will only initialize transverse modes, corresponding to polarization states $\lambda = 1,2$. When we explicitly discuss the longitudinal polarization, we initialize only $\lambda=3,$ and no transverse modes. 

%%%%%%%%%%%%%%%%%%%%%%%%%
%\begin{figure}[t]
%	\centering
%	\includegraphics[width=0.8\textwidth]{Pictures/Mode_rho_F_w_multi.pdf}
%	\caption{Statistical and spectral functions of \fig\ref{fig_rho_F_dt_multi} in frequency domain with $\Q \dtmax = 200$.}
%	\label{fig_rho_F_w_multi}
%\end{figure}
%%%%%%%%%%%%%%%%%%%%%%%%%%

The retarded propagator is then given by 
\begin{align}
 G_{R}(t, \tpert, \mbf p)  
 &=\frac{1}{\left(N^2-1\right)N_\lambda} \sum_\lambda\left(v_i^{(\lambda)}(\mbf p)\right)^* G_{R,ik}^{~~bb}(t, \tpert, \mbf p) v_k^{(\lambda)}(\mbf p) \nonumber \\
 &= \frac{1}{\left(N^2-1\right) N_\lambda V}\, \left\langle \left(j^i_{0, b}(\mbf p)\right)^* \langle \hat{a}_i^b(t, \mbf p)\rangle  \right\rangle_\uj.
\end{align}
Similarly the time derivative of the propagator is obtained as 
\begin{align}
 \dot{G}_{R}(t, \tpert, \mbf p) = \frac{1}{\left(N^2-1\right) N_\lambda V} \left\langle \left(j^k_{0, b}(\mbf p)\right)^* \langle \hat{e}^k_b(t, \mbf p)\rangle  \right\rangle_\uj.
\end{align}
The initial condition for the retarded propagator at  $t \rightarrow \tpert$ are
\begin{align}
 \label{eq_GR_init}
 \lim_{t \rightarrow \tpert} G_R(t, \tpert, \mbf p) &= 0 \nonumber \\
 \lim_{t \rightarrow \tpert}\dot{G}_{R}(t, \tpert, \mbf p) &= 1.
\end{align}

%%%%%%%%%%%%%%%%%%%%%%%%%
\begin{figure}[t]
	\centering
	\includegraphics[width=0.8\textwidth]{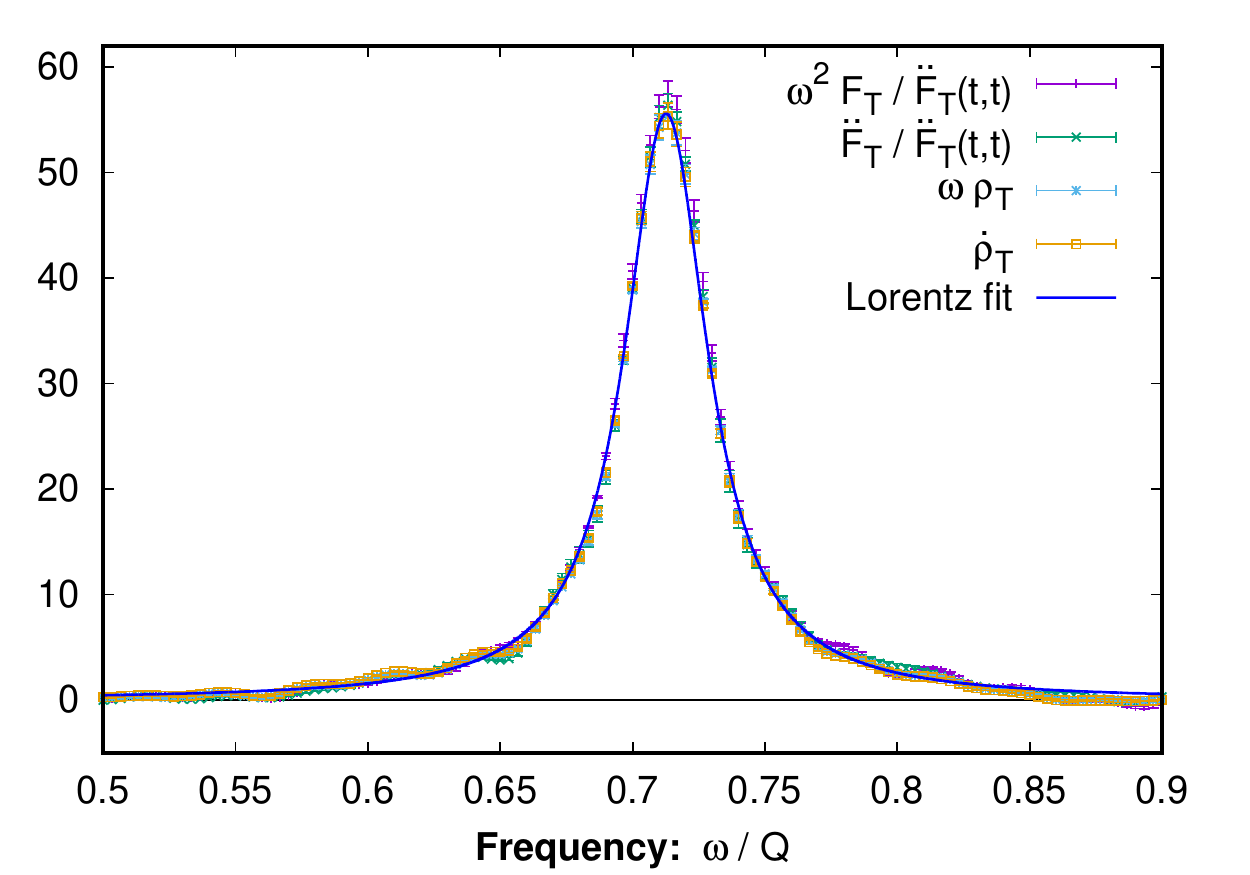}
	\caption{Here we compare the transverse statistical and spectral functions in frequency space for $p = 0.7\, \Q$ to the fit function of a Lorentzian form given by  \equref{eq_lor_curve}.}
	\label{fig_lorentz}
\end{figure}
%%%%%%%%%%%%%%%%%%%%%%%%%%

To summarize: we have obtained a practical algorithm for extracting the retarded propagator in linear response theory:
\begin{enumerate}
\item Choose the source $j$ according to \equref{eq_j_init}. The source should act instantaneously on a single time step.
\item Evolve the system of background field and linearized fluctuations in time. 
\item Correlate $e$ or $a$ with the source. This allows one to obtain the time derivative of the retarded propagator ($j e$ correlator) or the retarded propagator itself ($j a$ correlator).  
\end{enumerate}

\section{Results}
\begin{figure}[t]
	\centering
	\includegraphics[width=0.49\textwidth]{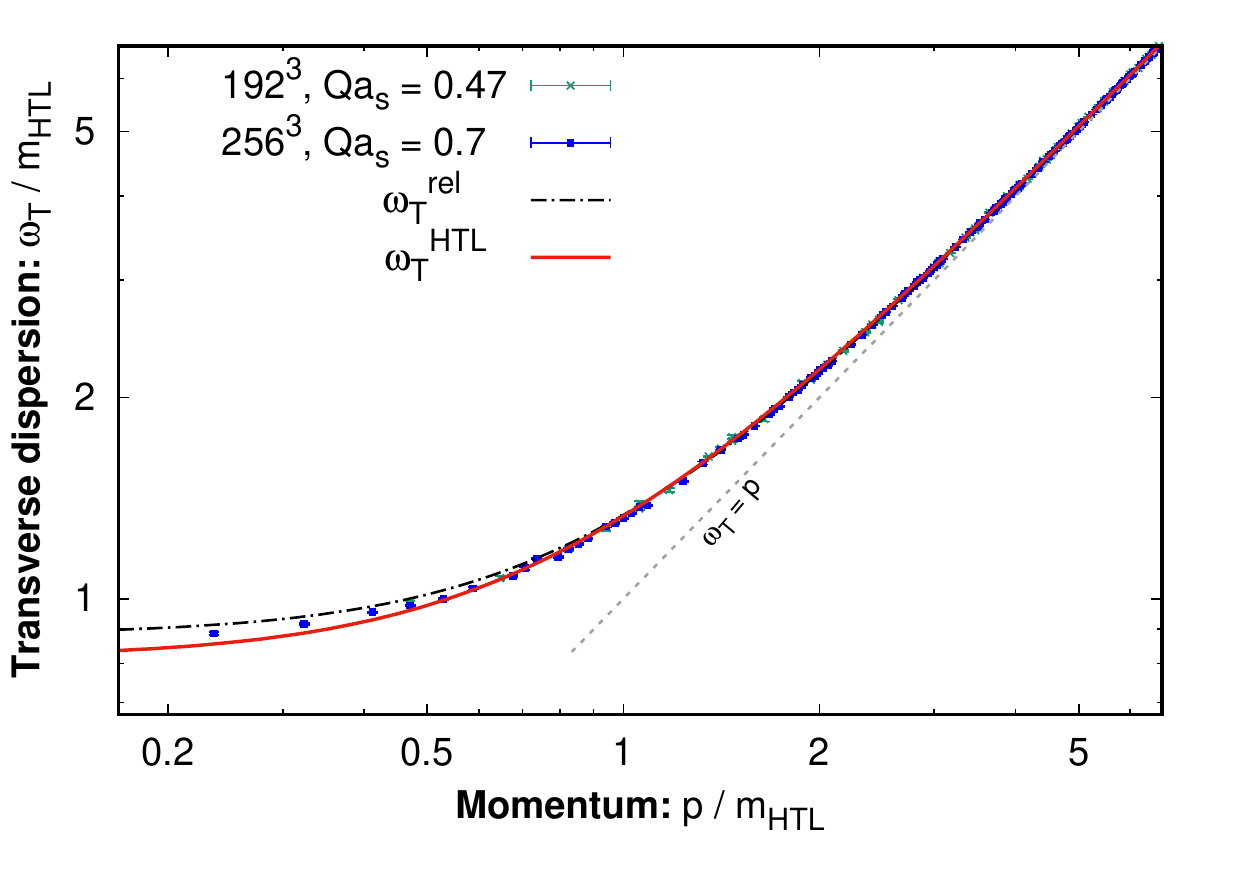}
	\includegraphics[width=0.49\textwidth]{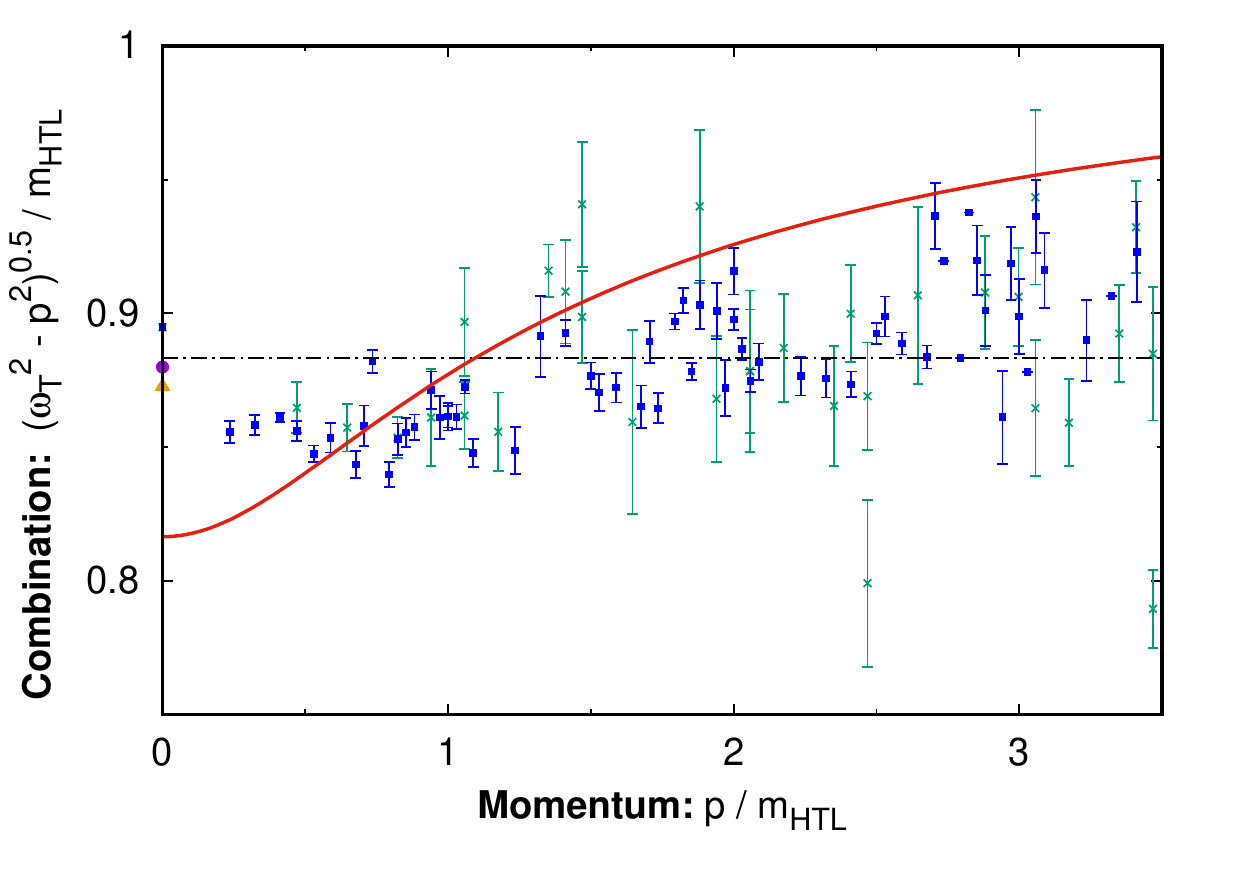}
	\caption{Left: The transverse dispersion relation $\omega_T(p)$ is shown here for two different discretizations. We obtain the dispersion relation by locating the maximum of the quasiparticle peak in $\dot{\rho}_T(\tpert,\omega,p)$ for every momentum. The red curve shows the solution of the HTL curve using the mass value of $m_{\mathrm{HTL}} = 0.149 Q$.  The dashed black curve shows the relativistic dispersion relation for comparison. The massless dispersion relation is also shown as a grey dashed curve and labeled as $\omega_T = p$. Right: Here we measure the deviation from the relativistic dispersion relation using $\sqrt{\omega_T^2 - p^2}$ using the same data on a linear plot. The dashed lines shows the relativistic dispersion relation. The orange triangle and purple circle correspond to zero mode frequencies obtained using the simulations with $\Q a_s = 0.7$ and $\Q a_s = 0.47$ with longitudinal polarization only. }
	\label{fig_disprel_trans}
\end{figure}

\subsection{Comparison of spectral function to the Lorentzian form}
%Before we proceed to compare spectral and statistical correlation functions, we would like to point out an important detail concerning the gauge fixing procedure for unequal time correlation functions. For equal time correlation functions we performed the Coulomb gauge fixing procedure on ever time step on which we performed measurements, and we will continue to do so also in this chapter. However, for unequal time correlation functions this procedure is no longer applicable, since Coulomb gauges at different time slices are not the same gauges. Thus we will fix the Coulomb gauge only when we introduce the fluctuations. In practice we have observed that fixing the gauge for unequal time correlation functions introduces gauge artifacts in the spectral function and other unequal time correlation functions.
Because we have $\tcent \gg \ \Delta t$ and the typical frequencies we are interested in are $\omega \propto \nicefrac{1}{\Delta t},$  we have $\tcent^{-1} \ll \omega$ and the correlation functions change much faster as a function of $\Delta t$ than $\tcent$. This means that we can approximate the time derivatives of the spectral and statistical functions as $\dot{\rho} \rightarrow \omega \rho$ and similarly for the spectral function. %We will also later verify this with our data. 
% A comparison of spectral and statistical functions is shown in \fig \ref{fig_rho_F_dt_multi} at time $Q t_{\mathrm{pert}} = 1500.$ We can see that the two are in good agreement and they both feature damped oscillatory behavior.

%It is instructive to look at the same data also in frequency domain. The Fourier transform is shown in \fig \ref{fig_rho_F_w_multi}. Unsurprisingly, we observe that the data also coincides in the momentum space. The position of the peak in this plot corresponds to the frequency of the mode. As we can see, for all the modes the peak is located above $\omega = p$, which is the physical region. 
 
Figure \ref{fig_lorentz} compares the Lorentzian form of the spectral function to our data for $p = 0.7 Q$. We have also introduced curves $\omega^2 F_T / \ddot{F}_T(t,t,p)$ and $\omega \rho_T$. These curves coincide with each other %and the previous ones, further 
justifying the assumption that the time derivatives can be approximated with frequency factors. We also find a good agreement with the Lorentzian shape \equref{eq_lor_curve}. This establishes the fact that our system indeed has quasiparticle excitations. We will study their properties in the following subsections in more detail.

\subsection{Results for transverse dispersion relation, damping rate and Landau cut}
The results for the transverse dispersion relation $\omega_T$ are shown in \fig \ref{fig_disprel_trans}. The plot shows the extracted dispersion relation for two sets of discretization parameters, giving similar results. The dispersion relation is extracted by finding the maximum of the quasiparticle peak in $\dot{\rho}_T(\tpert,\omega,p)$. The data can be compared to the numerically solved HTL dispersion relation, where the only free parameter is the mass scale, which we estimate using \equref{eq_mass_formula} and \equref{eq_distr_def} to be $m_\HTL = 0.149 \, \Q.$ We also show a curve with simple relativistic dispersion relation, which approximates the HTL dispersion relation at large $p$. This is denoted by $\omega_T^{\mathrm{rel}} = \sqrt{p^2+m_\rel^2},$ and the mass $m_\rel = 0.132\, \Q$ is obtained by fitting the curve to the data. It turns out that both functional forms offer a good description of the data. The fact that the data is overall well described using the mass scale arising from the self-consistent HTL approach also justifies its use in \fig \ref{fig_disprel_trans} and the following figures.

In the right panel of \fig \ref{fig_disprel_trans} we have quantified the deviations from the relativistic dispersion relation by measuring $\sqrt{\omega_T^2 - p^2},$ which should remain constant in the case of the relativistic dispersion relation. We observe that at low momentum the dispersion relation of the data is smaller than the mass predicted by the relativistic dispersion relation, and in the higher momentum regime the measured mass is larger. 

Performing the fits also permits us to measure the values of the mass scales. The plasmon frequency (corresponding to the plasmon mass) is measured by numerically finding the maximum of the spectral function for $p=0$ excitations in frequency space. These points (corresponding to different discretizations) are also shown in the right hand side of \fig \ref{fig_disprel_trans}. The result is 
\begin{align}
 \label{eq_wplas_data}
 \wplas^\fit(\Q t = 1500) / \Q = 0.132 \pm 0.002. 
\end{align}
This value is larger than the one predicted by HTL $\wplas^\HTL = 0.122\,\Q,$ but it is the same as the mass obtained using the relativistic fit. 

\begin{figure}[t]
	\centering
	\includegraphics[width=0.8\textwidth]{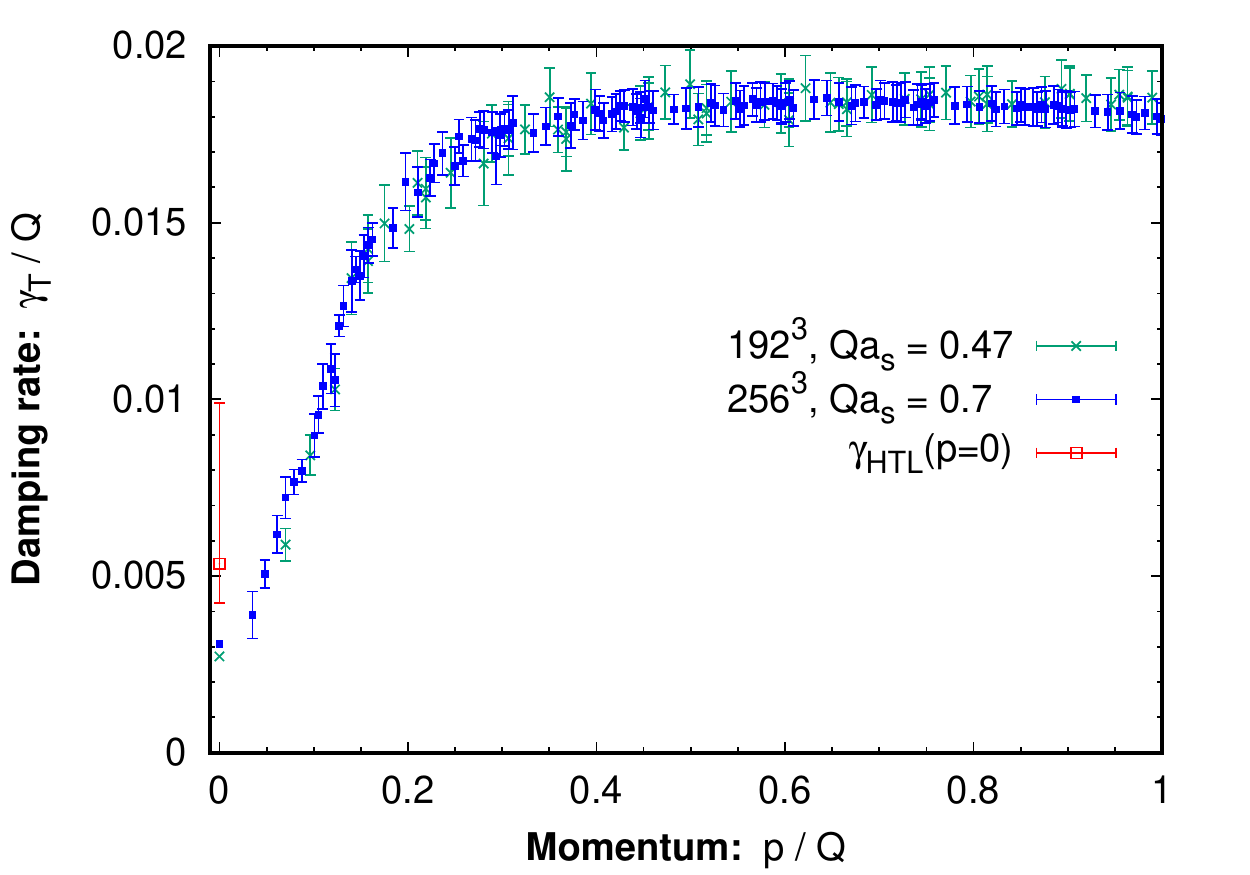}
	\caption{The damping rate for the transverse modes $\gamma_T\left(p \right)$ obtained by fitting damped oscillation to the spectral function $\dot{\rho}\left( t_{\mathrm{pert}} , \delta t , p \right)$ in the time domain. The time range used in the fit is $\Q \Delta t \leq 70$. }
	\label{fig_damp_rate}
\end{figure}

We have also measured the asymptotic mass scale by first numerically solving the HTL dispersion relation and by then fitting the data to the interpolation function of the solution. The error estimates are obtained by varying the minimum momentum in the fit between $p_{\mrm{min}} = 2m_\HTL - 4m_\HTL$ and by taking the maxima of $\rho_T$ and $\dot{\rho}_T$ into account when determining the systematic errors of the fit. The fit procedure also introduces an error, and the resulting value for the mass scale is 
\begin{align}
 \label{eq_extracted_mass}
 m_\fit = 0.138 \pm 0.002\; (\mrm{sys}) \pm 0.0015\; (\mrm{fit}). 
\end{align}
We observe that the mass scale obtained in this way is smaller than the one given by the self-consistent HTL formula \equref{eq_mass_formula} and on the other hand it is larger than the value $m_\rel$ we obtained using the relativistic fit. This is consistent with the right hand side of \fig \ref{fig_disprel_trans} where the datapoints are consistently below the HTL curve at higher momenta, but also above the curve corresponding to the relativistic dispersion relation.

We can also look for deviations from the leading order HTL formalism by looking at the ratios of the mass scales. The prediction made by HTL is $\wplas^\HTL / m_\HTL = \sqrt{2/3} \approx 0.8165.$ For the relativistic dispersion relation we have by construction $\wplas^\rel / m_\rel = 1$. Using the values we obtained by fitting we find 
\begin{align}
 \label{eq_wplas_m_ratio}
 \frac{\wplas^\fit}{m_\fit}(\Q t = 1500) = 0.957 \pm 0.028.
\end{align}
This value is clearly closer to 1 than the HTL expectation. 

%%%%%%%%%%%%%%%%%%%%%%%%%
\begin{figure}[t]
	\centering
	\includegraphics[width=0.6\textwidth]{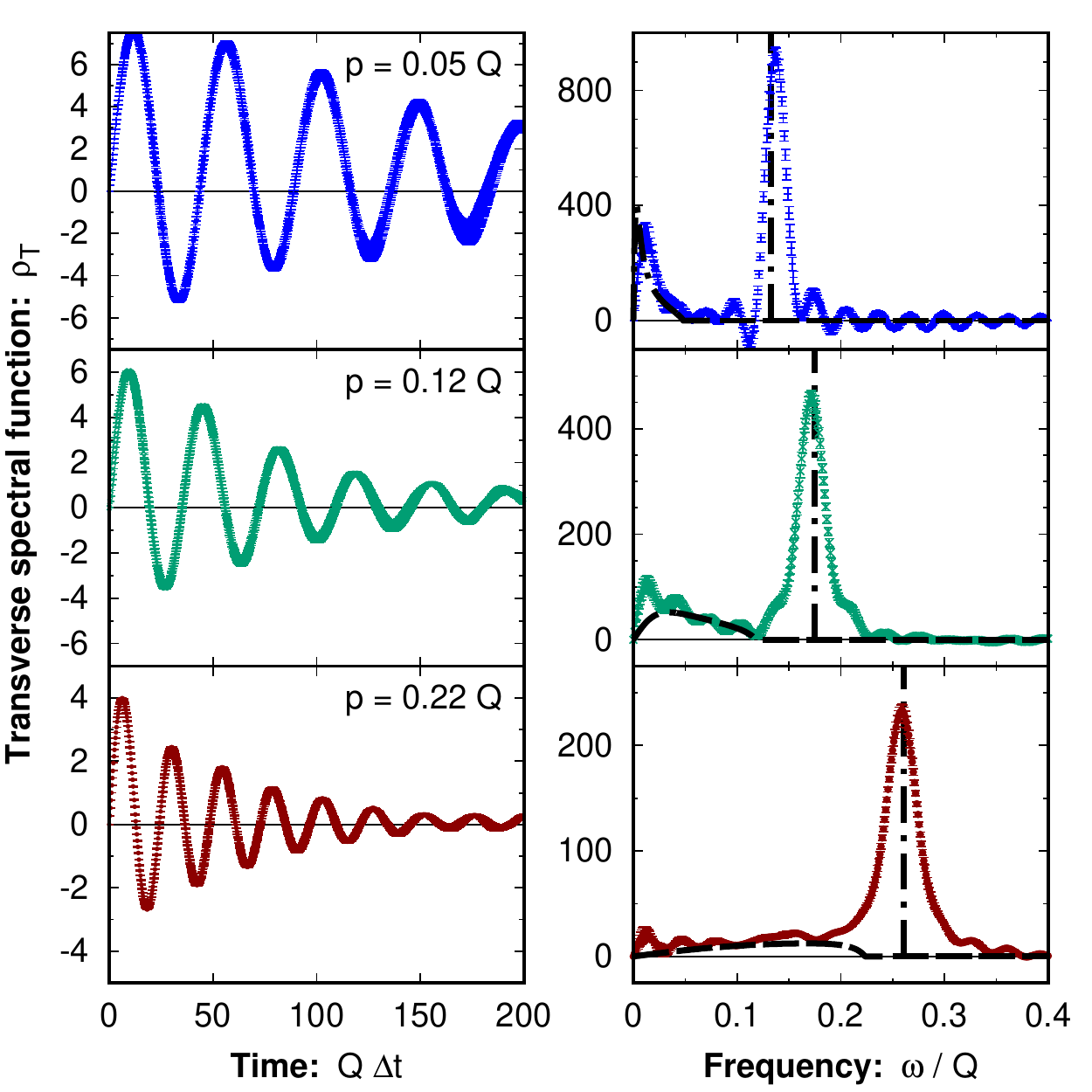}
	\caption{The left hand side features the spectral function in time domain for various small momenta $p$. The right hand side shows the corresponding spectral functions in the frequency domain with the maximum time $Q \Delta t_{\mathrm{max}} = 200$ used in the Fourier transformation. The dashed black lines are the corresponding solutions to the HTL spectral function solved using the value $m_{\mathrm{HTL}} = 0.149Q$ as the mass parameter. The nonzero region below the quasiparticle peak in the HTL solution corresponds to the Landau cut. }
	\label{fig_Landau_cut}
\end{figure}

Since we are able to measure the oscillations in the spectral function in the time domain, or the width of the quasiparticle peak in the frequency domain, we are also able to measure the damping rate $\gamma_T$ of the transverse quasiparticle excitations. To our knowledge, the damping rate has only been computed for the mode $p=0$ \cite{Braaten:1990it}. Thus our approach will enable us to  study the momentum dependence of the damping rate for the first time. In practice it turns out to be easier to measure the damping rate in the time domain, since the finite time window in the Fourier transform leads to deviations from the Lorentzian form in the momentum space. We can also extract the frequency of each mode while doing the fit, and we have checked that the values we get are consistent with data shown in \fig \ref{fig_disprel_trans}.

The results for the damping rate are shown in the  \fig \ref{fig_damp_rate} for two different discretizations yielding the same results within uncertainties. We observe that the damping rate increases linearly in $p$ for small enough momenta $p \leq 0.15 Q$.  At larger momenta ($p > 0.3 Q$) our results are consistent with a constant damping rate. 

The $p=0$ mode deserves special attention, since its damping rate has also been estimated analytically \cite{Braaten:1990it}. 
\begin{align}
 \label{eq_gamma_0}
 \gamma_\HTL(0) = 6.63538\, \frac{g^2 N_c T_*}{24\pi}\,,
\end{align}
with the replacement $T \mapsto T_*.$ Here $T_*$ is the effective temperature given by the ratio of the integrals
\begin{align}
 T_*(t) = \mathcal{I}(t) / \mathcal{J}(t),
\end{align}
where
\begin{align}
 \mathcal{I}(t) &= \frac{1}{2} \int \frac{\ud^3 p}{(2\pi)^3}\, f(t,p) \left(f(t,p) + 1 \right) \nonumber \\
 \mathcal{J}(t) &= \int \frac{\ud^3 p}{(2\pi)^3}\, \frac{f(t,p)}{\sqrt{m_\HTL^2 + p^2}}.
\end{align}
The errorbars in the estimate are given by using different definitions of the distribution function. Our result is roughly consistent with the HTL prediction. More precise analysis on the consistency is complicated by the fact that our results for the $p=0$ mode do not include error bars since the fitting was done using averaged data. %Our results for the damping rate at $p=0$ and the slope at $Qt = 1500$ are 
%\begin{align}
% \label{eq_gT_porp}
% \frac{\ud \gamma_T}{\ud p} &= 0.086 \pm 0.002, \nonumber \\
% \gamma_T(p \rightarrow 0) / \Q &= (8.5 \pm 2.3) \cdot 10^{-4}. 
%\end{align}
%In order to extract the damping rate at $p=0$ we have to rely on extrapolation, since measuring it would be computationally costly. 

Finally we want to study the shape of the spectral function in more detail. The spectral function consists of two parts, the quasiparticle peak, and the continuum of excitations at lower frequencies which are also known as the Landau cut. The form of the Landau cut for transverse excitations is given by \equref{eq:transverselandau}. The solution of the HTL spectral function with the Landau cut and our results for the spectral function are shown in \fig \ref{fig_Landau_cut}. In figure \ref{fig_lorentz} we had the time-derivative of the spectral function on the y-axis, corresponding roughly to the spectral function times the frequency. The multiplication by the frequency makes the low frequency structures indistinguishable, which is why in \fig \ref{fig_Landau_cut} we have left out the multiplication in order to zoom into the Landau cut region. We find that the low frequency excitations agree with the leading order HTL curves. However, at small frequencies one observes small deviations which we interpret as arising from the finite time window used in the Fourier transform. Thus the HTL framework seems to provide a good description of the Landau cut region also in a system which is far from equilibrium.

%%%%%%%%%%%%%%%%%%%%%%%%%
\begin{figure}[t]
	\centering
	\includegraphics[scale=0.6]{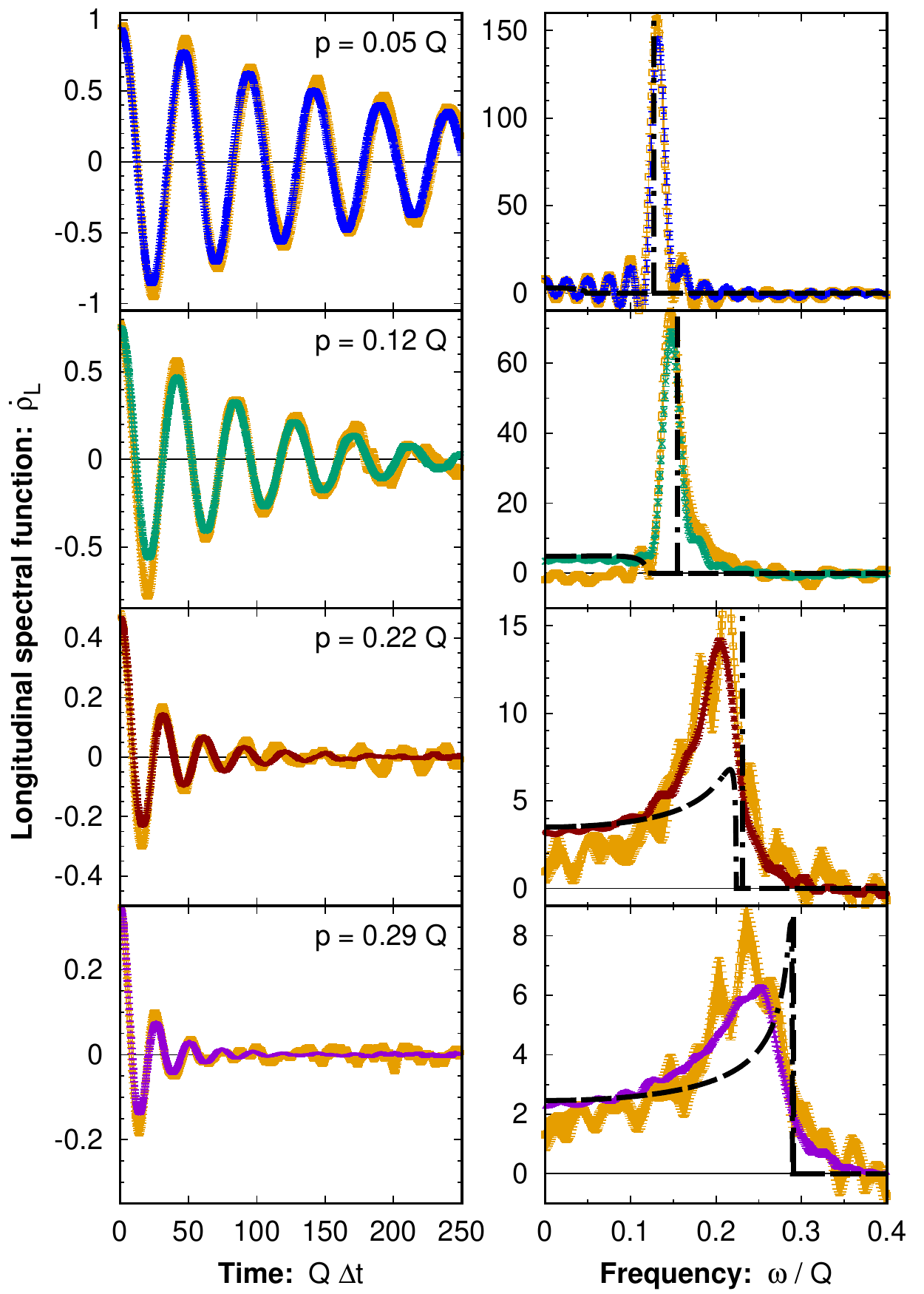}
	\caption{The longitudinal spectral function $\dot{\rho}_L$ rescaled by a factor $p^2/\omega^2$ in the time and frequency domain for various momenta $p.$ The extraction using linear response is shown in green, blue, brown and purple curves. The yellow curve corresponds to the normalized longitudinal statistical correlation function $\ddot{F}_L / \ddot{F}_L(t,\Delta t = 0,p)\; \dot{\rho}_L^\HTL(t,\Delta t = 0,p).$ The black dashed line corresponds to HTL estimate of the time derivative of the longitudinal spectral function $\dot{\rho}_{\mathrm{HTL}}.$ We use maximum $\Delta t = 250$ for the Fourier transform. }
	\label{fig_rho_dt_w_long}
\end{figure}
%%%%%%%%%%%%%%%%%%%%%%%%%%

\subsection{Results for longitudinal excitations}
Next we are going to take a look at results concerning the same observables for longitudinal modes. Figure \ref{fig_rho_dt_w_long} shows the longitudinal spectral function in the time and frequency domains. In the time domain we observe qualitatively similar damped oscillations as in the transverse case. However, it seems that the oscillations are more strongly damped than in the transverse case, especially for larger momenta. The Fourier transforms are shown on the right hand side of the \fig \ref{fig_rho_dt_w_long}.  For small momenta $\lesssim m$ we observe clearly two distinct structures. The Landau cut in the regime of low energy excitations and the quasiparticle peak. At larger momenta the dispersion relation approaches the light cone exponentially as can be seen from the high momentum estimate of the longitudinal dispersion relation given by \equref{eq_wHTL_Taylor_trans}. Ultimately the quasiparticle peak and the Landau cut become almost numerically indistinguishable. In practice, for momenta above $p \gtrsim 0.29\, \Q$ we can only see the Landau cut.

We can also extract the longitudinal dispersion relation in a similar fashion as we extracted the transverse dispersion relation. However, since we had problems resolving the quasiparticle peaks in the spectral function for the higher momenta, we try to increase our resolution by subtracting the Landau cut from the data. In this way we obtain an estimate for the longitudinal dispersion relation. However it is important to remember that the subtraction is not precise in the high momentum region. 

\begin{figure}[t]
	\centering
	\includegraphics[scale=0.7]{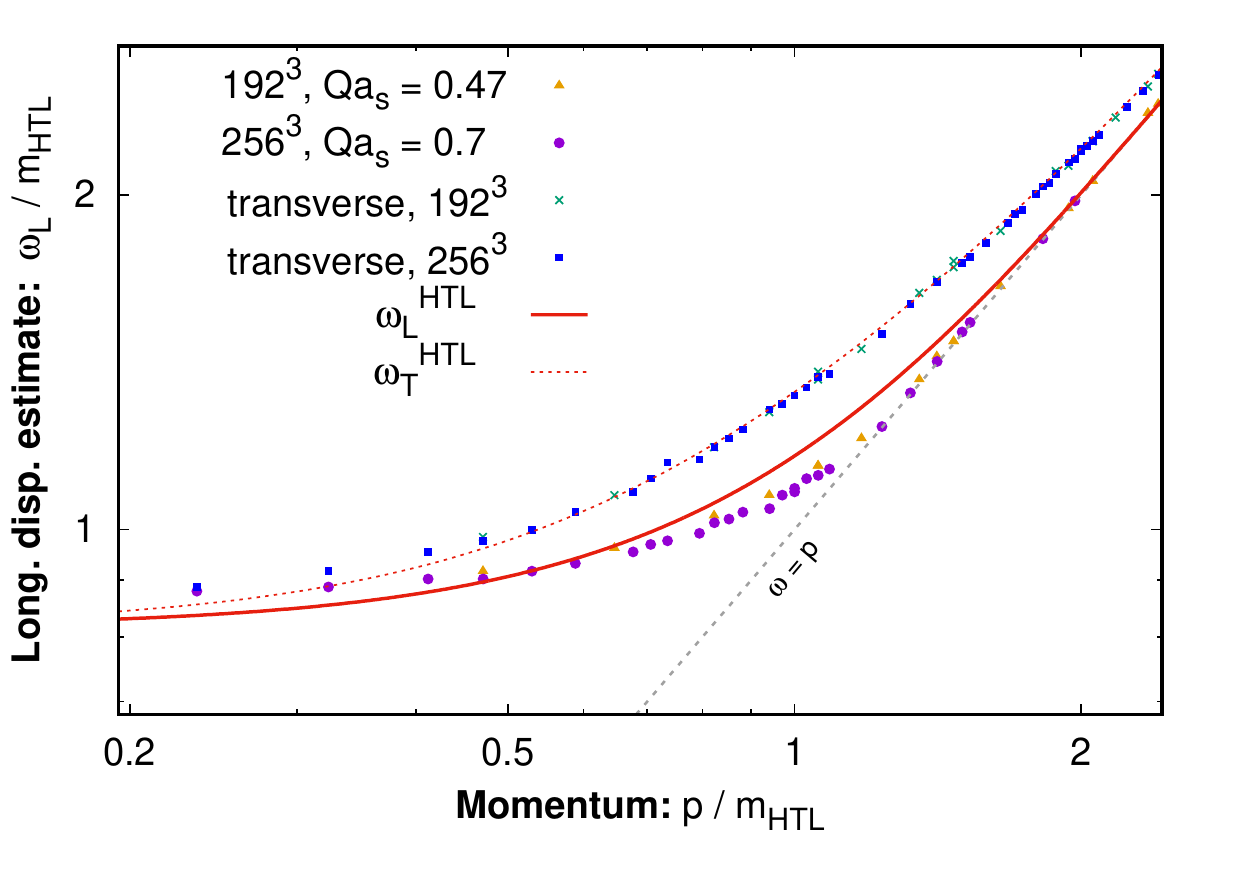}
	\caption{The measured dispersion relation of the longitudinal quasiparticles is shown along with the HTL expectation. The dispersion relation is obtained by finding the maximum of $\dot{\rho}_L(\tpert,\omega,p) \, p^2/\omega^2$ on log a scale after subtracting the expected Landau cut contribution. We also show the transverse data and the HTL dispersion relation.}
	\label{fig_disprel_long}
\end{figure}

The results for the longitudinal dispersion relation are shown in \fig \ref{fig_disprel_long}. We show two different discretizations for the longitudinal dispersion relation, the HTL curve for the longitudinal excitations and the transverse results for comparison. We observe that the longitudinal and transverse dispersion relations coincide when $p \rightarrow 0$ as they should. The longitudinal dispersion relation approaches the light cone at large $p$, but deviates from the HTL curve in the region where $ p \sim Q$. The reason for this behavior might be that the HTL Landau cut differs from the Landau cut of the data, which could influence our results. 

We can also extract the damping rate of longitudinal excitations. However, due to issues related to the Landau cut this is only possible for small momenta ($\nicefrac{p}{Q} < 0.14.$). We find that the longitudinal damping rate is consistent with the transverse damping rate within uncertainties.
%%%%%%%%%%%%%%%%%%%%%%%%%
%%%%%%%%%%%%%%%%%%%%%%%%%%

\subsection{Insensitivity to initial parameters}
\begin{figure}[t]
	\centering
	\includegraphics[scale=0.72]{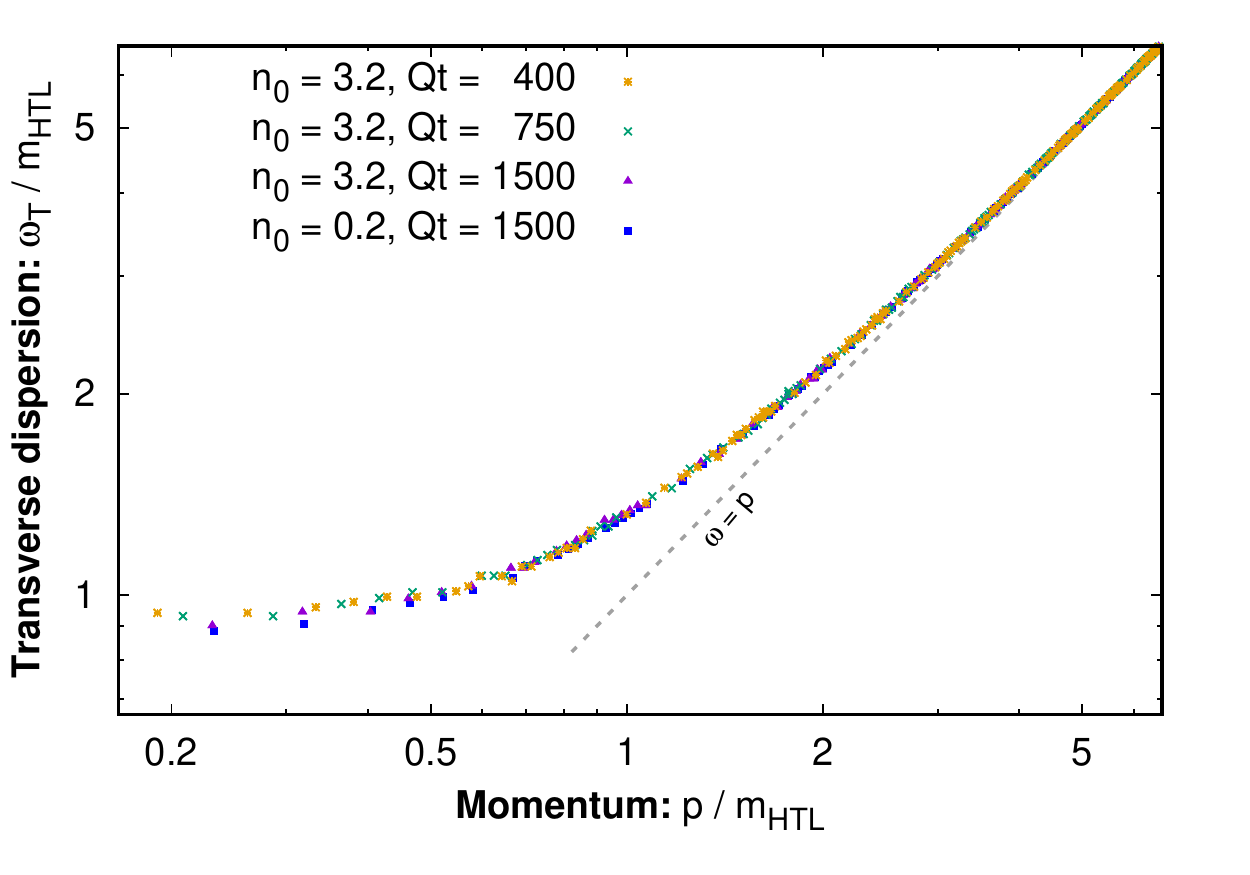}
	\caption{Dispersion relation for transverse excitations for different perturbation initialization times $\tpert$ and occupation numbers $n_0$. The frequency for each mode is obtained by finding the maximum of $\dot{\rho}_T(\tpert,\omega,p)$. }
	\label{fig_disprel_trans_params}
\end{figure}

We have also checked the sensitivity of our results to the initial parameters. The occupation number and time dependence of the dispersion relation of the transverse quasiparticles is shown in \fig \ref{fig_disprel_trans_params}. We have introduced the fluctuations at three different times $Q \tpert = 400, 750$ and $1500$ and two different occupation numbers, $n_0 = 3.2$ and $0.2$. It seems that our results are independent of time and occupation number.

%%%%%%%%%%%%%%%%%%%%%%%%%
%%%%%%%%%%%%%%%%%%%%%%%%%%

%A similar check can be done on the damping rate as well. The results are shown in \fig \ref{fig_damp_rate_params}. The damping rate turns out to be more sensitive to the variation of the initial parameters. In \fig \ref{fig_damp_rate_params} the curve with $n_0 = 0.2$ has to bu multiplied by a factor of $1.45$ in order to match it to the curve with $n_0 = 3.2$. Thus the damping rate also depends to some extent on the occupation number. We also observe a nontrivial dependence on time. The damping rate grows linearly in $p$ all the way until $p \approx m_{\mathrm{HTL}}$ for all simulations. Above the mass scale the damping rate plateaus also for different initiation time for the perturbations. Thus the functional form is similar for all sets of initial parameters, but the exact details depend on the initial parameters. 

%%%%%%%%%%%%%%%%%%%%%%%%%
%\begin{figure}[t]
%	\centering
%	\includegraphics[scale=0.72]{Pictures/Damping_rate_trans_params_2.pdf}
%	\caption{Transverse damping rate for different sets of initial simulation parameters. Here we vary the initialization time of the perturbatin $\tpert$ and the occupation number $n_0$. The damping rate is extracted by fitting damped oscillation to the spectral function $\dot{\rho}_T(\tpert,\Delta t, p)$ for a time range of $\Q \Delta t \leq 50$. The curve with $n_0=0.2$ has been multiplied by $1.45$ in order to make it overlap with the curve with $n_0=3.2$.}
%	\label{fig_damp_rate_params}
%\end{figure}
%%%%%%%%%%%%%%%%%%%%%%%%%%

\subsection{Relationship to the plasmon mass measurements}
In \ch \ref{plasmon} we tried to measure the dispersion relation using the ratio of the squares of the electric field and the gauge field using \equref{eq:huonodispersiorelaatio} and \equref{eq:dispersiorelaatio}. Both of these assumed that the quasiparticles have a well defined dispersion relation (in the language of the spectral functions this means that the spectral function is dominated by a single Lorentzian quasiparticle peak). There we observed that this observable was not in agreement with other measurements. We observed that \equref{eq:huonodispersiorelaatio} undershot  and \equref{eq:dispersiorelaatio} overshot the results given by the other methods. Figures \ref{fig_Landau_cut} and  \ref{fig_rho_dt_w_long} provide the information about the possible excitation spectra we were missing in \ch \ref{plasmon}. The transverse excitations indeed seem to have a clear quasiparticle peak. However, the quasiparticle excitations are not the only possible excitations, since we also have a continuum of low frequency excitations available in the Landau cut region. It is possible that the presence of these excitations have contaminated the dispersion relation measurements which we did in \ch \ref{plasmon}. In the case of longitudinal excitations the contribution coming from the Landau cut is even larger, and at large momentum the Landau cut and the quasiparticle peak became almost inseparable. This casts doubts on the reliability of the longitudinal dispersion relation extracted using \equref{eq:huonodispersiorelaatio} and \equref{eq:dispersiorelaatio}.  However, one has to keep in mind that the connection between our results concerning the spectral function and dispersion relation can not be directly compared to the results concerning the plasmon mass in \ch \ref{plasmon}, since there the reference scales are measured differently. In order to understand why the plasmon mass scales given by \equref{eq:huonodispersiorelaatio} and \equref{eq:dispersiorelaatio} seemed to be more consistent with the other methods in two dimensions than in three dimensions, it would be necessary to extract the spectral function also in the two-dimensional case. 

Our results on the damping rate shown in \fig \ref{fig_damp_rate} reinforce the conclusion we made about the size of the damping rate in \ch \ref{plasmon} compared to the plasmon frequency when we argued that the effect of the damping rate is negligible in \equref{eq:huonodispersiorelaatio} and \equref{eq:dispersiorelaatio}. Approximating $m_{\mathrm{HTL}} \approx 0.15,$ one finds that the damping rate plateaus at roughly $\nicefrac{\gamma_T}{m_{\mathrm{HTL}}} 	\approx 0.12$ and thus we always have $\omega_T \gg \gamma_T.$ 

\section{Summary of spectral properties}
Here we will briefly summarize our main results in this chapter. 
\begin{itemize}
\item We observe a clear quasiparticle in the spectral functions. This indicates that quasiparticles indeed exist in our system.
\item We measure the ratio of the asymptotic mass $m$ and plasmon frequency $\omega_{pl}$. The HTL prediction is $\nicefrac{\omega_{pl}^{\HTL}}{m^{\HTL}} = \sqrt{\nicefrac{2}{3}}$ and our numerical result is slightly larger $\nicefrac{\omega_{pl}}{m} = 0.957 \pm 0.028.$ 
\item The transverse damping rate seems to rise linearly for small ($\nicefrac{p}{Q} < 0.2$) momenta. For larger momenta ($\nicefrac{p}{Q} > 0.3$) our results are consistent with a constant damping rate.  The damping rate does not seem to depend on polarization. However this result should be interpreted with caution since longitudinal data is available only for small momenta.
\item The damping rate at zero momentum is roughly consistent with the analytical HTL prediction. However, precise analysis on the consistency is complicated because we were unable to obtain errorbars for the numerical results at zero momentum.
\item For transverse and longitudinal modes the spectral function exhibits a clear Landau cut.
\item We measure the dispersion relation of the transverse modes. We observe good overall agreement with HTL. For longitudinal excitations we observe more discrepancies, most likely due to the more noisy signal.
\item The longitudinal spectral function is more noisy than the transverse one. The quasiparticle peak and the Landau cut become almost indistinguishable at larger ($\nicefrac{p}{Q} > 0.2$) momenta. We are able to extract the dispersion relation after subtracting the HTL Landau cut.
\end{itemize}

\chapter{Conclusions and outlook}
\label{ch:outlook}
In papers \cite{Lappi:2016ato} and \cite{Lappi:2017ckt} we compared three different methods to measure the plasmon mass scale for overoccupied classical nonequilibrium systems in two and three dimensions. The methods were the effective dispersion relation (DR), the hard thermal loop formula (HTL) and the uniform electric field method (UE). The three-dimensional simulation was done in a fixed box, and it served as a warmup for the two-dimensional simulation, which more accurately mimics the 2+1 dimensional boost invariant system produced in an ultrarelativistic heavy-ion collision. We observed that the values given by the DR method are larger than the ones given by the other methods. The HTL method works well in three dimensions, but in two dimensions one has to be careful with the definition of the occupation number. The UE method works well in both two-dimensional and three-dimensional cases, but is computationally very expensive. In three dimensions we observed that the UE and HTL methods can be brought into agreement in the continuum limit. 

In paper \cite{Kurkela:2016mhu} we developed a method, which can be used to simulate linearized fluctuations on top of a classical Yang-Mills background. Our method conserves Gauss' law by construction. We have also numerically tested that the linearization works correctly and verified the conservation of Gauss' law. Our method was based on Hamiltonian formalism. We also presented an algorithm which corresponds to Lagrangian formalism, and we would expect this to conserve Gauss' law in a more straightforward manner. This expectation is based on the fact that Gauss' law is one of the Lagrangian equations of motion, and thus it should be conserved by construction.

In paper \cite{Boguslavski:2018beu} we studied linear response of classical gluodynamics in a fixed box by applying the formalism developed in \cite{Kurkela:2016mhu}.  We focused on the spectral properties, i.e. dispersion relation, spectral function and damping rate of the transverse and longitudinal modes of the gluon plasma. We compared our results to the predictions of the HTL perturbation theory. The results are mostly consistent with the predictions of the HTL for the transverse excitations. For the longitudinal modes we observed slight deviations from the dispersion relation predicted by HTL. These might be caused by problems in resolving the quasiparticle peak and the Landau cut. We also measured the ratio of the plasmon mass and the asymptotic mass and observed a slight deviation from the HTL expectation.

In this PhD thesis we have studied the quasiparticle properties in classical gluodynamics. In three dimensions the agreement between the plasmon mass scale given by the HTL formula and the UE method already gave us strong hints that classical overoccupied gauge theory can be understood in a quasiparticle picture. When we studied spectral properties of this system the existence of quasiparticles was firmly established by the clear measurement of the quasiparticle peak in the spectral function. %In two dimensional studies concerning the plasmon mass we could not draw the same conclusion, due to the uncertainty concerning the existence of the kinetic theory in the corresponding system. However, once we employ linear response theory in two-dimensional systems we can address the existence of quasiparticles by studying the shape of the spectral function, and we aim to address this in future work.

In two-dimensional systems we clearly identified the plasmon mass scale \cite{Lappi:2017ckt}. However, its interpretation in terms of HTL and kinetic theory is less obvious. This points towards an interesting direction for future work: a similar study of spectral properties of the two-dimensional system. This could help us to understand to what extent the quasiparticle description is valid for two dimensional system.

We have been able to establish that our results on the spectral properties of overoccupied gluodynamics are in good overall agreement with leading order HTL perturbation theory. However, we expect this formalism to be valid also in situations in which HTL is no longer valid. Our future goals include to study spectral properties of overoccupied classical gluodynamical systems out of equilibrium in anisotropic geometry in the short term. It would also be interesting to study expanding geometries, which are more relevant in the framework of ultrarelativistic heavy-ion collisions. 

A linear response analysis of the plasma quasiparticle properties is not, however, the only application of our linearized fluctuation setup. We also expect to be able to study the time-evolution of unstable quantum fluctuations, which are expected to play a major role in ultrarelativistic heavy-ion collisions. They might also contribute significantly to the pressure isotropization process of the strongly interacting matter created in ultrarelativistic heavy-ion collisions \cite{Gelis:2013rba}.
\bibliography{../../../LatexBibliografia/Megabib.bib}

\end{document}